\documentclass{article}
\usepackage{graphicx} 
\usepackage[a4paper,
            bindingoffset=0.2in,
            left=0.5in,
            right=0.5in,
            top=1in,
            bottom=1in,
            footskip=.25in]{geometry}

\usepackage{subfigure} 
\usepackage{amsmath}
\usepackage[english]{babel}
\usepackage{amsfonts}
\usepackage{bbm}
\usepackage{shuffle}
\usepackage{amsthm, mathrsfs}
\usepackage{cancel}
\usepackage[dvipsnames]{xcolor}
\usepackage{varwidth}
\usepackage{enumitem}
\usepackage{mathtools}
\usepackage{ marvosym }
\usepackage{tcolorbox}
\usepackage{listings}
\usepackage[round]{natbib}
\usepackage{mathpazo}
\usepackage{tikz}
\usepackage{booktabs}
\usetikzlibrary{positioning}
\RequirePackage[colorlinks,citecolor=blue!70!green,urlcolor=blue, linkcolor=blue!70!green]{hyperref}

\numberwithin{equation}{section}

\hypersetup{
    colorlinks=true,
    linkcolor=blue,
    filecolor=magenta,      
    urlcolor=cyan,
    pdftitle={A joint framework for SPX, VIX and VXX},
    pdfpagemode=FullScreen,
    }
\theoremstyle{plain}
\newtheorem{theorem}{Theorem}[section]
\newtheorem{lemma}[theorem]{Lemma}
\newtheorem{corollary}[theorem]{Corollary}
\newtheorem{proposition}[theorem]{Proposition}

\theoremstyle{definition}
\newtheorem{definition}[theorem]{Definition}
\newtheorem{example}[theorem]{Example}

\newtheorem{remark}[theorem]{Remark}

\newcommand{\E}{\mathbb{E}}

\newcommand{\dd}[1]{\,{d}{#1}}

\newcommand{\mo}{\text{model}}
\newcommand{\wh}[1]
{\,\widehat{#1}}
\newcommand{\R}{\mathbb{R}}
\newcommand{\Fcal}{\mathcal{F}}
\newcommand{\Kcal}{\mathcal{K}}
\newcommand{\Lcal}{\mathcal{L}}

\newcommand{\Tcal}{\mathcal{T}}
\newcommand{\ve}{\mathrm{vec}}
\newcommand{\X}{\mathbb{X}}
\newcommand{\Y}{\mathbb{Y}}
\newcommand{\Q}{\mathbb Q}
\newcommand{\SPX}{\mathrm{SPX}}
\newcommand{\VIX}{\mathrm{VIX}}
\newcommand{\VXX}{\mathrm{VXX}}
\newcommand{\T}{\mathrm{T}}

\newcommand{\mkt}{\text{mkt}}

\title{A joint framework for SPX, VIX and VXX}
\author{%
  Martino Grasselli\thanks{%
    University of Padova,
    Department of Mathematics ``Tullio Levi-Civita'',
    Via Trieste 63, 35121 Padova, Italy, and De Vinci Higher Education, De Vinci Research Center, Paris, France.
    Email: grassell@math.unipd.it.
  }
  \and
  Andrea Stanghellini\thanks{%
    University of Verona, Department of Economics,
    Via Cantarane 24, 37129 Verona, Italy.
    Emails: andrea.stanghellini@univr.it (Andrea Stanghellini);
    sara.svalutoferro@univr.it (Sara Svaluto-Ferro).
  }
  \and
  Sara Svaluto-Ferro\footnotemark[2]
}

\begin{document}

\maketitle
\begin{abstract}
Options on VXX bring the dynamics of the VIX futures curve into the joint SPX--VIX calibration problem. We construct a common model by specifying the VIX and then recovering instantaneous SPX variance from the requirement that its risk-neutral conditional expected average over the next thirty days equals the squared VIX. With truncated signatures of time-extended polynomial diffusions, this recovery reduces to finite-dimensional linear algebra and gives explicit access to the futures underlying VXX. The same specification therefore governs equity variance and rolling volatility exposure. To accommodate flexible specifications while respecting variance nonnegativity, we develop a regularized model and quantify the pricing errors of its tractable VIX and VXX proxies. Applications to short-dated option smiles illustrate joint calibration across the three markets, with Monte Carlo estimates of the proxy error bounds well below quoted bid--ask spreads.
\end{abstract}



\tableofcontents

\section{Introduction}
Options on the S\&P 500 index ($\SPX$), the Cboe Volatility Index ($\VIX$), and the iPath S\&P 500 VIX Short-Term Futures ETN ($\VXX$) price different aspects of the same volatility risk. The VIX links equity options to risk-neutral expected future variance, while VXX turns exposure to the volatility futures curve into a traded investment product. Their option markets therefore impose distinct constraints on a common set of dynamics. A joint model must accommodate these constraints while preserving the relations that connect the equity index, the volatility index, and the rolling futures strategy.

The additional modeling content of VXX comes from the roll. The note tracks a strategy in the first two VIX futures contracts, with weights adjusted to maintain short-term exposure; see \cite{BarclaysVXX:2018}. Its evolution depends on the level and shape of the futures curve and on the joint dynamics of the contracts entering the strategy. Consequently, the information relevant to VXX options extends beyond the marginal VIX distributions at individual maturities. The empirical findings of \cite{Lu:23}, who documents that VXX option information is not fully spanned by VIX options, support this distinction. Incorporating VXX into a joint calibration thus makes the dynamics of rolling volatility exposure an additional part of the modeling problem.

Our approach places the tractable specification at the level of the VIX. We represent the index as a linear functional of the truncated signature of a time-extended polynomial diffusion and recover instantaneous SPX variance from the requirement that its risk-neutral conditional expected average over the next thirty days equals the squared VIX. The same signature coordinates determine the VIX futures entering the VXX roll. This common representation allows the implications of a volatility-index specification for equity variance and rolling volatility exposure to be studied within a single analytical framework.

The joint SPX--VIX calibration problem has generated a broad range of models, motivated in particular by the need to reconcile the pronounced downside skew of SPX options with the increasing implied-volatility smile of VIX options. Affine and jump specifications have been studied in \cite{CK:13,BB:14}, while \cite{PS:14} and \cite{FS:18} develop regime-switching and stochastic vol-of-vol models, respectively. Continuous models include quadratic rough Heston in \cite{GJR:20}, the quintic Ornstein--Uhlenbeck model in \cite{AJIL:22}, and the path-dependent specifications in \cite{GL:23}; \cite{GM:22} study neural calibration. A complementary line of work formulates calibration directly under consistency constraints, using optimal transport in \cite{GLOW:22} and dispersion-constrained martingale Schr\"odinger problems in \cite{Guyon:24}. These contributions provide different ways of matching the two option markets while respecting the relation between equity returns and future variance.

The literature on VIX derivatives and volatility-linked notes addresses a further part of this structure. Direct models for VIX futures and options provide tractable descriptions of the volatility-index market; see \cite{LC:10,MS:13}. The investment and trading properties of volatility exchange-traded products are studied in \cite{AlexanderKorovilas:13,AvellanedaPapanicolaou:17}. \cite{GW:20} derive an analytical link between VIX and VXX, and \cite{GMP:24} develop a framework for their joint option calibration. More recently, \cite{Lu:23} examines the role of volatility clustering, and \cite{Lu:25} uses a discrete-time analytical framework to incorporate information from high-frequency VIX data. These works show how the futures curve and the rolling rule shape VXX dynamics. The present paper embeds this relationship in a model that also retains the connection to SPX variance and options.

Recovering equity variance from prescribed volatility-index dynamics is itself a consistency problem. \cite{BG:15} derive restrictions for jointly modeling VIX futures and equity derivatives, and \cite{P:22} formulates and studies the inverse problem of recovering stochastic volatility from a VIX market model. The latter work identifies both solvability and nonnegativity of the recovered variance as central requirements. Our contribution is a finite-dimensional signature realization of this inverse construction, coupled with the VXX rolling strategy and quantitative control of the approximations used in calibration.

The mathematical structure comes from combining polynomial processes with path signatures. Affine models and their polynomial extensions offer tractable conditional moments and, in the affine case, transform methods; see \cite{DFS:03,CKT:12,FL:16}. Signatures,
rooted in Chen's work on iterated integrals
\cite{Chen:54} and subsequently developed within rough-path
theory \cite{Lyons:98,LCL:07,FV:10},
encode a path through its iterated integrals and provide
a systematic family of coordinates for path-dependent models. Their financial applications include pricing and hedging in \cite{LNA:20}, optimal stopping in \cite{BHR:23}, and model calibration in \cite{CGS:23}. The affine and polynomial structure of signature dynamics is developed in \cite{CST:23}. For polynomial primary diffusions, truncated signatures retain a finite-dimensional polynomial structure, and their conditional expectations can be computed by matrix exponentials.

Our starting point within this framework is the joint SPX--VIX signature model of \cite{CGMS:24}. When SPX volatility is a linear signature functional, the shuffle identity expresses its square in the same algebra, and conditional averaging gives an explicit linear representation of the squared VIX. The VIX itself is then obtained by taking a square root. This construction is financially consistent, but the square root generally prevents VIX futures from retaining the same finite linear form. An exact shuffle-square representation with a nonnegative factor, or an appropriate approximation, can recover that form; the additional step becomes especially relevant when futures must be evaluated along the VXX roll. This is the tractability issue that motivates specifying the VIX directly. The comparison with the quintic Ornstein--Uhlenbeck model of \cite{AJIL:22} is instructive: both approaches exploit nonlinear functions of tractable factors, whereas signatures also supply coordinates for path dependence. In a direct volatility specification, variance is nonnegative as a square; in the inverse specification, nonnegativity of the recovered variance must be addressed separately.

The main analytical gain is that the relations linking the three markets remain explicit as the VIX specification is enriched. The reconstruction theorem
(Theorem~3.14) reduces the recovery of SPX variance to a finite matrix inversion, and the spectral criterion in Lemma~3.17 characterizes when this inversion is possible. Under the nonnegativity conditions, the resulting model satisfies the VIX definition exactly and yields explicit VIX futures and a pathwise representation of the VXX roll. The leverage-saturation construction of Proposition~3.11 completes the equity specification by ensuring that the discounted SPX price is a true martingale without imposing a deterministic upper bound on volatility.

The approximation analysis extends the reach of this construction to specifications for which global nonnegativity is difficult to enforce. We construct a regularized equity model and bound the discrepancy between its tractable pricing proxies and the corresponding quantities in the regularized model. For VXX, the analysis follows the effect of futures-price errors through both the current futures basket and the accumulated roll. 
The resulting bounds for VIX and VXX call prices (Corollary~\ref{cor:option-price-errors}) are uniform in the strike and allow the pricing impact of the approximation to be assessed against market bid--ask spreads. This allows the flexibility of the signature specification to be retained together with quantitative control of the pricing consequences of regularization.

The same algebra also makes richer specifications computationally accessible. Building on the separation between simulated features and parameter optimization in \cite{CGMS:24}, conditional-expectation operators and path features are prepared in advance. Restricting the calculations to suitable generator-invariant subspaces preserves the required conditional-averaging identities exactly. In the enriched Ornstein--Uhlenbeck example, the required calculations can be performed in dimension 216 instead of using all 5461 coordinates of the complete level-six signature. Selected higher-order features can therefore be incorporated while retaining a tractable representation of the links among the three markets.

The short-maturity calibrations show the benefit of this flexibility. In the Ornstein--Uhlenbeck specification, adding two signature coordinates and jointly recalibrating the model steepens the VIX smile and reduces the underestimation of its upper wing while retaining the SPX downside skew and the increasing VXX profile. The calibrated prices are evaluated on independent Monte Carlo paths. For both this specification and the model combining Ornstein--Uhlenbeck and CIR factors, Monte Carlo estimates of the VIX and VXX option-price approximation bounds lie well below the corresponding quoted bid--ask spreads.

The paper is organized as follows. Section~\ref{Notation and basic properties} recalls the signature and polynomial-process tools. Section~\ref{The model} develops the volatility-based and VIX-based constructions, their VXX representations, and the regularized extension with approximation bounds. Section~\ref{Calibration to option prices} presents the pricing and calibration procedure and the numerical illustrations.

\section{Notation and basic properties}\label{Notation and basic properties}
\subsection{Tensor algebra and signature}
We start this section by introducing the basic notions needed to properly define the signature of an $\R^d$-valued semimartingale. Similar introductions can be found in \cite{CGMS:24}, \cite{CGS:23},  or \cite{BHR:23}.

Let $n,d\in\mathbb{N}$ with $n\neq 0$. We define recursively the $n$-fold tensor product of $\R^d$ as
$(\R^d)^{\otimes 0}:=\R$, and $(\R^d)^{\otimes n}:=(\R^d)^{\otimes (n-1)}\otimes \R^d.$
The extended tensor algebra of $\R^d$ is defined as
$$T\big((\R^d)\big):=\{{\bf a}=(a_0,\dots,a_n,\dots)\colon a_n\in(\R^d)^{\otimes n}\}.$$
Similarly, the truncated tensor algebra of order $n\in\mathbb{N}$ is given by
$$T^n(\R^d):=\{{\bf a}\in T\big((\R^d)\big)\colon a_k=0,\,\forall k>n\}.$$
We observe that the truncated tensor algebra of order $n$ of $\R^d$ has dimension $d_n=1+n$ for $d=1$,
\begin{equation}\label{d_n}
    d_n=1+d+\dots+d^n=\frac{d^{n+1}-1}{d-1},\qquad d\geq 2
\end{equation}
and we set
$T(\R^d):=\bigcup_{n\in\mathbb{N}}T^n(\R^d)$.
    Observe that the extended tensor algebra $T((\R^d))$ is a collection of infinite sequences, while the tensor algebra $T(\R^d)$ is a collection of sequences that are nonzero up to some level $m\in\mathbb{N}$.
    
For each ${\bf a},{\bf b}\in T\big((\R^d)\big)$, $\lambda\in\R$, and $n\in \mathbb N$ we define
$
        {\bf a}+{\bf b}:=(a_0+b_0,\ldots,a_n+b_n,\ldots),
        \lambda\cdot {\bf a}:=(\lambda a_0,\ldots,\lambda a_n,\ldots),$ ${\bf a}^{\leq n}:=(a_0,\ldots, a_n)$, and  $
        {\bf a}\otimes {\bf b}:=(c_0,\ldots,c_n,\ldots)$ for $
        c_n=\sum_{k=0}^na_k\otimes b_{n-k}.
$
Observe that
    $(T\big((\R^d)\big),+,\cdot,\otimes)$ is a real, non-commutative algebra.

We proceed by introducing multi-indices to facilitate a more effective representation of the subsequent operations. Fix a multi-index $I=(i_1,\ldots,i_n)$ with $i_j\in\{1,\ldots,d\}$. We define the order (i.e., the length) of the multi-index by $|I|=n$, and we set $|J|=0$ for the empty multi-index $J=\emptyset$. We also define the truncated multi-indices $I'=(i_1,\ldots,i_{n-1})$ and $I''=(i_1,\ldots,i_{n-2})$, for $n\geq 1$ and $n\geq 2$, respectively. We also set $e_{I'}=0$ and $e_{I''}=0$ for each $I$ with $|I|=0$ and $|I|\leq1$, respectively.

For a multi-index $I$, we set
$e_I:=e_{i_1}\otimes\dots\otimes e_{i_n},$
where $e_{i_j}$ is the $i_j$-th vector of the canonical basis of $\R^d$; this implies that $\{e_I:|I|=n\}$ is an orthonormal basis of $(\R^d)^{\otimes n}$. Denoting by $e_\emptyset$ the basis element corresponding to $(\R^d)^{\otimes 0}$, we derive a coordinate representation for all elements ${\bf a}\in T((\R^d))$ as
$${\bf a}=\sum_{|I|\geq 0}a_Ie_I,\qquad a_I\in\R.$$
For each ${\bf a}\in T(\R^d)$ and ${\bf b}\in T\big((\R^d)\big)$ we set
$\langle {\bf a},{\bf b}\rangle:=\sum_{|I|\geq 0} a_Ib_I.$ This in particular implies that $a_I=\langle e_I,{\bf a}\rangle$.

Consider a filtered probability space $(\Omega,\mathcal{F},(\mathcal{F}_t)_{t\geq 0},\mathbb{P})$. We now give the definition of the signature of an $\R^d$-valued continuous semimartingale.
\begin{definition}
    Let $(X_t)_{t\geq 0}$ be a continuous $\R^d$-valued semimartingale. 
    The signature of $X$ is the $T((\R^d))$-valued process 
    $(s,t)\mapsto \X_{s,t}$ whose components are recursively defined by
    \[
        \langle e_\emptyset,\X_{s,t}\rangle := 1,
        \qquad
        \langle e_I,\X_{s,t}\rangle
        = \int_s^t \langle e_{I'},\X_{s,u}\rangle \circ \dd{X}^{i_n}_u,
    \]
    for every multi-index $I=(i_1,\ldots,i_n)$ with $n\geq 1$, where
     $\circ$ denotes Stratonovich integration.
    When $s=0$, we write $\X_t:=\X_{0,t}$. We denote the signature truncated
    at level $n$ by $\X_{s,t}^{\leq n}$, and, when $s=0$, by $\X_t^{\leq n}$.

    Finally, we denote by $\wh{\X}_{s,t}$ and $\wh{\X}_{s,t}^{\leq n}$,
    respectively, the signature and the signature truncated at level $n$ of
    the time extension $(\wh{X}_t)_{t\geq 0}$ of $X$, defined by
    \[
        \wh{X}_t^0 := t,
        \qquad
        \wh{X}_t^i := X_t^i,
        \quad i=1,\ldots,d.
    \]
\end{definition}
    Note that the signatures of $X$ and of $X-c$ coincide for all $c\in\R^d$.
    Using the recursive formula, one can represent the signature of $X$ by
    $$\X_t=\bigg(\underbrace{1}_{\text{level 0}},\underbrace{\int_0^t1\circ\dd{X}_s^1,\ldots,\int_0^t1\circ\dd{X}_s^d}_{\text{level 1}},\underbrace{\int_0^t\bigg(\int_0^s 1\circ\dd{X}_u^1\bigg)\circ\dd{X}_s^1,\ldots,\int_0^t\bigg(\int_0^s 1\circ\dd{X}_u^d\bigg)\circ\dd{X}_s^d}_{\text{level 2}},\ldots\bigg).$$
A well-known and extremely useful property of the signature is that every polynomial function in the signature has a linear representation. For the precise statement we first
need to introduce the following concept (see Definition~I.2.12 in \cite{sigbook} or also Definition 2.4 in \cite{LNA:20}).

For $n,m\in\mathbb{N}$, we let  $S^{n,m}$ be the set of all permutations $\sigma\in S_{n+m}$ that respect the order of the first $n$ elements and of the last $m$ elements:
    \begin{equation}\label{shuffleset}
        \begin{aligned}
            S^{n,m}=\{\sigma\in S_{n+m}\colon\sigma(1)<\cdots<\sigma(n)\,\text{and }\sigma(n+1)<\cdots<\sigma(n+m)\}.
        \end{aligned}
    \end{equation}
    For a multi-index $I$ set \(
\sigma(I):=
(i_{\sigma(1)},\ldots,i_{\sigma(n)}).
\)

\begin{definition}
    We define the \emph{shuffle product} $\shuffle : T(\R^d)\times T(\R^d)\to T(\R^d)$ as the bilinear product determined by $e_\emptyset \shuffle e_I= e_I \shuffle e_\emptyset= e_I$
    and
    $$e_I\shuffle e_J=\sum_{\sigma\in S^{|I|,|J|}}e_{\sigma^{-1}(IJ)},$$
    for every $I$ and $J$. Consequently, 
    $${\bf a}\shuffle {\bf b}=\sum_{|I|,|J|\geq 0}a_I b_J\,(e_I\shuffle e_J).$$
    Equipped with the shuffle product, $T(\R^d)$ is a commutative  algebra.
\end{definition}
The importance of the shuffle product is due to the following property of the signature.
\begin{proposition}
    Let $(X_t)_{t\geq 0}$ be a continuous $\R^d$-valued semimartingale. Then $\langle e_I,\X_t\rangle\langle e_J,\X_t\rangle=\langle e_I\shuffle e_J,\X_t\rangle$ for any two multi-indices $I,J$.
\end{proposition}

\subsection{Polynomial processes}
Motivated by the need to reconcile analytical tractability with modeling flexibility in mathematical finance, polynomial processes—Markov processes for which the conditional expectation of any polynomial in the state variable remains a polynomial of at most the same  degree—were rigorously characterized by \cite{CKT:12} and subsequently unified within a comprehensive diffusion framework by \cite{FL:16}.

    Let $(X_t)_{t\geq 0}$ be a weak solution of the SDE
\begin{equation}\label{Polynomial diffusion}
    \dd{X}_t=b(X_t)\dd{t}+\sqrt{a(X_t)}\dd{W}_t,\hspace{1cm} X_0=x_0
\end{equation}
where $W_t$ is a $d$-dimensional Brownian motion and $b:\R^d\to\R^d$ and $a:\R^d\to \mathbb S^d$ are polynomials of degree at most 1 and 2, respectively.
 By Lemma~4.1 and Lemma~4.2 in \cite{CGMS:24}, the coefficients admit the representation
$b_j(X_t)=\langle\mathbf{b_j},\X_t^{\leq 1}\rangle$ and $a_{ij} (X_t)=\langle\mathbf{a_{ij}},\X_t^{\leq 2}\rangle$ for some explicitly constructed $\mathbf{b_j}\in T^1(\R^d)$ and $\mathbf{a_{ij}}\in T^2(\R^d)$.
Moreover, the truncated signature $(\X^{\leq n}_t)_{t\geq 0}$ is still a polynomial diffusion, and for each $|I|\leq n$  we have
\begin{equation}\label{sig-dynamics}
    \langle e_I,\X_t^{\leq n}\rangle= 1_{\{I=\emptyset\}}+\int_0^t\langle Le_I,\X_s^{\leq n}\rangle\dd{s}+\int_0^t\langle e_{I'},\X^{\leq n}_s\rangle\sigma_{i_{|I|}\cdot}(X_s)\dd{W}_s,
\end{equation}
where $\sigma(X_t)$ denotes the matrix square root of $a(X_t)$ and $L:T(\R^d)\to T(\R^d)$, called the dual operator corresponding to $\X$, satisfies $L(T^n(\R^d))\subseteq T^n(\R^d)$ and is given by
\begin{equation}\label{L-operator}
    Le_I=e_{I'}\shuffle\mathbf{b}_{i_{|I|}}+\frac{1}{2}e_{I''}\shuffle {\mathbf a}_{i_{|I|-1}i_{|I|}}.
\end{equation}
Fix an injective labeling function $\mathcal{L}:\{I:|I|\leq n\}\to \{1,\ldots,d_n\}$. For ${\bf u}\in T^n(\R^d)$, set
$$\ve({\bf u})=\sum_{|I|\leq n}e_{\mathcal L (I)} u_I\in \R^{d_n}.$$ 
Let $G$ be the matrix representation with respect to this basis order, i.e., set
$$
 G_{\mathcal{L}(I),\mathcal{L}(J)}:=\eta_{JI}
\qquad \text{ for $\eta$ such that }\qquad Le_I=\sum_{|J|\leq n}\eta_{IJ}e_J.
 $$
 Observe that for each  ${\bf u}\in T^n(\R^d)$ the matrix representative $G$ of ${L}$ satisfies 
$\ve(L{\bf u})=G\ve({\bf u})$.
Finally, we  report here  Theorem~4.4 in \cite{CGMS:24}, providing a representation of the expected signature of a polynomial process.
\begin{theorem}\label{thm1}
    Let $(Y_t)_{t\in [0,\overline{T}]}$ be a polynomial diffusion of the form \eqref{Polynomial diffusion}, denote by $\Fcal$ its natural filtration, and let $G$ be the $d_n\times d_n$ matrix representative of the dual operator corresponding to $\Y$. Then, for each $T,t\geq 0$ such that $T+t \leq \overline{T}$, we have
    \begin{equation}
        \E[\ve(\Y^{\leq n}_{T+t})|\mathcal{F}_T]=e^{tG^\mathrm{T}}\ve(\Y^{\leq n}_T)
    \end{equation}
    or equivalently, for each $|I|\leq n$
\begin{equation}\label{expected signature}
        \E[\langle e_I,\Y^{\leq n}_{T+t}\rangle|\mathcal{F}_T]=\sum_{|J|\leq n} \Big(e^{tG^\mathrm{T}}\Big)_{\mathcal{L}(I),\mathcal{L}(J)}\langle e_J,\Y^{\leq n}_T\rangle.
    \end{equation}
\end{theorem}

\section{The model}\label{The model}
Suppose that the dynamics of the discounted, dividend-adjusted price process of the S\&P 500 index under the risk-neutral measure are given by
\begin{equation}
    \label{Asset dynamic} dS_t=S_t\sigma_t^S d B_t
\end{equation}
for a Brownian motion $B$ and a stochastic process $\sigma^S$. This, in particular, implies that
$$S_t=S_0\exp\bigg(\int_0^t\sigma_u^S\dd{B}_u-\frac{1}{2}\int_0^t(\sigma_u^S)^2\dd{u}\bigg).$$
Following \cite{CW:06} (see also \cite{CGMS:24}), we define the VIX index in terms of the risk-neutral expected integrated variance as follows.
\begin{definition}\label{def1}
The value of the $\VIX$ index (expressed in decimal form, i.e., divided by \(100\)) at time $t$ is given by
\begin{equation}\label{VIX}
    \VIX_t:=\sqrt{\E_\Q\bigg[\frac{1}{\Delta}\int_t^{t+\Delta}(\sigma_u^S)^2\dd{u}|\mathcal{F}_t\bigg]}.
\end{equation}
Time is measured in years and $\Delta = 30/365$. Consistently, the futures price at time $t$ of the $\VIX$ with maturity $T$ is given by 
\begin{equation}\label{futures}
    F_{t}(T):=\E_\Q[\VIX_{T}|\mathcal{F}_t].
\end{equation}
\end{definition}

We now introduce the dynamics of the $\VXX$ following the pricing supplement of the VXX Prospectus of Barclays. For reference, see (8) in \cite{GW:20}, with a minor variation in the definition of $\lambda_t$, or Section 4.1 in \cite{GMP:24}. Recall that VIX futures expire once per month.
\begin{definition}\label{def_vxx}
Let $T_{-1}<T_0\le0<T_1<T_2<\cdots$ denote consecutive VIX futures expiration dates. We assume that $\sum_{i\ge1}\lambda_t^iF_t(T_i)>0$ throughout the pricing horizon. If time zero is a roll date, initial values are understood as right limits. Then, we set
$$\lambda_t^i= \min\left( \frac{t - T_{i-2}}{T_{i-1}-T_{i-2}},\frac{T_i - t}{T_i-T_{i-1}} \right)^+,\qquad i=1,2,3,...$$
The dynamics of $\VXX$, discounted and excluding fees, under the risk-neutral measure $\mathbb{Q}$ are then given by
$$\frac{d \VXX_t}{\VXX_t}=\frac{\sum_{i=1}^\infty\lambda_t^idF_{t}(T_i)}{\sum_{i=1}^\infty\lambda_t^iF_{t}(T_i)}
=\sum_{i=1}^\infty
\frac{( t-T_{i-1})dF_{t}(T_{i+1})
+(T_i - t)dF_{t}(T_i)}
{(t-T_{i-1})F_{t}(T_{i+1})
+(T_i - t)F_{t}(T_i)}
\mathbbm{1}_{t\in(T_{i-1},T_i]},$$
where $F_t(T)$ is given by \eqref{futures}.
\end{definition}

Observe that for each fixed $t$, the coefficient $\lambda_t^i$ is nonzero for at most two values of $i$. Before proceeding with the main theorem of this section, we need a technical lemma.

\begin{lemma}\label{lemVXX}
Let $X$ and $Y$ be continuous $d$-dimensional semimartingales, with $Y$ of finite variation, and assume that $Y_t^\T X_t>0$ on the time interval considered. Write $X=X^M+X^{FV}$, where $X^M$ is a continuous local martingale and $X^{FV}$ is of finite variation.
     Then
     $$\frac {dZ_s}{Z_s}=\frac{Y_s^\T}{Y_s^\T X_s}dX_s^M,\qquad \text{if and only if }\qquad {Z_t}={Z_0}\frac{Y_t^\T X_t}{Y_0^\T X_0}\exp\Big(-\int_0^t \frac{X_s^\T dY_s+Y_s^\T dX_s^{FV}}{Y_s^\T X_s}\Big).$$
\end{lemma}
\begin{proof}
By It\^o's formula, we know that
$$Y_t^\T X_t=Y_0^\T X_0+\int_0^t Y_s^\T \dd{X}_s^M+\int_0^t Y_s^\T \dd{X}_s^{FV}+\int_0^t X_s^\T \dd{Y}_s.$$
$(\Longrightarrow)$ Set $R_t=\int_0^t\frac {d(Y_s^\T X_s)}{Y_s^\T X_s}$ and note that $Y_t^\T X_t=Y_0^\T X_0\mathcal E(R)_t$, where $\mathcal E$ denotes the stochastic exponential
$$\mathcal E(S)_t=\exp(S_t-\frac 1 2 [S]_t).$$
We can then compute   
\begin{align*}
    \frac{Z_t}{Z_0}&=\mathcal E\Big(\int_0^\cdot \frac{Y_s^\T}{Y_s^\T X_s}dX_s^M\Big)_t=\mathcal E\Big(R-\int_0^\cdot \frac{Y_s^\T}{Y_s^\T X_s} \dd{X}_s^{FV}-\int_0^\cdot \frac{X_s^\T}{Y_s^\T X_s} \dd{Y}_s\Big)_t\\
    &=\mathcal E(R)_t\mathcal E\Big(-\int_0^\cdot \frac{Y_s^\T}{Y_s^\T X_s} \dd{X}_s^{FV}-\int_0^\cdot \frac{X_s^\T }{Y_s^\T X_s}\dd{Y}_s\Big)_t=\frac{Y_t^\T X_t}{Y_0^\T X_0}\exp\Big(-\int_0^\cdot \frac{Y_s^\T}{Y_s^\T X_s} \dd{X}_s^{FV}-\int_0^\cdot \frac{X_s^\T}{Y_s^\T X_s} \dd{Y}_s\Big)_t,
\end{align*}
proving the claim.\\
$(\Longleftarrow)$ Applying  It\^o to $Z_t=\frac{Z_0}{Y_0^\T X_0}\,(Y_t^\T X_t)\exp\!\left(-\int_0^t\frac{X_s^\T dY_s+Y_s^\T dX_s^{FV}}{Y_s^\T X_s}\right)$, we obtain
$$\begin{aligned}
  \frac{Y_0^\T X_0}{Z_0} {\dd{Z}_t}&=\dd{(Y^\T_tX_t)}\exp\bigg(-\int_0^t \frac{X_s^\T \dd{Y}_s+Y_s^\T \dd{X}_s^{FV}}{Y_s^\T X_s}\bigg)+Y^\T_tX_t\dd{\bigg(\exp\bigg(-\int_0^t \frac{X_s^\T \dd{Y}_s+Y_s^\T\dd{X}_s^{FV}}{Y_s^\T X_s}\bigg)\bigg)}\\
    &=\frac{Y_0^\T X_0}{Z_0}Z_t\bigg(\dfrac{\dd{(Y^\T_tX_t)}}{Y_t^\T X_t}-\frac{X_t^\T \dd{Y}_t+Y_t^\T \dd{X}_t^{FV}}{Y_t^\T X_t}\bigg)=\frac{Y_0^\T X_0}{Z_0}Z_t\dfrac{Y_t^\T}{Y_t^\T X_t}\dd{X}_t^M,
\end{aligned}$$
concluding the proof.
\end{proof}

\begin{proposition}\label{corVXX}
Let $F_t(T)$ denote the futures price at time $t$ of the $\VIX$ with maturity $T$ as in \eqref{futures}.
Set 
$$P_t:=\sum_{i=1}^\infty
\Big((t-T_{i-1})F_{t}(T_{i+1})
+(T_i - t)F_{t}(T_i)\Big)
\mathbbm{1}_{t\in(T_{i-1},T_i]}.$$
Then, the $\VXX$-dynamics as in Definition~\ref{def_vxx} satisfy
    $$\VXX_t={\VXX_0}\frac{P_t\tau_0}{P_0\tau_t}
    \times \exp\Big(\sum_{i=1}^\infty\int_{(t\land T_{i-1})\lor 0}^{t\land T_i} \frac{ F_s(T_i) - F_s(T_{i+1}) }{P_s}ds\Big),$$
      where $\tau_t=\sum_{i=1}^\infty (T_i-T_{i-1})\mathbbm{1}_{t\in(T_{i-1},T_i]}$.
\end{proposition}
\begin{proof}
    Fix a finite time horizon and retain the finite collection of futures contracts that can enter the roll before that horizon. We then apply Lemma~\ref{lemVXX} to $Y_t=(\lambda_t^1,\lambda_t^2,\ldots)$, $X_t=(F_t(T_1),F_t(T_2),\ldots)$, and $Z_t=\VXX_t$, noticing that $X$ is a local martingale by definition. We would indeed get that 
    \begin{align*}
        Y_t^\T X_t&=\sum_{i=1}^\infty\frac{(t-T_{i-1})F_{t}(T_{i+1})
+(T_i - t)F_{t}(T_i)}{T_i-T_{i-1}}
\mathbbm{1}_{t\in(T_{i-1},T_i]}=\sum_{i=1}^\infty\frac{P_t}{T_i-T_{i-1}}
\mathbbm{1}_{t\in(T_{i-1},T_i]}\\
X_s^\T dY_s&= \sum_{i=1}^\infty\frac{F_{s}(T_{i+1})
-F_{s}(T_i)}{T_i-T_{i-1}}
\mathbbm{1}_{s\in(T_{i-1},T_i]}ds
    \end{align*} and hence
    $$\VXX_t\frac{Y_0^\T X_0}{\VXX_0}=(Y_t^\T X_t)\exp\Big(-\int_0^t \frac{X_s^\T dY_s}{Y_s^\T X_s}\Big)=(Y_t^\T X_t)\exp\Big(-\int_0^t \sum_{i=1}^\infty\frac{{F_{s}(T_{i+1})
-F_{s}(T_i)}}{P_s}\mathbbm{1}_{s\in(T_{i-1},T_i]}ds\Big).$$
\end{proof}
Finally, for the reader's convenience, we investigate the provided representation for $T\leq T_2$.
The following two representations of the VXX dynamics in Definition 3.2 hold for $0\le T\le T_1$ and $T_1<T\le T_2$, respectively.
\begin{corollary}\label{cormatvxx}
 Consider the setting of Proposition~\ref{corVXX} and assume that $T_i-T_{i-1}=\tau$ for each $i$. 
    The $\VXX$-dynamics as in Definition~\ref{def_vxx} for $T\leq T_1$ and $T\in (T_1,T_2]$ satisfy
$$\VXX_T=\VXX_0\;
\frac{(T-T_0)\,F_T(T_2)+(T_1-T)\,F_T(T_1)}
{-T_0\,F_0(T_2)+T_1\,F_0(T_1)}
\;\exp\!\left(\int_0^T\frac{F_s(T_1)-F_s(T_2)}{(s-T_0)F_s(T_2)+(T_1-s)F_s(T_1)}\,ds\right),$$
and
\begin{align*}
    \VXX_T&=\VXX_0\;
\frac{(T-T_1)\,F_T(T_3)+(T_2-T)\,F_T(T_2)}
{-T_0\,F_0(T_2)+T_1\,F_0(T_1)}
\;\exp\!\left(\int_0^{T_1}\frac{F_s(T_1)-F_s(T_2)}{(s-T_0)F_s(T_2)+(T_1-s)F_s(T_1)}\,ds\right)\\
&\qquad\times\;\exp\!\left(\int_{T_1}^{T}\frac{F_s(T_2)-F_s(T_3)}{(s-T_1)F_s(T_3)+(T_2-s)F_s(T_2)}\,ds\right),
\end{align*}
respectively.
\end{corollary}

\subsection{Volatility-based signature models}
\cite{CGMS:24} proposed a model where both $\sigma^S$ and $\VIX^2$ can be represented as a linear combination of the time-extended signature of a polynomial process. As this is the starting point for the present model, we report the main constructions for the reader's convenience.
Let $(X_t)_{t\geq 0}$ be a polynomial diffusion with dynamics
$$\dd{X}_t=b(X_t)\dd{t}+\sqrt{a(X_t)}\dd{W}_t,\hspace{1cm} X_0=x_0$$
as in \eqref{Polynomial diffusion} and denote by $\widehat \X_t$ its time-extended signature at time $t$. 
We consider the following model for a stock price $(S_t)_{t\in [0,\overline T]}$ under the risk-neutral measure $\mathbb Q$.
\begin{definition}\label{volbased}
We say that a stochastic process $(S_t)_{t\in [0,\overline T]}$ follows a \emph{volatility-based signature model} if
    \begin{equation}\label{vol-based dyn}
        \begin{aligned}
            dS_t=S_t\sigma_t ^S(\ell)\dd{B}_t\qquad \text{for}\qquad \sigma_t^S(\ell)&=\langle\ell,\wh{\X}^{\leq n}_t\rangle
        \end{aligned}
    \end{equation}
    and thus
    $$S_T=S_0\exp\bigg(\int_0^T\langle\ell,\wh{\X}^{\leq n}_t\rangle\dd{B}_t-\dfrac{1}{2}\int_0^T\langle\ell,\wh{\X}^{\leq n}_t\rangle^2\dd{t}\bigg),$$
    for some $\ell\in T^{n}(\R^{d+1})$.
\end{definition}
 We propose here an adaptation of Theorem~5.1 in \cite{CGMS:24}. Set $d_n$ as in \eqref{d_n}
and fix  an injective labeling function $\mathcal{L}:\{I:|I|\leq 2n\}\to\{1,\ldots,(d+1)_{2n}\}$.
\begin{theorem}\label{thm2}
    Let $(S_t)_{t\in[0,\overline T]}$ be a volatility-based signature model satisfying  \eqref{vol-based dyn}. Let $G$ be the $(d+1)_{2n}$-dimensional matrix representative of the dual operator corresponding to $\wh{\X}^{\leq 2n}$. Then,
    \begin{equation}
        \E_\Q\bigg[\int_0^t (\sigma_s^S(\ell))^2\dd{s}\bigg]<\infty
    \end{equation}
    holds for every $t\geq 0$ and 
    \begin{equation}
        \VIX_t(\ell)=\sqrt{\frac{1}{\Delta}\ve(\ell)^\mathrm{T}Q(t,\Delta)\ve(\ell)},
    \end{equation}
    for each $t\in [0,\overline T-\Delta]$, where
    \begin{equation}
        Q_{\mathcal{L}(I),\mathcal{L}(J)}(t,\Delta)=\int_0^{\Delta}\ve(e_I\shuffle e_J)^\T e^{sG^\T}\ve(\wh{\X}^{\leq 2n}_t)\dd{s}=\ve(e_I\shuffle e_J)^\T A\ve(\wh{\X}^{\leq 2n}_t)
        \end{equation}
for $A:=\int_0^\Delta e^{sG^\T}\dd{s}$.
    
\end{theorem}

\begin{remark}
    In order to compute the dynamics of $\VIX$ futures, and thus of the $\VXX$, through the explicit representation of the expected signature we would need to write the $\VIX$ process (and not just its square) as a linear combination of the components of $\widehat \X$.  
The formula for $\VIX$ is simple but, at this stage, does not allow us to derive explicit formulas to compute the dynamics of $\VXX$. We observe that if we are able to write
$\VIX^2_T=\langle\ell\shuffle\ell,\wh{\X}^{\leq 2n}_T\rangle$ for some $\ell\in T^n(\R^{d+1})$, we would get $\VIX^2_T=\langle\ell,\wh{\X}^{\leq n}_T\rangle\langle\ell,\wh{\X}^{\leq n}_T\rangle$ and thus $\VIX_T=|\langle\ell,\wh{\X}^{\leq n}_T\rangle |$.
\end{remark}

\begin{theorem}\label{thm3}
    Let $(S_t)_{t\in [0,\overline T]}$ follow a volatility-based signature model, fix an injective labeling function $\mathcal{L}:\{I:|I|\leq 2n\}\to\{1,\ldots,(d+1)_{2n}\}$, and let $G$ be the $(d+1)_{2n}$-dimensional matrix representative of the dual operator corresponding to $\wh{\X}^{\leq 2n}$. Then, for each $t\in [0,\overline T-\Delta]$
     \begin{equation}
        \begin{aligned}
            \VIX_t(\ell)&=\sqrt{\langle h_{\VIX^2}(\ell),\wh{\X}_t^{\leq 2n}\rangle}
            \qquad\text{for}\qquad
            \ve(h_{\VIX^2}(\ell))^\T=\dfrac{1}{\Delta}\ve(\ell\shuffle\ell)^\T A,
                    \end{aligned}
    \end{equation}
    with $A:=\int_0^\Delta e^{sG^\T}\dd{s}$.
Choosing $h_\VIX(\ell)$ such that $|\VIX_T(\ell)-\langle h_\VIX(\ell),\wh{\X}_T^{\leq n}\rangle|\leq \eta_T(\ell)$ for some random variable $\eta_T(\ell)$, we also get
   $$|F_t(T, \ell)-\langle h_{F,T-t}(\ell),\wh{\X}^{\leq n}_t\rangle|\leq \E_\Q[\eta_T(\ell)|\Fcal_t                                  ]\qquad\text{for}\qquad
            \ve(h_{F,\delta}(\ell))^\T=\ve(h_\VIX(\ell))^\T e^{\delta G^\T},$$
            for each $t\leq T\leq \overline T-\Delta$.
\end{theorem}
\begin{proof}
    The representations of $\VIX_t(\ell)$ and $F_t(T, \ell)$ follow from Theorem~\ref{thm2} and Theorem~\ref{thm1}, respectively. The bound follows from Lemma~\ref{errFut} below.
\end{proof}

To construct $h_{\VIX}(\ell)$ so as to obtain an accurate approximation,
\[
\VIX_t(\ell)
=
\sqrt{\langle h_{\VIX^2}(\ell),\widehat{\X}_t^{\leq 2n}\rangle}
\approx
\langle h_{\VIX}(\ell),\widehat{\X}_t^{\leq n}\rangle,
\]
we considered two alternative strategies.

First, we exploit the fact that, for every
$h\in T((\R^d))$ with $h_{\emptyset}>0$, 
there exists $\ell\in T((\R^d))$ such that
$
\ell\shuffle\ell=h.
$
See Appendix~\ref{srshuffle} for further details. However, imposing the
nonnegativity constraint
$
\langle h_{\VIX}(\ell),\widehat{\X}_t^{\leq n}\rangle\geq 0
$
in this setting is nontrivial.
Alternatively, one can choose $h_{\VIX}(\ell)$ to minimize the $L^2$ error
between
\[
\langle h_{\VIX^2}(\ell),\widehat{\X}_t^{\leq 2n}\rangle
\quad\text{and}\quad
\langle h_{\VIX}(\ell),\widehat{\X}_t^{\leq n}\rangle^2.
\]
This approach is effective for computing $\VXX$ prices for a fixed set of
parameters $\ell$. However, it is too computationally demanding for the
joint calibration of all three surfaces.

Alternatively, one can consider an approach based on the Chen relation, which guarantees that, for each $t\leq T_1$,
$$\langle h_{\VIX^2}(\ell),\widehat \X_{T_1}^{\leq 2n}\rangle=\sum_I \sum_{e_{I_1}\otimes e_{I_2}=e_I}h_{\VIX^2}(\ell)_I\langle e_{I_1},\widehat \X^{\leq 2n}_{t}\rangle\langle e_{I_2},\widehat \X^{\leq 2n}_{t,T_1}\rangle.$$
Assuming, for example, that $X$ is a Lévy process and thus has independent and stationary increments, we can then write
$$\E_\Q[\sqrt{\VIX_{T_1}(\ell)^2}|\widehat\X^{\leq 2n}_t={\bf x}]=u({\bf x},T_1-t,\ell)$$
for 
$$u({\bf x},T_1-t,\ell)=\E_\Q[\sqrt{\sum_I \sum_{e_{I_1}\otimes e_{I_2}=e_I}h_{\VIX^2}(\ell)_I\langle e_{I_1},{\bf x}\rangle\langle e_{I_2},\widehat \X^{\leq 2n}_{T_1-t}\rangle}].$$
Approximating the map $u$ (for example with a neural network) on a reasonable domain would lead  to an alternative modeling framework. Recall that in the model of \cite{CGMS:24}, the authors chose a 3-dimensional process as the primary process, resulting in an 85-dimensional time-extended signature. Even when exploiting the polynomial structure of the truncated signature and removing deterministic terms, this still requires handling a regression variable \({\bf x}\) of dimension 7. This dimension should then be added to the dimension of $\ell$, which is $85$.

\subsection{$\VIX$-based signature model}
This approach is the converse of the previous one: we start by defining the VIX as a linear combination of the signature  
$
    \VIX_t(\ell)=\langle\ell,\wh{\X}_t^{\leq n}\rangle.
$
We consider the following model.
\begin{definition}\label{defvixbase}
We say that $(S_t)_{t\in [0,\overline T]}$ follows a $\VIX$-based signature model if
    \begin{equation}
            dS_t=S_t\sigma_t ^S(\ell)\dd{B}_t\qquad \text{for}\qquad \sigma_t^S(\ell)^2=\langle h(\ell),\wh{\X}^{\leq 2n}_t\rangle
    \end{equation}
for some   $h(\ell)\in T^{2n}(\R^{d+1})$    and $\VIX_t(\ell)$, defined according to Definition~\ref{def1}, satisfies
    \begin{equation}\label{VIX model 2 definition}
    \VIX_t(\ell)=\langle \ell,\widehat \X^{\leq n}_t\rangle,
    \end{equation}
    for each $t\in [0, \overline T-\Delta]$ for some $\ell\in T^n(\R^{d+1})$.
    In this case we thus get
    $$S_T=S_0\exp\bigg(\int_0^T\sqrt{\langle h(\ell),\wh{\X}^{\leq 2n}_t\rangle}\dd{B}_t-\dfrac{1}{2}\int_0^T\langle h(\ell),\wh{\X}^{\leq 2n}_t\rangle\dd{t}\bigg).$$
\end{definition}

For pricing purposes, it is essential to verify that the price process
\((S_t)_{t\in[0,\overline T]}\) is a true martingale, rather than merely a
local martingale. Martingality criteria for stochastic volatility models
have been established in several settings; see, for instance,
\cite{AGS:25} for an example in the signature setting. Here, we adopt a constructive approach by introducing a
\emph{leverage saturation mechanism}, which attenuates the correlation
between the price and the volatility drivers at high volatility levels.
The role of correlation in martingality and moment bounds is studied
in \cite{LM:07}, while a state-dependent correlation is used in
\cite{FS:23} for the joint calibration of short-maturity SPX and VIX
smiles. A related attenuation appears in the Jacobi stochastic
volatility model of \cite{AFP:18}, where the effective correlation
vanishes as the variance approaches its upper boundary.
The following proposition shows that leverage saturation guarantees
true martingality by bounding the volatility loading on the shared
Brownian drivers, without imposing a deterministic upper bound on
the volatility process.

\begin{proposition}
\label{prop:true-martingale}
Let \(W=(W^1,\ldots,W^d)\) be a vector of correlated Brownian motions with
covariance matrix \(R\), that is,
$
d[W]_t=R\,dt,
$
where \(R\) is positive definite. Consider a continuous process
\((\sigma_t)_{t\in[0,\overline T]}\), adapted to the filtration generated by
\(W\), and let \(W^\perp\) be a Brownian motion independent of \(W\).
Assume that
\[
\int_0^{\overline T}\sigma_t^2\,dt<\infty
\qquad\text{a.s.}
\]
Fix \(\rho\in\mathbb R^d\) such that
$
\rho^\T R^{-1}\rho\leq 1.
$
For some \(L>0\), define
\[
\chi_L(x):=
\begin{cases}
1, & 0\leq x\leq L,\\[1mm]
L/x, & x>L,
\end{cases}
\]
and set
\begin{equation}
\label{eq:martingale-preserving-correlation}
B_t^{L}
=
\int_0^t
\chi_L(|\sigma_s|)\rho^\T R^{-1}dW_s
+
\int_0^t
\sqrt{1-\chi_L(|\sigma_s|)^2\rho^\T R^{-1}\rho}dW_s^\perp.
\end{equation}
 Then \(B^L\) is a standard Brownian motion, and the process
\[
S_t^{L}
=
S_0
\exp\left(
\int_0^t\sigma_s\,dB_s^{L}
-\frac12\int_0^t\sigma_s^2\,ds
\right)
\]
is a strictly positive true martingale on \([0,\overline T]\). In particular,
$
\mathbb E_{\mathbb Q}[S_t^{L}]=S_0,
\qquad 0\leq t\leq\overline T.
$
\end{proposition}

\begin{proof}
Since \(W^\perp\) is independent of \(W\), the quadratic variation of
\(B^L\) is
\begin{align*}
d[ B^L]_t
&=
\chi_L(|\sigma_t|)^2
\rho^\T R^{-1} R R^{-1}\rho\,dt
+
\left(1-\chi_L(|\sigma_t|)^2\rho^\T R^{-1}\rho\right)dt=dt.
\end{align*}
Therefore, Lévy's characterization implies that \(B^L\) is a standard
Brownian motion. Moreover,
$
d[ B^L,W^i]_t
=
\chi_L(|\sigma_t|)\rho_idt.
$

Define the two local martingales
\[
M_t^W
:=
\int_0^t
\sigma_s\chi_L(|\sigma_s|)
\rho^\T R^{-1}\,dW_s
\qquad\text{
and
}\qquad
M_t^\perp
:=
\int_0^t
\sigma_s
\sqrt{1-\chi_L(|\sigma_s|)^2\rho^\T R^{-1}\rho}\,dW_s^\perp.
\]
Since \([M^W,M^\perp]=0\), we have
$
\mathcal E(M^W+M^\perp)
=
\mathcal E(M^W)\mathcal E(M^\perp).
$
We now show that both factors are true martingales. By construction,
$
\sigma_t\chi_L(|\sigma_t|)=\operatorname{sgn}(\sigma_t)(|\sigma_t|\wedge L),
$
and therefore
\[
[M^W]_{\overline T}
=
\int_0^{\overline T}
\sigma_t^2\chi_L(|\sigma_t|)^2
\rho^\T R^{-1}\rho\,dt
\leq
L^2\rho^\T R^{-1}\rho\,\overline T.
\]
Consequently, Novikov's condition holds for \(M^W\), and
\(\mathcal E(M^W)\) is a true martingale.

On the other hand, conditionally on the path of \(W\), the integrand in
\(M^\perp\) is deterministic with respect to \(W^\perp\). Hence,
$
\mathbb E_{\mathbb Q}
[
\mathcal E(M^\perp)_{\overline T}
|
\mathcal F_{\overline T}^{W}
]
=1.$
It follows that
\begin{align*}
\mathbb E_{\mathbb Q}
\left[
\mathcal E(M^W+M^\perp)_{\overline T}
\right]
&=
\mathbb E_{\mathbb Q}
\left[
\mathcal E(M^W)_{\overline T}
\mathbb E_{\mathbb Q}
\left[
\left.
\mathcal E(M^\perp)_{\overline T}
\right|
\mathcal F_{\overline T}^{W}
\right]
\right]=
\mathbb E_{\mathbb Q}
\left[
\mathcal E(M^W)_{\overline T}
\right]
=1.
\end{align*}
Thus,
$
S^L/S_0=\mathcal E(M^W+M^\perp)
$
is a true martingale.
\end{proof}

The parameter \(L\) determines the saturation level of the leverage effect.
The instantaneous factor--equity correlations coincide with the baseline
correlations \(\rho\) in ordinary volatility regimes and are attenuated in
states of extreme volatility. Indeed,
$
\sigma_t{d[B^{L},W^i]_t}
=
\rho_i\operatorname{sgn}(\sigma_t)(|\sigma_t|\wedge L)dt.
$

\begin{corollary}
Let \(S\) follow a \(\VIX\)-based signature model driven by a primary
process \(X\), adapted to a vector of Brownian motions \(W\). Fix \(R\) and
\(\rho\) as in Proposition~\ref{prop:true-martingale}, and suppose that the
Brownian motion \(B\) driving the price is given by
\eqref{eq:martingale-preserving-correlation} for some \(L>0\). Then \(S\)
is a true martingale and
$
\mathbb E_{\mathbb Q}[S_t]=S_0,
$ for each $t\in[0,\overline T]$.
\end{corollary}
\begin{remark}\label{rem_mod1vsmod2}
    The difference between a signature model  based, respectively,  on volatility and on $\VIX$ is subtle. In both cases,  the dynamics of $\VIX^2$ and of $\sigma(\ell)^2$ are assumed to be representable as a linear combination of the time-extended signature of an underlying process $X$. The difference lies in which of the two processes is \emph{also} assumed to be the square of a linear combination of the time-extended signature of an underlying process $X$. For volatility-based signature models this is the case for $\sigma(\ell)^2$; for $\VIX$-based signature models, it is the case for $\VIX^2$.
\end{remark}

The recovery of instantaneous variance from a prescribed $\VIX$ specification is an inverse problem for a time-averaged transition semigroup, as studied for instance by \cite{P:22}. The following theorem provides a finite-dimensional formulation of this problem in the signature setting.
\begin{theorem}\label{thmvixbased}
    Let $(S_t)_{t\in [0,\overline T]}$ follow a $\VIX$-based signature model, fix an injective labeling function $\mathcal{L}:\{I:|I|\leq 2n\}\to\{1,\ldots,(d+1)_{2n}\}$, and let $G$ be the $(d+1)_{2n}$-dimensional matrix representative of the dual operator corresponding to $\wh{\X}^{\leq 2n}$. Then
    $$F_t(T, \ell)=\langle h_{F,T-t}(\ell),\wh{\X}^{\leq n}_t\rangle\qquad\text{for}\qquad
            \ve(h_{F,\delta}(\ell))^\T=\ve(\ell)^\T e^{\delta G^\T},$$
           for each $t\leq T\leq \overline T-\Delta$.
    Assuming that $A:=\int_0^\Delta e^{sG^\T}\dd{s}$ is invertible, we can also choose $h(\ell)$ such that
    $ \ve(h(\ell))^\T=\Delta\cdot\ve(\ell\shuffle\ell)^\T A^{-1},$
    and thus
    $$\sigma_t^S(\ell)^2
    =\Delta\cdot\ve(\ell\shuffle\ell)^\T A^{-1}\ve(\wh{\X}^{\leq 2n}_t),$$
    for each $t\in [0,\overline T]$.
\end{theorem}
\begin{proof}
     The representation of $F_t(T, \ell)$ follows as in the proof of Theorem~\ref{thm3}. 
  For the representation of $h(\ell)$, the definition of the VIX and the conditional-expectation formula in Theorem~\ref{thm1} yield 
         \begin{equation}
        \begin{aligned}
            \VIX_t(\ell)^2&=\langle h_{\VIX^2}(\ell),\wh{\X}_t^{\leq 2n}\rangle
            \qquad\text{for}\qquad
            \ve(h_{\VIX^2}(\ell))^\T=\dfrac{1}{\Delta}\ve(h(\ell))^\T A.
                    \end{aligned}
    \end{equation}
By definition and the shuffle property, it holds that 
$\VIX^2_t(\ell)=\langle\ell,\wh{\X}^{\leq n}_t\rangle^2=\langle\ell\shuffle\ell,\wh{\X}_t^{\leq 2n}\rangle.$
Defining $h(\ell)$ such that 
$$\dfrac{1}{\Delta}\ve(h(\ell))^\T A=\ve(\ell\shuffle\ell)^\T$$
and thus, assuming that $A$ is invertible, 
$ \ve(h(\ell))^\T=\Delta\cdot\ve(\ell\shuffle\ell)^\T A^{-1},$
the claim then follows from Theorem~\ref{thm1} and the definition of the VIX.
\end{proof}

\begin{remark}
    Note that for $\VIX$-based signature models we are implicitly assuming that $\ell$  is chosen such that both 
    \begin{equation}\label{eqnnonneg}
    \VIX_t(\ell)\geq 0\qquad\text{and}\qquad \sigma_t^S(\ell)^2\geq0.
    \end{equation}
    As soon as these conditions are satisfied and \begin{equation}\label{eqninvA}
        \text{the matrix $A$ is invertible},
    \end{equation} the dynamics of $\sigma^S(\ell)^2$, $\VIX(\ell)$ and $F_t(T,\ell)$ are given in closed form.

        The simplest (and in some cases only) way to verify that a linear combination of signature terms is nonnegative is to check if it can be written as a sum of squares of other linear combinations. If some of its components are nonnegative, as in the case of $X$ being a geometric Brownian motion or a CIR process, one can also include some odd terms. For example, setting
    $\ell=\sum_{i=1}^m h_i\shuffle h_i+h_0e_0,$
    for some $h_0\geq 0$, one would get
    $$\VIX_t=\sum_{i=1}^m \langle h_i,\widehat\X_t^{\leq n}\rangle^2+h_0 t.$$
\end{remark}

\begin{remark}
By Proposition~\ref{corVXX} we know that setting
            $$P_t(\ell):=\sum_{i=1}^\infty
\Big((t-T_{i-1})F_{t}(T_{i+1},\ell)
+(T_i - t)F_{t}(T_i,\ell)\Big)
\mathbbm{1}_{t\in(T_{i-1},T_i]},$$
we get
    $$\frac{\VXX_t(\ell)}{\VXX_0(\ell)}=\frac{P_t(\ell)\tau_0}{P_0(\ell)\tau_t}
    \times \exp\Big(\sum_{i=1}^\infty\int_{t\land T_{i-1}\lor 0}^{t\land T_i} \frac{ F_s(T_i,\ell) - F_s(T_{i+1},\ell) }{(s-T_{i-1})F_{s}(T_{i+1},\ell)
+(T_i - s)F_{s}(T_i,\ell)}\dd s\Big).$$
When $S$ follows a $\VIX$-based signature model, one can insert the explicit form of the futures and these expressions reduce to
$$P_t(\ell):= \ve(\ell)^\T\sum_{i=1}^\infty
\Big((t-T_{i-1})e^{(T_{i+1}-t)G^\T}
+(T_i - t)e^{(T_{i}-t)G^\T}\Big)
\mathbbm{1}_{t\in(T_{i-1},T_i]}\ve(\widehat \X_t^{\leq n}).$$
Consequently, we get
    $$\frac{\VXX_t(\ell)}{\VXX_0(\ell)}=\frac{P_t(\ell)\tau_0}{P_0(\ell)\tau_t}
    \times \exp\Big(\sum_{i=1}^\infty\int_{t\land T_{i-1}\lor 0}^{t\land T_i} \frac{  \ve(\ell)^\T
(e^{(T_{i}-s)G^\T}-e^{(T_{i+1}-s)G^\T})\ve(\widehat \X_s^{\leq n})}{P_s(\ell)}ds\Big).$$
The practical advantage of these expressions over the original formulation given by Proposition~\ref{corVXX} is however unclear.
\end{remark}

\begin{lemma}\label{leminv}
    Let $\lambda_1,\ldots,\lambda_m$ be the eigenvalues of $G$. Then $A=\int_0^\Delta e^{tG^\T}\dd t$ is invertible if and only if $\lambda_j\neq 2\pi k\mathrm{i}/\Delta$ for $j=1,\ldots,m$ and $k\in \mathbb Z\setminus\{0\}$.
\end{lemma}
\begin{proof}
Up to a change of basis, $G^\T$ can be written as an upper triangular
matrix whose diagonal entries are $\lambda_1,\ldots,\lambda_m$.
Consequently, $e^{tG^\T}$ is also upper triangular in the same basis,
with diagonal entries
$
e^{t\lambda_1},\ldots,e^{t\lambda_m}.
$
It follows that $A$ is upper triangular in this basis, with diagonal
entries
\[
\int_0^\Delta e^{t\lambda_j}\dd t
=
\begin{cases}
\dfrac{e^{\lambda_j\Delta}-1}{\lambda_j},
& \lambda_j\neq 0,\\[2mm]
\Delta,
& \lambda_j=0.
\end{cases}
\]

Hence $A$ is invertible if and only if all these quantities are nonzero.
For $\lambda_j=0$ the corresponding value is $\Delta>0$. If
$\lambda_j\neq0$, then
$
(e^{\lambda_j\Delta}-1)/{\lambda_j}=0
$
if and only if
$
e^{\lambda_j\Delta}=1,
$
which is equivalent to
$
\lambda_j=\frac{2\pi k\mathrm{i}}{\Delta}
$
for some $k\in\mathbb Z\setminus\{0\}$.
This proves the claim.
\end{proof}

\begin{remark}\label{dimRed}
Let $\mathcal I\subseteq\{I:|I|\le n\}$ be a finite set of words containing the empty word, and set
$D_q:=\sum_{j=0}^{q}(d+1)^j.$
For $\ell=\sum_{I\in\mathcal I}\ell_I e_I$, consider the signed functional $\langle\ell,\widehat{\mathbb X}_T^{\le n}\rangle$. Its conditional expectation defines the VIX futures proxy, before stopping. Define the shuffle-product space
\begin{equation}
    \mathcal P^2(\mathcal I):=\operatorname{span}\{e_I\shuffle e_J:I,J\in\mathcal I\}\subseteq T^{2n}(\mathbb R^{d+1}).
\end{equation}
This space can have substantially lower dimension than $T^{2n}(\mathbb R^{d+1})$; in particular, $\ell\shuffle\ell\in\mathcal P^2(\mathcal I)$ and
$|\langle\ell,\widehat{\mathbb X}_t^{\le n}\rangle|^2=\langle\ell\shuffle\ell,\widehat{\mathbb X}_t^{\le 2n}\rangle.$
Since $\emptyset\in\mathcal I$, every polynomial of degree at most two in the coordinates $\langle e_I,\widehat{\mathbb X}_t\rangle$, $I\in\mathcal I$, admits a representation $\langle b,\widehat{\mathbb X}_t^{\le 2n}\rangle$ for some $b\in\mathcal P^2(\mathcal I)$. Let now $G\in\mathbb R^{D_{2n}\times D_{2n}}$ be the matrix representation of the dual operator $L$ on $T^{2n}(\mathbb R^{d+1})$, with
$\ve(L u)=G\ve(u).$
For $b\in\mathcal P^2(\mathcal I)$, write $\mathbf b:=\operatorname{vec}(b)$, and let $y\in\mathbb R^{D_{2n}}$. Let $M\in\mathbb R^{s\times D_{2n}}$ have rows $v_1^\T,\ldots,v_s^\T$ spanning $\ve(\mathcal P^2(\mathcal I))$. Under the invariance condition
$$L\bigl(\mathcal P^2(\mathcal I)\bigr)\subseteq\mathcal P^2(\mathcal I),$$
we compute $\mathbf b^\T\exp(tG^\T)y$ using the restriction of $G$ to this invariant subspace.

\begin{enumerate}
\item \textbf{SVD of $M$.} Let $r=\operatorname{rank}(M)$ and compute the compact singular value decomposition
$M=U_r\Sigma_r Q^\T,$
where
$U_r\in\mathbb R^{s\times r},$ $\Sigma_r\in\mathbb R^{r\times r},$ and $Q\in\mathbb R^{D_{2n}\times r}.$
The columns of $U_r$ and $Q$ are orthonormal, and $\Sigma_r$ contains the positive singular values of $M$; in particular,
$Q^\T Q=I_r$ and $\operatorname{range}(Q)=\ve\bigl(\mathcal P^2(\mathcal I)\bigr).$
\item \textbf{Restriction of $G$.} Define the reduced matrix 
$\widetilde G:=Q^\T GQ\in\mathbb R^{r\times r}.$
Under the invariance condition
$ L(\mathcal P^2(\mathcal I))
\subseteq\mathcal P^2(\mathcal I)$, we have
$GQ=Q\widetilde G.$
Thus $\widetilde G$ represents the restriction of $G$ to the coefficient subspace of interest in the orthonormal basis given by the columns of $Q$.
\item \textbf{Small-scale matrix exponential.} We compute
$\exp(t\widetilde G^\T)\in\mathbb R^{r\times r}.$
The matrix exponential is therefore evaluated in dimension $r$, yielding a computational saving when
$r\ll D_{2n}$.
\item \textbf{Projection and reconstruction.} Set
$\mathbf b_r:=Q^\T\mathbf b,$ and $ y_r:=Q^\T y.$
Since $\mathbf b\in\operatorname{range}(Q)$, we have $\mathbf b=Q\mathbf b_r$. Consequently,
$$\mathbf b^\T e^{tG^\T}y=\mathbf b_r^\T e^{t\widetilde G^\T}y_r=\mathbf b^\T Q e^{t\widetilde G^\T}Q^\T y.$$
This identity is exact under the stated invariance condition. An example satisfying this condition is the complete word set $\mathcal I=\{I:|I|\le n\}.$
For selected enrichments, let $n\ge1$ and consider
$$\mathcal I=\{I:|I|\le n-1\}\cup\mathcal J,\qquad\mathcal J\subseteq\{J:|J|=n\}.$$
Write
$V_{\mathcal I}:=\operatorname{span}\{e_I:I\in\mathcal I\}.$
Sufficient conditions for invariance are
$ L e_J\in V_{\mathcal I}$ for every $J\in\mathcal J,$
together with constant covariance entries $a_{i j_n}$ and $a_{j_n i}$ for every $J=(j_1,\ldots,j_n)\in\mathcal J$ and every letter $i$. Indeed, products involving only words of length at
most $n-1$ are covered by the complete-word-set case. For nonempty words $I=(i_1,\ldots,i_p)$ and $J=(j_1,\ldots,j_q)$ in $\mathcal I$, with at least one belonging to $\mathcal J$, the generator product rule gives
$$ L(e_I\shuffle e_J)=( L e_I)\shuffle e_J+e_I\shuffle( L e_J)+\frac{1}{2}\bigl(a_{i_pj_q}+a_{j_qi_p}\bigr)(e_{I'}\shuffle e_{J'}).$$
The first two terms belong to $\mathcal P^2(\mathcal I)$ by generator closure. The last term also belongs to this space, since its covariance coefficient is constant and both prefixes $I'$ and $J'$ have length at most $n-1$. Products involving the empty word follow directly from generator closure. Hence
$ L\bigl(\mathcal P^2(\mathcal I)\bigr)\subseteq\mathcal P^2(\mathcal I).$
\end{enumerate}
This procedure allows us to reduce, consistently, both the dimension of the stored tensors and consequently the time needed to compute the products during the optimization. For example, choosing $\mathcal I=\{I\colon |I|\leq 2\}$ using an alphabet of size $d+1=4$ we are able to reduce the dimensionality from $d_4=341$ to $r=181$. Choosing $\mathcal I=\{I\colon |I|\leq 2\}\cup\{(1,1,1)\}$ and $d+1=4$ we reduce the dimensionality from $d_6=5461$ to $r=198$. For the specification containing both \((1,1,1)\) and \((0,0,0)\), the implementation uses a generator-invariant space of dimension 216.
\end{remark}

We focus now on the one-dimensional case. The proof of the following lemma and the subsequent corollary can be found in Appendix~\ref{app1d}.
\begin{lemma}\label{lem1d}
Consider the process $(X_t)_{t\in[0,\overline T]}$ given by $dX_t=b\dd t+\sqrt{\bar a+
\alpha X_t}\dd W_t$,
where $W$ is  a 1-dimensional Brownian motion. Let $G$ be the $5$-dimensional matrix representative of the dual operator corresponding to $\X$ and set $A=\int_0^\Delta e^{tG^\T}\dd{t}$. Let $S$ follow a $\VIX$-based signature model. Set 
$$\VIX_t=\ell_\emptyset+\ell_1(X_t-X_0)+\ell_{11}\dfrac{(X_t-X_0)^2}{2}.$$
Then $A$ is invertible and, setting $a=\bar a+\alpha X_0$, we get
\begin{align*}
    F_t(T,\ell)&=\VIX_t+ \frac{\left(T - t\right) \left(4 b \ell_{1} + \ell_{11} \cdot \left(2 a + b \left(T - t\right) \left(\alpha + 2 b\right)\right)\right)}{4}+  \frac{\ell_{11} \left((T-t) \alpha + 2 (T-t) b\right)}{2}(X_t-X_0)\\
    \sigma_t^S(\ell)^2
&=\VIX_t^2 +
\frac{\Delta^{2} a \ell_1 \ell_{11} \left(\alpha + 2 b\right)}{4} + \frac{\Delta^{2} \ell_{11}^{2} \left(- 3 \Delta^{2} \alpha^{3} b - 11 \Delta^{2} \alpha^{2} b^{2} - 12 \Delta^{2} \alpha b^{3} - 4 \Delta^{2} b^{4} + 60 a^{2}\right)}{480}\\
&\qquad\qquad- \Delta b \ell_{\emptyset} \ell_1 + \frac{\Delta \left( \ell_{\emptyset} \ell_{11} + \ell_1^{2}\right) \left(\Delta \alpha b + 2 \Delta b^{2} - 6 a\right)}{12}\\
&+(X_t-X_0)\Big(\frac{\Delta^{2} a \ell_{11}^{2} \left(\alpha + b\right)}{2} + \frac{\Delta \ell_1 \ell_{11} \left(\Delta \alpha^{2} + 3 \Delta \alpha b + 2 \Delta b^{2} - 6 a\right)}{4} - \frac{\Delta \left(\alpha + 2 b\right) \left( \ell_{\emptyset} \ell_{11} +  \ell_1^{2}\right)}{2}\Big)\\
&- \frac{(X_t-X_0)^2}2\Big(3 \Delta \ell_1 \ell_{11} \left(\alpha + b\right) - \frac{\Delta \ell_{11}^{2} \cdot \left(3 \Delta \alpha^{2} + 5 \Delta \alpha b + 2 \Delta b^{2} - 6 a\right)}{4}\Big)\\
&-\frac{(X_t-X_0)^3}6 \frac{3 \Delta \ell_{11}^{2} \cdot \left(3 \alpha + 2 b\right)}{2},
\end{align*}
whenever the latter is nonnegative.
\end{lemma}

\begin{corollary}\label{corPippo}
    Consider the model proposed by Lemma~\ref{lem1d} and set $b=\alpha=0$,  $\bar a=1$, and $X_0=0$. Then  $X_t=W_t$ and $\VIX_t=\ell_\emptyset+\ell_1X_t+\ell_{11}\dfrac{X_t^2}{2}$. The expressions, assuming $T_{i+1}-T_i=\tau$, simplify to
    \begin{align*}
        F_t(T,\ell)&=\VIX_t+\frac{1}{2}\ell_{11}(T-t),\\
        \sigma_t^S(\ell)^2&= \VIX_t^2+
        \frac{\Delta^{2} \ell_{11}^{2}}{8} - \frac{\Delta \left(2 \ell_{1}^{2} + 2 \ell_{\emptyset} \ell_{11}\right)}{4}- \frac{3 \Delta \ell_{1} \ell_{11}}{2}X_t- \frac{3 \Delta \ell_{11}^{2}}{2}\frac{X_t^2} 2,\\
       \VXX_t &=\VXX_0\times\frac{2\VIX_t+\ell_{11}\tau}{2\ell_{\emptyset}+\ell_{11}\tau}\times\exp\bigg(-\int_0^t\frac{\ell_{11}}{2\VIX_s+\ell_{11}\tau}\dd{s}\bigg).
    \end{align*}
    In this case, for $\ell_{11}>0$, nonnegativity of $\VIX_t$ and $\sigma_t^S(\ell)^2$ can be guaranteed by imposing the tight condition
    \begin{equation}
        \frac{\ell_\emptyset}{\ell_{11}}\geq \frac{\ell_1^2}{2\ell_{11}^2} +\frac{7\Delta}{16}.
    \end{equation}
    Similarly, choosing $\bar a=b=0$ and $\alpha=1$, we get $dX_t=\sqrt {X_t} \dd W_t$ and the expressions read
\begin{align*}
F_t(T,\ell)
&=
\VIX_t
+\frac{\ell_{11}}{2}(T-t)
X_t,
\\
\sigma_t^S(\ell)^2
&=\VIX_t^2
+\frac{\Delta X_t}{4}
\Bigg[
\Delta\ell_{11}
\left(
\ell_1-\ell_{11}X_0
+\frac{3}{2}\ell_{11}X_t
\right)
-2\left(
\left(\ell_1-\ell_{11}X_0+\ell_{11}X_t\right)^2
+\ell_{11}\VIX_t
\right)
\Bigg]\\
\VXX_t
&=
\VXX_0
\frac{
2\VIX_t+
\ell_{11}X_t\tau
}{
2\ell_\emptyset+\ell_{11}X_0\tau
}
\exp\left(
-\int_0^t
\frac{
\ell_{11}X_s
}{
2\VIX_s+
\ell_{11}\tau X_s
}
\dd s
\right).
\end{align*}
        In this case, nonnegativity of $\VIX_t$ and $\sigma_t^S(\ell)^2$ can be guaranteed by imposing one of the following sufficient conditions: $\ell_{11}=0$ and 
        \begin{equation}
            \ell_\emptyset\geq\frac{\Delta \ell_1}{8}+\ell_1X_0, \qquad\ell_1\geq0,
        \end{equation}
        or $\ell_{11}>0$ with
\begin{equation}
\begin{gathered}
\frac{
\ell_\emptyset}{\ell_{11}}\geq\frac{\ell_1}{\ell_{11}}X_0-\frac{1}{2}X_0^2
+
\max\Bigg\{
\frac{
\Delta\left(\frac{\ell_1}{\ell_{11}}-X_0\right)
\left[
2\left(\frac{\ell_1}{\ell_{11}}-X_0\right)-\Delta
\right]
}{
2\left[
4\left(\frac{\ell_1}{\ell_{11}}-X_0\right)-\Delta
\right]
},
\frac{3\Delta^2}{16}
-
\left(
\frac{\ell_1}{\ell_{11}}-X_0-\frac{3\Delta}{4}
\right)^2
\Bigg\},\qquad
\frac{\ell_1}{\ell_{11}}\geq X_0+\frac{3\Delta}{4}.
\end{gathered}
\end{equation}
\end{corollary}

\begin{example}\label{ex1}
    Assume that $X:=(B_t)_{t\geq 0}$ is given by a 1-dimensional Brownian motion. Then a random sample of the introduced objects is represented in Figure~\ref{fig:bm_traj}.
\begin{figure}[h!]
  \centering
  \subfigure{\includegraphics[width=0.48\textwidth]{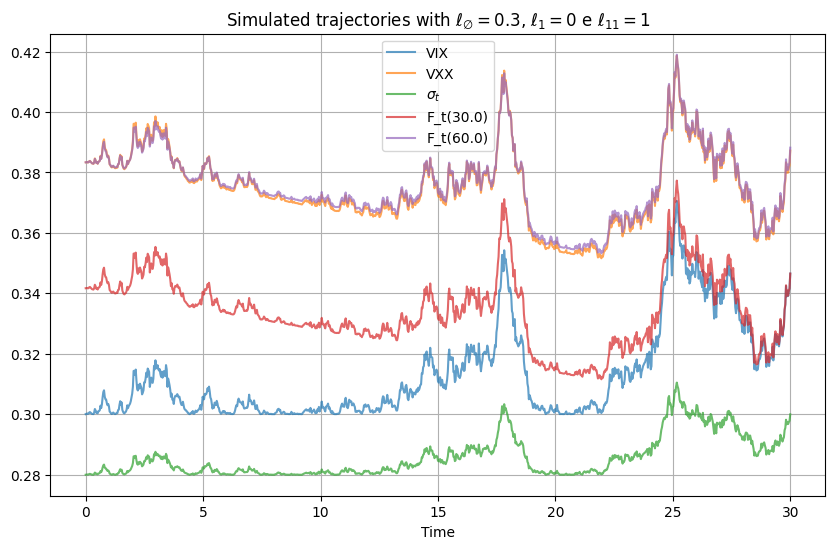}}
  \hfill
  \subfigure{\includegraphics[width=0.48\textwidth]{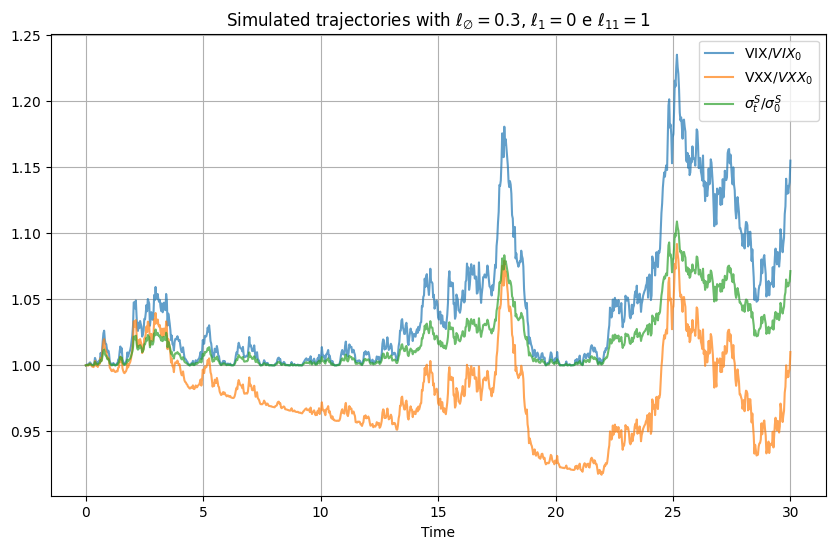}}
  \caption{Simulated trajectories with a Brownian primary process}
  \label{fig:bm_traj}
\end{figure}
    \end{example}

\subsection{Relaxing nonnegativity assumptions}\label{relaxing}
   {In practice, assuming nonnegativity of $\langle h(\ell),\wh{\X}^{\leq 2n}_t\rangle$ and $\langle \ell, \wh{\X}^{\leq n}_t\rangle$, which are the expressions corresponding to $\sigma_t^S(\ell)^2$ and $\VIX_t$, respectively, could be too restrictive. Alternatively, one can proceed by defining a slightly different model.
   \begin{definition}\label{regVIXbased}
       We say that $(S_t)_{t\in [0,\overline T]}$ follows a regularized $\VIX$-based signature model if
    \begin{equation}\label{Ssigma2}
            dS_t=S_t\sigma_t ^S(\ell)\dd{B}_t\qquad \text{for}\qquad \sigma_t^S(\ell)^2=\langle h(\ell),\wh{\X}^{\leq 2n}_t\rangle^+
    \end{equation}
for some $h(\ell)\in T^{2n}(\R^{d+1})$   such that
$$\langle\ell,\widehat \X^{\leq n}_t\rangle^2={\E_\Q\bigg[\frac{1}{\Delta}\int_t^{t+\Delta}\langle h(\ell),\wh{\X}^{\leq 2n}_u\rangle\dd{u}|\mathcal{F}_t\bigg]}.$$
The corresponding dynamics of the $\VIX$ and the corresponding futures are given by Definition~\ref{def1} and read
$$
\VIX_t:=\sqrt{\E_\Q\bigg[\frac{1}{\Delta}\int_t^{t+\Delta}\langle h(\ell),\wh{\X}^{\leq 2n}_u\rangle^+\dd{u}|\mathcal{F}_t\bigg]}
\qquad\text{and}\qquad
    F_{t}(T):=\E_\Q[\VIX_{T}|\mathcal{F}_t],$$
    respectively. 
In this case we again get
    $$S_t=S_0\exp\bigg(\int_0^t\sqrt{\langle h(\ell),\wh{\X}^{\leq 2n}_s\rangle^+}\dd{B}_s-\dfrac{1}{2}\int_0^t\langle h(\ell),\wh{\X}^{\leq 2n}_s\rangle^+\dd{s}\bigg).$$ 
We then set 
$\widetilde{\VIX}_t(\ell):=\langle \ell,\widehat \X^{\leq n}_t\rangle$ and consider $|\widetilde{\VIX}_t(\ell)|$ as a proxy for $\VIX_t$ and $\E_\Q[\widetilde{\VIX}_T(\ell)|\Fcal_t]$ as a proxy for $F_t(T)$.
   \end{definition}

Also in this case choosing $B^L$ as in Proposition~\ref{prop:true-martingale} leads to the true martingality of the price process $S$.
   
 \begin{corollary}
Let \(S\) follow a regularized \(\VIX\)-based signature model driven by a primary
process \(X\), adapted to a vector of Brownian motions \(W\). Fix \(R\) and
\(\rho\) as in Proposition~\ref{prop:true-martingale}, and suppose that the
Brownian motion \(B\) driving the price is given by
\eqref{eq:martingale-preserving-correlation} for some \(L>0\). Then \(S\)
is a true martingale and
$
\mathbb E_{\mathbb Q}[S_t]=S_0,
$ for each $t\in[0,\overline T]$.
\end{corollary}

   Note that, in principle, no nonnegativity assumptions on $\langle\ell,\widehat \X_t^{\leq n}\rangle$ are imposed. The rest of the section is devoted to quantifying the error introduced by replacing $\VIX_t(\ell)$ and $F_t(T)$ with their corresponding tractable proxies $|\langle \ell,\widehat \X^{\leq n}_t\rangle|$ and $\E_\Q[\widetilde{\VIX}_T(\ell)|\Fcal_t]$.
\begin{lemma}\label{errVIX}
Let
$
\widetilde{\VIX}_t(\ell)
:=\langle \ell,\widehat \X_t^{\leq n}\rangle
$
and choose the deterministic parameter
\[
\gamma_t(\ell):=\sqrt{
\mathbb E_{\mathbb Q}\!\left[
\frac1\Delta\int_t^{t+\Delta}
\langle h(\ell),\widehat{\X}_u^{\leq 2n}\rangle^-\,du
\right]}.
\]
Suppose that \(\gamma_t(\ell)>0\) and define $\varepsilon_t(\ell):=\E_\Q[\zeta_t(\ell)|\Fcal_t]$ for
\[
\zeta_t(\ell):=
\frac{
\displaystyle\frac1\Delta\int_t^{t+\Delta}
\langle h(\ell),\widehat{\X}_u^{\leq 2n}\rangle^-\,du
+
\left(
\sqrt{\widetilde{\VIX}_t(\ell)^2
      +\gamma_t(\ell)^2}
-\left|\widetilde{\VIX}_t(\ell)\right|
\right)^2
}{
2\sqrt{\widetilde{\VIX}_t(\ell)^2
      +\gamma_t(\ell)^2}
}.
\]
Then
$
0\leq
\VIX_t(\ell)-|\widetilde{\VIX}_t(\ell)|
\leq\varepsilon_t(\ell),
$
or, equivalently,
\[
2\bigl(\widetilde{\VIX}_t(\ell)\bigr)^-
\leq
\VIX_t(\ell)-\widetilde{\VIX}_t(\ell)
\leq
2\bigl(\widetilde{\VIX}_t(\ell)\bigr)^-
+\varepsilon_t(\ell).
\]
If \(\gamma_t(\ell)=0\), both conclusions hold with
\(\varepsilon_t(\ell)=\zeta_t(\ell)=0\).
\end{lemma}

\begin{proof}
Set
\[
\delta_t(\ell)^2:=
\mathbb E_{\mathbb Q}\!\left[
\left.
\frac1\Delta\int_t^{t+\Delta}
\langle h(\ell),\widehat{\X}_u^{\leq 2n}\rangle^-\,du
\right|\mathcal F_t
\right].
\]
By assumption,
$
\VIX_t(\ell)^2
=\widetilde{\VIX}_t(\ell)^2+\delta_t(\ell)^2.
$
Suppose first that \(\gamma_t(\ell)>0\), and write
\[
s_t:=
\sqrt{\widetilde{\VIX}_t(\ell)^2+\gamma_t(\ell)^2}.
\]
The inequality \(2s_t\sqrt z\leq s_t^2+z\), applied with
\(z=\widetilde{\VIX}_t(\ell)^2+\delta_t(\ell)^2=\VIX_t(\ell)^2\), gives
\begin{align*}
\VIX_t(\ell)-|\widetilde{\VIX}_t(\ell)|
&=\sqrt z-|\widetilde{\VIX}_t(\ell)|
\leq
\frac{\delta_t(\ell)^2+
\bigl(s_t-|\widetilde{\VIX}_t(\ell)|\bigr)^2}
{2s_t}
=\varepsilon_t(\ell),
\end{align*}
where the last equality follows because \(s_t\) and
\(\widetilde{\VIX}_t(\ell)\) are \(\mathcal F_t\)-measurable.
The lower bound follows from \(\VIX_t(\ell)\geq
|\widetilde{\VIX}_t(\ell)|\). Moreover,
\[
\VIX_t(\ell)-\widetilde{\VIX}_t(\ell)
=
\VIX_t(\ell)-|\widetilde{\VIX}_t(\ell)|
+2\bigl(\widetilde{\VIX}_t(\ell)\bigr)^-,
\]
which proves the second pair of bounds.

If \(\gamma_t(\ell)=0\), the nonnegative integrand in its
definition vanishes almost surely. Hence
\(\delta_t(\ell)=0\) almost surely and
\(\VIX_t(\ell)=|\widetilde{\VIX}_t(\ell)|\), as claimed.
\end{proof}

The pointwise choice that makes the upper bound an equality is
\(\gamma_t(\ell)=\delta_t(\ell)\), but it requires computing a
separate conditional expectation. The choice in the lemma is
deterministic and has the same unconditional quadratic scale:
by the tower property,
$
\gamma_t(\ell)^2
=\mathbb E_{\mathbb Q}\!\left[\delta_t(\ell)^2\right].
$
It therefore provides a scale for the regularization while
leaving only one conditional expectation in the error bound.
}

{ 
Similarly, one can construct an approximation of  VIX futures prices and $\VXX$ values, and then estimate the corresponding error.

\begin{definition}\label{deflowm}
    Given a tractable proxy $\widetilde\VIX_t(\ell)$ for  $\VIX_t(\ell)$, set $\widetilde F_t(T):=\E_\Q[\widetilde \VIX_T(\ell)|\mathcal F_t]$. For a fixed $m\in (0,\widetilde F_0(T))$ we then define
    \begin{align*}
    \tau_m(T)&:=\inf\{t\in [0,\overline T-\Delta]\colon \widetilde F_t(T)\leq m\}\\
        \widetilde F_t(T)^m&:=\widetilde F_{t\land \tau_m(T)}(T)\\
        \widetilde P_t^m&:=\sum_{i=1}^\infty
\Big((t-T_{i-1})\widetilde F_{t}(T_{i+1})^m
+(T_i - t)\widetilde F_{t}(T_i)^m\Big)
\mathbbm{1}_{t\in(T_{i-1},T_i]},\\
        \widetilde\VXX_t^m&:={\VXX_0}\frac{\widetilde P_t^m\tau_0}{\widetilde P_0^m\tau_t}
    \times \exp\Big(\sum_{i=1}^\infty\int_{(t\land T_{i-1})\lor 0}^{t\land T_i} \frac{ \widetilde F_s(T_i)^m - \widetilde F_s(T_{i+1})^m }{\widetilde P_s^m}ds\Big),
    \end{align*}
      for $\tau_t=\sum_{i=1}^\infty (T_i-T_{i-1})\mathbbm{1}_{t\in(T_{i-1},T_i]}$.
\end{definition}

\begin{lemma}\label{errFut}
    Suppose that $\underline \eta_T(\ell)\leq \VIX_T(\ell)-\widetilde \VIX_T(\ell)\leq \overline \eta_T(\ell)$ and set
    $$
    \underline\eta_{t,T}(\ell):=\E_\Q[\underline \eta_T(\ell)|\Fcal_t]\qquad\text{and}\qquad  \overline\eta_{t,T}(\ell):=\E_\Q[\overline \eta_T(\ell)|\Fcal_t].
$$
Then $\underline F_t(T)\leq F_t(T)\leq \overline F_t(T)$ and 
    $ |F_t(T)-\widetilde F_t(T)^m|\leq \rho_{t,T}(\ell),$
    for
    \[
\rho_{t,T}(\ell)
:=
\max\left\{
\left|\widetilde F_t(T)^m-\underline F_t(T)\right|,
\left|
\overline F_t(T)
-\widetilde F_t(T)^m
\right|
\right\},
\]
where
$\underline F_t(T):=\E_\Q[\widetilde \VIX_T(\ell)|\mathcal F_t]+\underline\eta_{t,T}(\ell)$ and $\E_\Q[\widetilde \VIX_T(\ell)|\mathcal F_t]+ \overline\eta_{t,T}(\ell)=:\overline F_t(T).$
\end{lemma}
\begin{proof}
Note that the tower property yields
\[
\underline\eta_{t,T}(\ell)\leq F_t(T)-\E_\Q[\widetilde \VIX_T(\ell)|\mathcal F_t]
\leq\overline\eta_{t,T}(\ell)
\]
and thus
$\underline F_t(T)\leq F_t(T)\leq \overline F_t(T).$
These bounds and the triangle inequality imply
the claim.
\end{proof}

\begin{lemma}\label{errVXX}
    Suppose that $\underline \eta_T(\ell)\leq \VIX_T(\ell)-\widetilde \VIX_T(\ell)\leq \overline \eta_T(\ell)$ for each relevant maturity $T_i$. Set
    $
p_s^\ast:=\min\{P_s,\widetilde P_s^m\}
$ and
assume that 
$\inf_{0\leq s\leq T}p_s^*>0$,
and  define $$D_s
:=
(s-T_{j-1})\rho_{s,T_{j+1}}(\ell)
+
(T_j-s)\rho_{s,T_j}(\ell),\qquad s\in(T_{j-1},T_{j}].$$
    Then $\left|
\log({\VXX_T}/{\widetilde{\VXX}_T^m})
\right|
\leq
B_T^{m}(\ell)$ where $B_T^m(\ell)=\beta_T+\beta_0+\sum_{j=1}^\infty\int_{(T\land T_{j-1})\lor 0}^{T\land T_j}
\beta^{j}_sds$
for 
\begin{align}
\beta_T
&:=\frac{D_T}{p_T^\ast},\qquad
\beta_0:=
\frac{D_0}{p_0^\ast},\qquad\text{and}\qquad
\beta_s^j:=
\frac{
\rho_{s,T_j}(\ell)
+
\rho_{s,T_{j+1}}(\ell)
}{
 P_s
}
+
\frac{
\left|
\widetilde F_s(T_j)^m-\widetilde F_s(T_{j+1})^m
\right|D_s
}{
 P_s\,\widetilde P_s^m
}.
\end{align}
Consequently, 
$
\E_\Q[|\VXX_T-\widetilde{\VXX}_T^m|]
\leq 
\VXX_0-\E_\Q[\widetilde{\VXX}_T^m]
+2\E_\Q[\widetilde{\VXX}_T^mB_T^{m}(\ell)].
$
    \end{lemma}
    \begin{proof}
Set $P$ and $\widetilde P^m$ as in 
 Proposition~\ref{corVXX} and Definition~\ref{deflowm}, respectively.
Note that, by construction,
$
\widetilde P_s^m\geq \tau_s m.
$
Note that 
$
F_s(T)\geq\underline F_s(T),
$
and
\begin{equation}\label{PovP}
     |P_s-\widetilde P_s^m|\leq
D_s.
 \end{equation}
Taking the logarithm of the ratio between $\VXX_T$ and $\widetilde \VXX_T^m$  gives
\[
\begin{aligned}
\log\frac{\VXX_T}{\widetilde{\VXX}_T^m}
={}&
\log\frac{P_T}{\widetilde P_T^m}
-
\log\frac{P_0}{\widetilde P_0^m}
+
\sum_{j=1}^\infty\int_{(T\land T_{j-1})\lor 0}^{T\land T_j}
\left[
\frac{F_s(T_j)-F_s(T_{j+1})}{P_s}
-
\frac{
\widetilde F_s(T_j)^m-\widetilde F_s(T_{j+1})^m
}{
\widetilde P_s^m
}
\right]ds.
\end{aligned}
\]
Note that since
$
|\log (x/y)|
\leq{|x-y|}/{\min\{x,y\}}
$ for each $x,y>0$,
we obtain
\[
\left|
\log\frac{P_s}{\widetilde P_s^m}
\right|
\leq
\frac{D_s}{p_s^\ast}.
\]
Moreover, by \eqref{PovP} we have
\[
\begin{aligned}
\left|
\frac{F_s(T_j)-F_s(T_{j+1})}{P_s}
-
\frac{
\widetilde F_s(T_j)^m-\widetilde F_s(T_{j+1})^m
}{
\widetilde P_s^m
}
\right|
&\leq
\frac{
|F_s(T_j)-\widetilde F_s(T_j)^m|
+
|F_s(T_{j+1})-\widetilde F_s(T_{j+1})^m|
}{
P_s
}\\
&\qquad
+
\left|
\widetilde F_s(T_j)^m-\widetilde F_s(T_{j+1})^m
\right|
\frac{|P_s-\widetilde P_s^m|}
{P_s\widetilde P_s^m}\\
&\leq
\frac{
\rho_{s,T_j}(\ell)
+
\rho_{s,T_{j+1}}(\ell)
}{
 P_s
}
+
\frac{
\left|
\widetilde F_s(T_j)^m-\widetilde F_s(T_{j+1})^m
\right|D_s
}{
 P_s\,\widetilde P_s^m
}.
\end{aligned}
\]
For the conclusion,  note that
\[
|a-b|
=a-b
+2(b-a)^+
=a-b
+2b\left(1-e^{-\log(b/a)}\right)
1_{\{b>a\}}\leq a-b
+2b|\log(b/a)|,
\]
where the inequality follows from \(1-e^{-x}\leq x\), for \(x\geq0\). The claim follows by choosing $a=\VXX_T$ and $b=\widetilde{\VXX}_T^m$ and recalling that $(\VXX_t)_{t\in [0,T]}$, being a nonnegative local martingale, is also a supermartingale and thus satisfies $\E_\Q[\VXX_T]\leq \VXX_0$.
\end{proof}

\begin{corollary}\label{CorVXX} Fix $T\in(T_{i^*-1},T_{i^*}]$ and set $\widetilde \VIX_t(\ell)$, $\gamma_t(\ell)$ and $\zeta_t(\ell)$ as in Lemma~\ref{errVIX} and $\beta_T,\beta_0$ and $\beta_s^j$ as in Lemma~\ref{errVXX}. Set $\overline \zeta_t(\ell):=\zeta_t(\ell)+2(\widetilde{\VIX}_t(\ell))^-,$ fix $m>0$ such that $m<\E_\Q[\widetilde \VIX_{T_i}(\ell)]$ for each $i\in \{1,\ldots, i^*+1\},$
and on the set $A_t:=\{\tau_m(T_1),\ldots,\tau_m(T_{i^*+1})>t\}$ define
\begin{align*}
\overline \beta_0&:=\frac{-T_0\E_\Q[\overline \zeta_{T_2}(\ell)]
+
T_1\E_\Q[\overline \zeta_{T_{1}}(\ell)]}{\widetilde P^m_0},\\
 \overline \beta_T&:=\frac{(T-T_{i^*-1})\overline \zeta_{T_{i^*+1}}(\ell)+(T_{i^*}-T)\overline \zeta_{T_{i^*}}(\ell)}{\widetilde P^m_T},\\
\overline \beta_s^j&:=\frac{
\overline \zeta_{T_{j}}(\ell)
+
\overline \zeta_{T_{j+1}}(\ell)
}{
\widetilde P^m_s
}
+
\frac{
\left|
\widetilde F_s(T_j)-\widetilde F_s(T_{j+1})
\right|((s-T_{j-1})\overline \zeta_{T_{j+1}}(\ell)
+
(T_j-s)\overline \zeta_{T_j}(\ell))
}{
(\widetilde P^m_s)^2
}.
\end{align*}
Then $    \E_\Q[|\VXX_T-\widetilde{\VXX}_T^m|]\leq C_T^m(\ell)$ for
\begin{align*}
C_T^m(\ell)
&:=
\VXX_0-\E_\Q[\widetilde{\VXX}_T^m]
+2\E_\Q\Big[\widetilde{\VXX}_T^m(\overline \beta_T1_{A_T}+\beta_T1_{A_T^c}
+
\overline\beta_0)+
\sum_{j=1}^\infty\int_{(T\land T_{j-1})\lor 0}^{T\land T_j}
\widetilde{\VXX}_s^m(\overline \beta^j_s1_{A_s}+\beta^j_s1_{A_s^c})ds\Big].
\end{align*}
\end{corollary}
\begin{proof}
Note that by Lemma~\ref{errVIX} it holds $\underline \eta_{T_i}(\ell)\leq \VIX_{T_i}(\ell)-\widetilde \VIX_{T_i}(\ell)\leq \overline \eta_{T_i}(\ell)$ for $\underline\eta_{t,{T_i}}(\ell):=0$ and
$$
    \overline\eta_{t,{T_i}}(\ell):=\E_\Q[{\varepsilon_{T_i}(\ell)}+2(\widetilde{\VIX}_{T_i}(\ell))^-|\Fcal_t]=\E_\Q[\overline \zeta_{T_i}|\Fcal_t],
$$
for each $i\in\{1,\ldots, i^*+1\}$.
Observe also that  
$\beta_T,\beta_0,\beta_s^j$ are measurable with respect to $\mathcal F_T$, $\mathcal F_0$, and $\mathcal F_s$, respectively. By supermartingality of $\widetilde \VXX^m$ and Lemma~\ref{errVXX} we get
$$
\E_\Q[|\VXX_T-\widetilde{\VXX}_T^m|]
\leq 
\VXX_0-\E_\Q[\widetilde{\VXX}_T^m]
+2\Big(\E_\Q[\widetilde{\VXX}_T^m(\beta_T+\beta_0)]+\sum_{j=1}^\infty\int_{(T\land T_{j-1})\lor 0}^{T\land T_j}
\E_\Q[\widetilde{\VXX}_s^m\beta^{j}_s]ds\Big).
$$
As on $A_t$ it holds  $\underline F_t(T_i)=\widetilde F_t(T_i)^m=\E_\Q[\widetilde \VIX_{T_i}(\ell)|\mathcal F_t]$ and hence
$
\rho_{t,T_i}(\ell)
=
\overline\eta_{t,T_i}(\ell)
$ and 
$p^\star_t\geq \widetilde P_t^m,$
we can conclude that 
\begin{align*}
\beta_T
&=\frac{(T-T_{i^*-1})\E_\Q[\overline \zeta_{T_{i^*+1}}(\ell)|\Fcal_T]+
(T_{i^*}-T)\E_\Q[\overline \zeta_{T_{i^*}}(\ell)|\Fcal_T]}{p_T^\ast}\leq\E_\Q[\overline \beta_T|\Fcal_T],\qquad\text{on }A_T\\
\beta_s^j&=
\frac{
\E_\Q[\overline \zeta_{T_{j}}(\ell)|\Fcal_s]
+
\E_\Q[\overline \zeta_{T_{j+1}}(\ell)|\Fcal_s]
}{
 P_s
}
 +
\frac{
\left|
\widetilde F_s(T_j)-\widetilde F_s(T_{j+1})
\right|((s-T_{j-1})\E_\Q[\overline \zeta_{T_{j+1}}(\ell)|\Fcal_s]
+
(T_j-s)\E_\Q[\overline \zeta_{T_j}(\ell)|\Fcal_s])
}{
P_s\widetilde P^m_s
}\\
&\leq\E_\Q[\overline \beta_s^j|\Fcal_s],\qquad \text{on }A_s
\end{align*}
and the claim follows by Lemma~\ref{errVXX} and the tower property.
\end{proof}

The previous results bound the approximation errors at the level of
the underlying processes. We now translate them into bounds for the
corresponding option prices considered in Section~\ref{Calibration to option prices}.
Let
\[
\overline\pi_{\VIX}(T,K)
:=
e^{-rT}
\mathbb E_{\mathbb Q}
\left[
\left(
100|\langle\ell,\widehat \X_T^{\leq n}\rangle|-K
\right)^+
\right]
\]
denote the VIX call price computed using the absolute-value proxy.
Similarly, let
\[
\widetilde\pi_{\VXX}^{\,m}(T,K)
:=
e^{-rT}
\mathbb E_{\mathbb Q}
\left[
\left(
e^{(r-c)T}\widetilde{\VXX}_T^{\,m}-K
\right)^+
\right]
\]
denote the VXX call price computed using the tractable VXX proxy.

\begin{corollary}\label{cor:option-price-errors}
Assume the hypotheses of Lemma~\ref{errVXX} and that the quantities below are integrable.

\begin{enumerate}[label=\textup{(\roman*)}]
\item The error of the absolute-value VIX proxy satisfies
\begin{align}
\left|
\pi_{\VIX}(T,K)
-
\overline\pi_{\VIX}(T,K)
\right|
\leq
100e^{-rT}
\mathbb E_{\mathbb Q}
\left[
\varepsilon_T(\ell)
 \right].
\label{eq:VIX-option-price-error}
\end{align}
\item Suppose that the assumptions of Lemma~\ref{errVXX} and Corollary~\ref{CorVXX} hold with
the futures error bound
\(\rho_{t,T}(\ell)\). Then
\begin{equation}
\left|
\pi_{\VXX}(T,K)
-
\widetilde\pi_{\VXX}^{\,m}(T,K)
\right|
\leq
e^{-cT}
C_T^m(\ell).
\label{eq:VXX-option-price-error}
\end{equation}
\end{enumerate}
\end{corollary}

\begin{proof}
The map \(x\mapsto(x-K)^+\) is \(1\)-Lipschitz and vanishes for $x\leq K$. Therefore,
$$
\left|
(100\VIX_T-K)^+
-
(100|\langle\ell,\widehat \X_T^{\leq n}\rangle|-K)^+
\right|
\leq
100\left|
\VIX_T-|\langle\ell,\widehat \X_T^{\leq n}\rangle|
\right|.
$$
Lemma~\ref{errVIX} proves the first claim. The second follows similarly
from
\[
\left|
e^{-rT}(e^{(r-c)T}\VXX_T-K)^+
-
e^{-rT}(e^{(r-c)T}\widetilde{\VXX}_T^{\,m}-K)^+
\right|
\leq
e^{-cT}
\left|
\VXX_T-\widetilde{\VXX}_T^{\,m}
\right|
\]
and Corollary~\ref{CorVXX}.
\end{proof}
}

\section{Calibration to option prices}\label{Calibration to option prices}
This section is devoted to the calibration of the proposed models to option prices. Recall that we model the dynamics of the discounted, dividend-adjusted price process as \eqref{Asset dynamic}, i.e.
\begin{align*}
   \dd{S}_t&=S_t\sigma_t^S(\ell)\dd{B}_t,\hspace{1cm} S_0=s_0.
\end{align*}
Observe that $S$ is a local martingale under $\Q$.
We denote the undiscounted, unadjusted price process by $\widetilde{S}$, which is given by
$\widetilde S_t=e^{(r-q)t}S_t$,
where $r,q \geq 0$ denote the interest rate and the dividend yield, respectively, both of which we assume to be constant. 
\begin{definition}\label{FinProd}
We consider four types of financial products.

{\bf Call options on the SPX.} Denoting by $T$ the maturity and by $K$ the strike, the corresponding payoff and current price are given by
    $$(\widetilde S_T-K)^+\qquad\text{and}\qquad \pi_{\SPX}(T,K)=\E_\Q[e^{-rT}(\widetilde S_T-K)^+]=\E_\Q[e^{-rT}(e^{(r-q)T} S_T-K)^+],$$
    respectively.
    
  {\bf VIX Futures.}   As we are assuming deterministic interest rates, the price process $F_t(T)$ at time $t$ of a VIX futures contract with maturity $T$ coincides with the forward one. It is defined as the price fixed at time $t$, to be paid at time $T$, to get $\VIX_T$ at time $T$. It solves the equation
    $\E_\Q[(\VIX_T-F_t(T))e^{-rT}|\Fcal_t]=0$
    and is thus given by
$$F_t(T)=\E_\Q[\VIX_T|\Fcal_t].$$ 
Observe that $\VIX_T=\sqrt{\VIX_T^2}$ and $\VIX_T^2$ can be represented, under the usual model-free assumptions, in terms of a portfolio of SPX options (see \citealp{CW:06}, equation~(10), for the theoretical representation, and \citealp{CboeVIX2026}, Section~3.1, for the discretized formula used to calculate the index). Throughout, $\VIX_t$ and $F_t(T)$ are expressed in decimal
volatility units. Their quotations in VIX index points are
$100\VIX_t$ and $100F_t(T)$, respectively.

The current price $F_0(T)$ will be of particular importance.  Observe that the price to pay at time $t$ to get $F_T(T)$ at time $T$ is given by
$e^{-r(T-t)}F_t(T)$, which, once discounted with $e^{-rt}$, is again a $\Q$-martingale as expected. 

   {\bf Call options on the VIX} Denoting again by $T$ the maturity and by $K$ the strike expressed in index points, the corresponding payoff and price are given by
    $$(100F_T(T)-K)^+=(100\VIX_T-K)^+\qquad\text{and}\qquad \pi_{\VIX}(T,K)=\E_\Q[e^{-rT}(100\VIX_T-K)^+],$$
    respectively.
    
    {\bf American call options on  the VXX ETN.}  
Denoting again by $T$ the maturity, by $K$ the strike, and letting $\Theta_T$ denote the set of all stopping times $\theta\in [0, T]$, the corresponding payoff and price are given by
$$(\VXX_\theta e^{(r-c)\theta} -K)^+\qquad\text{and}\qquad 
\pi_{\VXX}^{\mathrm{Am}}(T,K)
=
\sup_{\theta\in\Theta_T}
\E_\Q\!\left[
e^{-r\theta}
\left(
e^{(r-c)\theta}\VXX_\theta-K
\right)^+
\right],
$$
respectively, where $c>0$ denotes a fixed annual fee continuously applied to the value of the VXX.
When $c=0$, the standard no-early-exercise argument for American call options
applies, under the usual integrability assumptions, and the American price
coincides with the price of the corresponding European option. In the present
setting, however, the management fee $c>0$ acts as a continuous negative carry
on the VXX position. Hence, the absence of an early-exercise premium is no
longer guaranteed.
Since the VXX options considered in our calibration have short maturities and
the annual fee is small  (according to the VXX's prospectus, $c=0.89\%$), we neglect the early-exercise premium and
use the corresponding European option price as an approximation:
\[
\pi_{\VXX}^{\mathrm{Am}}(T,K)
\approx
\pi_{\VXX}(T,K)
:=
\E_\Q\!\left[
e^{-rT}
\left(
e^{(r-c)T}\VXX_T-K
\right)^+
\right].
\]
\end{definition}
 Historical bid and ask quotes for SPX, VIX, and VXX options dated 24 February 2026 were obtained through Cboe DataShop under license. The implied volatilities were computed by the authors from these quotes. We use one option maturity for each underlying. The selected quotes cover the following strike ranges:
\begin{center}
\small
\begin{tabular}{lrrrr}
\toprule
Market & Maturity (days) & Number of strikes & Strike range & Moneyness range $(K/F^{\mkt}_{\cdot})$\\
\midrule
SPX & $24$ & $29$ & $[5610,7575]$ & $[81.28\%,109.75\%]$\\
VIX & $22$ & $13$ & $[16.5,28]$   & $[81.47\%,138.26\%]$\\
VXX & $24$ & $13$ & $[25,37]$    & $[87.55\%,129.58\%]$\\
\bottomrule
\end{tabular}
\end{center}
The table reports forward moneyness $K/F_A^\mkt(T)$ for each underlying $A$ at the option maturity $T$. We infer the forward from the put-call parity for $\SPX$, we use the quoted futures price for $\VIX$ and we set it to $\VXX_0e^{rT}$ for $\VXX$. The interest rate is $r=0.0367$, and the VIX futures maturities used for the VXX roll are 
\[
(T_0,T_1,T_2,T_3,T_4)=\frac{(-6,22,50,84,113)}{365},
\qquad \Delta=\frac{30}{365}.
\]
Since the VXX option expires after the first futures expiry, its construction includes the change in the rolling futures pair at day 22. We will require the raw variance $\langle h(\ell),\widehat{\mathbb X}_t^{\le2n}\rangle$ to be at least
$10^{-6}$ at every calibration grid point through day 84. The separate simulation for the VXX approximation bound extends
to $T_3+\Delta=114/365$ to cover the variance window starting at $T_3$.

We work throughout with the tractable proxies \(\overline{\VIX}_t(\ell):=|\langle\ell,\widehat{\mathbb X}_t^{\le n}\rangle|\), \(\widetilde F^m_t(T)\), and \(\widetilde{\VXX}^m_t\) for $m=0.01$, corresponding to one VIX index point. Accordingly, the reported calibration results concern the proxy system rather than the exact VIX, VIX-futures, and VXX quantities generated by the regularized model; the error bounds derived in Section~\ref{relaxing} quantify the corresponding approximation gap and will be estimated in both calibration examples.  We express the signature coefficients in decimal volatility units: the tractable VIX functional in index points is
$100|\langle\ell,\widehat \X_t\rangle|$.
For each market $A\in\{\SPX,\VIX,\VXX\}$, the parameter $\beta_A\in[0,1]$ balances the price penalties for deviations from the bid--ask interval and from the midquote, with larger values placing more weight on the former. The weights $\lambda_A$ determine the relative contributions of the three normalized market blocks to the joint calibration objective.  The loss function, block normalization and computational procedure are detailed in Appendix~\ref{QuantCorner}. The implemented code is available in the following GitHub repository:\\
\href{https://github.com/andreastanghellini/Joint-calibration-SPX-VIX-VXX}{https://github.com/andreastanghellini/Joint-calibration-SPX-VIX-VXX}.

\subsection{Ornstein--Uhlenbeck primary process}
We first investigate the VIX-based signature model with a three-dimensional Ornstein--Uhlenbeck primary process $X_t=(X_t^1,X_t^2,X_t^3)$. Under the risk-neutral measure, its components satisfy
\begin{equation}\label{eq:three-ou-primary}
    dX_t^i=\kappa_i(\theta_i-X_t^i)\,dt+\eta_i\,dW_t^i, \qquad i=1,2,3,
\end{equation}
where $W$ is a vector of correlated standard Brownian motions with $d[ W^i,W^j]_t=(R_{\mathrm{OU}})_{ij} \dd t$. 
We fix the primary-process hyperparameters following \cite{CGMS:24}:
\begin{center}
\begin{tabular}{crrrr}
\toprule
    Factor $i$ & $\kappa_i$ & $\theta_i$ & $\eta_i$ & $X_0^i$\\
\midrule
    1 & $0.1$ & $0.1$ & $0.7$ & $1$\\
    2 & $25$ & $4$ & $10$ & $0.08$\\
    3 & $10$ & $0.08$ & $5$ & $2$\\
\bottomrule
\end{tabular}
\end{center}
The first factor provides a slow mean-reversion scale, whereas the other two factors revert more rapidly. Their Brownian correlation matrix is
\begin{equation*}
    R_{\mathrm{OU}}=
    \begin{pmatrix}
        1&0.213&-0.576\\
        0.213&1&-0.044\\
        -0.576&-0.044&1
    \end{pmatrix}.
\end{equation*}
In particular, the primary-process instantaneous covariance entries are $a_{ij}=\eta_i\eta_j(R_{\mathrm{OU}})_{ij}$ and are constant, so that $X$ is a polynomial diffusion. The time-extended signature uses the alphabet $\{0,1,2,3\}$, with $0$ denoting time and $i$ denoting the $i$-th OU factor. 
 For the Brownian motion $B$ driving the SPX we follow \eqref{eq:martingale-preserving-correlation} choosing 
$$B_t=\int_0^t\chi_L(\sigma_s^S(\ell))\rho^\T R_{\mathrm{OU}}^{-1}dW_s+\int_0^t\sqrt{1-\chi_L(\sigma_s^S(\ell))^2\rho^\T R_{\mathrm{OU}}^{-1}\rho}dW_s^\perp,$$
for $L=100$. 

\paragraph{Calibration settings} The calibration, described in detail in Appendix~\ref{QuantCorner}, uses $N=10\,000$ Monte Carlo paths and four steps per day, with the following loss function parameters
$$(\lambda_{\mathrm{SPX}},\lambda_{\mathrm{VIX}},\lambda_{\mathrm{VXX}})=(0.35,0.5,0.15),\qquad(\beta_{\mathrm{SPX}},\beta_{\mathrm{VIX}},\beta_{\mathrm{VXX}})=(0.98,0.1,0.98)$$
The primary-process hyperparameters remain fixed, while the signature coefficients and the three correlation parameters $\rho$ are optimized jointly, the latter subject to \(\rho^\T R_{\mathrm{OU}}^{-1}\rho<1\).

\paragraph{Complete level-two specification.}
\begin{figure}[htbp]
    \centering
    \includegraphics[width=0.95\linewidth]{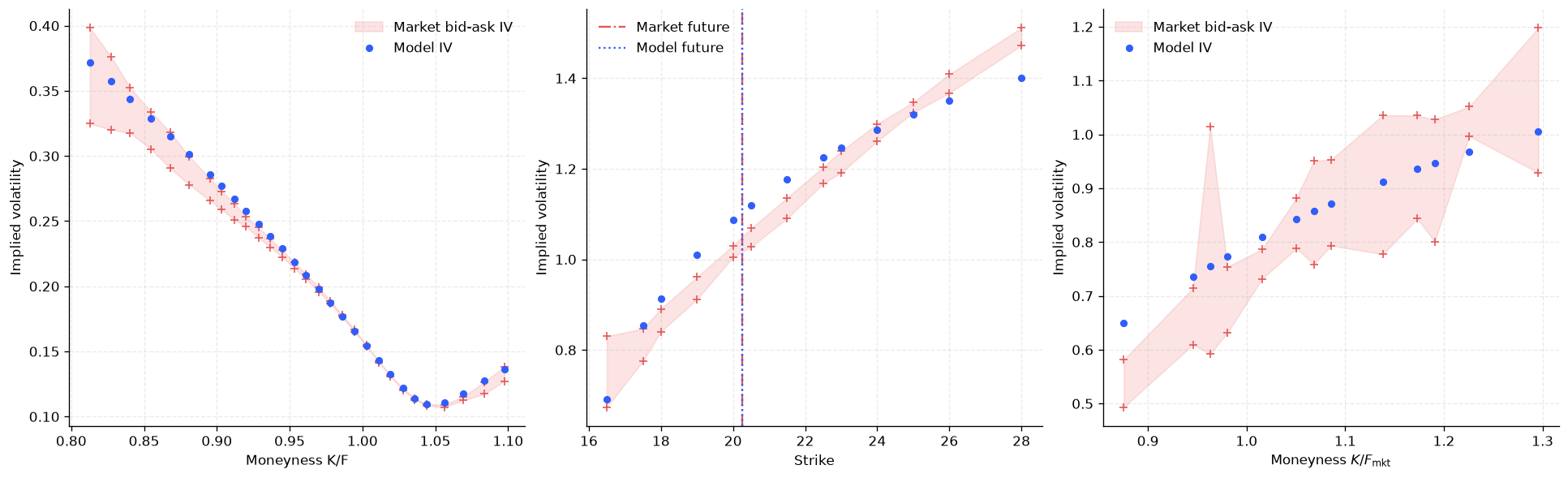}
    \caption{SPX, VIX and VXX implied volatilities for the complete level-two VIX functional, computed using $10^6$ independent Monte Carlo paths. Shaded regions show the quoted call bid--ask bands. The VIX panel also reports the model and market futures levels.}
    \label{fig:three-ou-level2-oos}
\end{figure}
The baseline VIX functional retains all signature words of length at most two, including the empty word, giving $1+4+4^2=21$ coefficients.
Its shuffle square reaches level four. Figure~\ref{fig:three-ou-level2-oos} shows the corresponding smiles computed over a sample size of $10^6$. The SPX curve follows the pronounced downside skew and the mild right-wing upturn, with model-implied volatilities generally within or close to the quoted bands. The VIX smile has the correct increasing shape, but its slope is too flat: the fit is relatively high at several low and intermediate strikes and underestimates the upper wing. The 22-day raw VIX futures proxy is $20.2511$, compared with $20.2524$ in the market, a relative difference of approximately $0.006\%$. The calibrated coefficient vector, ordered by word length and then lexicographically within each level, is 
\begin{equation}
\begin{aligned}
\ell_2^\star
&= \left(
\begin{gathered}
     0.190867,\ -0.215044,\ -0.365274,\  0.002658,\  0.055771,\ -0.637830,\\
     3.319946,\ -0.016356,\ -0.365003,\  3.414089,\  1.380427,\ -0.010032,\\
    -0.143587,\ -0.037989,\ -0.012147,\  0.000191,\  0.000124,\ -0.406777,\\
    -0.139771,\  0.000905,\  0.022677
\end{gathered}
\right)^{\mathrm{T}} \in \mathbb{R}^{21},\\
\rho_2^\star
&= \bigl(0.83893325,\ 0.08516936,\ -0.92248544\bigr)^{\mathrm{T}}.
\end{aligned}
\end{equation}
Observe that \((\rho_2^\star)^\T R_{\mathrm{OU}}^{-1}\rho_2^\star \approx 0.9944\), as required by Proposition~\ref{prop:true-martingale}.

\paragraph{Parsimonious level-three enrichment.}
To address the flat VIX upper wing, we retain the complete level-two functional and, following Remark~\ref{dimRed}, add two additional words: $(1,1,1)$, associated with the slow OU factor, and $(0,0,0)$, associated with time:
\begin{equation}
\begin{aligned}
 \ell&=\ell^{\leq2}+c_{111}e_{(1,1,1)}+c_{000}e_{(0,0,0)},\qquad
 \langle e_{(1,1,1)},\widehat \X_t\rangle
       &=\frac{(X_t^1-X_0^1)^3}{6},\qquad
 \langle e_{(0,0,0)},\widehat \X_t\rangle=\frac{t^3}{6}.
\end{aligned}
 \label{eq:three-ou-slow-cubic}
\end{equation}
Thus there are 23 VIX coefficients. This is a selected level-three enrichment, rather than a full level-three truncation, which would contain 85 coefficients. The slow-OU cubic introduces an odd nonlinear term into the signed signature functional. The time cubic adds the deterministic term $c_{000}T^3/6$ at maturity $T$.  The additional terms provide flexibility for the raw futures term structure used in the VXX roll. The combined impact on the smiles is assessed through joint recalibration. Both added words close under
the dual generator together with the complete level-two basis:
$$
 L e_{(1,1,1)}=-3\kappa_1 e_{(1,1,1)}+\kappa_1(\theta_1-X_0^1)e_{(1,1)}+\frac{\eta_1^2}{2}e_{(1)}
 =-0.3e_{(1,1,1)}-0.09e_{(1,1)}+0.245e_{(1)}\quad\text{and}\quad
 L e_{(0,0,0)}=e_{(0,0)}.
$$
The implementation retains a reduced generator-invariant space of dimension 216 containing the required shuffle contributions. This reduction preserves the identity linking the raw variance to the squared VIX proxy, without storing all 5461 coordinates of the complete level-six signature.
\begin{figure}[htbp]
\centering
\includegraphics[width=\linewidth]{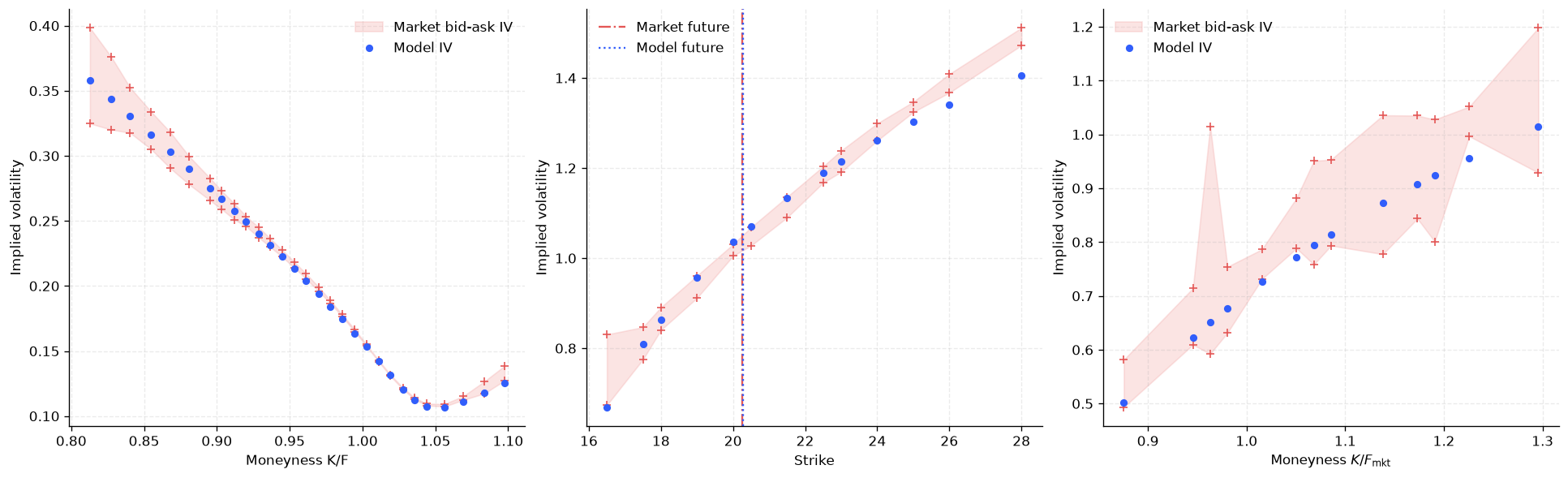}
\caption{SPX, VIX and VXX implied volatilities after adding the slow-OU word $(1,1,1)$ and the time word $(0,0,0)$ to the complete level-two VIX functional and jointly recalibrating, using $10^6$ independent Monte Carlo paths. The specification has 23 VIX coefficients.}
\label{fig:three-ou-slow-cubic-oos}
\end{figure}
Figure~\ref{fig:three-ou-slow-cubic-oos} reports implied-volatility smiles computed with the calibrated parameters held fixed on $10^6$ independent Monte Carlo trajectories. It shows that the SPX downside skew and right-wing upturn remain well reproduced. The VIX curve is steeper than in the level-two benchmark and the underestimation of the upper wing is reduced. Some deviations remain at low and intermediate strikes, and the highest-strike point remains below the quoted band. The two futures markers are nearly coincident. The VXX fit retains its increasing profile and remains broadly consistent with the comparatively wide market bands. These observations describe the fixed-parameter smiles computed on newly simulated paths.
The calibrated coefficients are ordered as the 21 level-two coefficients, followed by the $(1,1,1)$ and $(0,0,0)$ coefficients, respectively:
\begin{equation}
\begin{aligned}
\ell_3^\star&=\left(\begin{gathered}
        0.183160,\ -0.001962,\ -0.216225,\ -0.000552,\ 0.066644,\ 4.320270,\\
        1.799492,\ 0.010978,\ -0.323009,\ 1.683842,\ 1.814017,\ 0.005986,\\
        -0.002905,\ 0.010985,\ 0.011239,\ -0.000115,\ -0.000214,\ -0.380179,\\
        -0.007736,\ 0.000301,\ 0.035222,\ -5.407749,\ -155.189355
    \end{gathered}\right)^{\T}\in\mathbb{R}^{23},\\
\rho_3^\star&=\bigl(0.72476236,\ 0.08704893,\ -0.97668878\bigr)^{\T}.
\end{aligned}
\label{eq:three-ou-slow-cubic-fitted}
\end{equation}
Observe that $(\rho_3^\star)^\T R_{\mathrm{OU}}^{-1}\rho_3^\star\approx 0.9933\leq 1$, as required by Proposition~\ref{prop:true-martingale}.
Finally, Table~\ref{tab:ou call-price-error-bounds} reports Monte Carlo estimates of the VIX and VXX call-price errors introduced by working with tractable proxies, together with the corresponding market bid--ask spreads.

 \begin{table}[htbp]
\centering
\begin{tabular}{lccc}
\toprule
Underlying & Maturity (days) & Error bound & Spread (min--max)\\
\midrule
VIX & 22 & $4.46\cdot 10^{-7}$ & 0.04 - 0.15\\
VXX & 24 & $4.11\cdot 10^{-3}$ & 0.13 - 1.18\\
\bottomrule
\end{tabular}
\caption{Monte Carlo estimates of upper bounds on the absolute approximation errors in European call prices, expressed in price points. The VIX bound follows from equation \eqref{eq:VIX-option-price-error}, while the VXX bound follows from equation \eqref{eq:VXX-option-price-error}. Both bounds are estimated using $10^6$ trajectories with $i^*=2$. None of the relevant VIX futures proxies reached the floor $m=0.01$ on any simulated path or grid point. Consequently, the Monte Carlo estimate of the bound in Corollary~\ref{CorVXX} was computed with no sampled contribution from the \(A_s^c\)-terms.}
\label{tab:ou call-price-error-bounds}
\end{table}

\subsection{Numerical results with two OU factors and one CIR factor}
We next consider a mixed polynomial primary process  $X_t=(X_t^1,X_t^2,Y_t)$ obtained by replacing the third OU factor with an independent CIR process. The two OU factors retain their slow and fast mean-reverting scales while the CIR introduces a nonnegative state with state-dependent instantaneous variance. Under the risk-neutral measure, the dynamics are
\begin{align*}
    dX_t^i&=\kappa_i(\theta_i-X_t^i)\,dt+\eta_i\,dW_t^i,\qquad i=1,2\\
    \dd{Y}_t&=\kappa_3(\theta_3-Y_t)\,dt+\eta_3\sqrt{Y_t}\,dW_t^3.
\end{align*}
The fixed hyperparameters used in the implementation are
\begin{center}
\begin{tabular}{lrrrr}
\toprule
Factor & $\kappa_i$ & $\theta_i$ & $\eta_i$ & Initial state\\
\midrule
Slow OU & $0.1$ & $0.1$ & $0.7$ & $1$\\
Fast OU & $25$ & $4$ & $10$ & $0.08$\\
CIR     & $4$ & $1$ & $2$ & $1$\\
\bottomrule
\end{tabular}
\end{center}
The Brownian drivers have an instantaneous correlation matrix given by $d[ W^i,W^j]_t=R_{ij}dt$ for
\begin{equation*}
 R=\begin{pmatrix}
 1&0.213&0\\
 0.213&1&0\\
 0&0&1
 \end{pmatrix}.
\end{equation*}
The Feller ratio is $2\kappa_3\theta_3/\eta_3^2=2$, so that the continuous-time CIR process started at $Y_0=1$ remains strictly positive. Note that the CIR driver is chosen to be independent of the two OU drivers to preserve the polynomial structure.  For the Brownian motion $B$ driving the SPX we again follow \eqref{eq:martingale-preserving-correlation} choosing 
$$
B_t
=
\int_0^t
\chi_L(\sigma_s^S(\ell))\rho^\T R^{-1}dW_s
+
\int_0^t
\sqrt{1-\chi_L(\sigma_s^S(\ell))^2\rho^\T R^{-1}\rho}dW_s^\perp,
$$
for $L=100$.

\paragraph{Calibration settings} The calibration, described in detail in Appendix~\ref{QuantCorner}, uses $N=10\,000$ Monte Carlo paths, with four steps per day, with the following loss function parameters $$(\lambda_{\mathrm{SPX}},\lambda_{\mathrm{VIX}},\lambda_{\mathrm{VXX}})=(0.25,0.45,0.3),\qquad(\beta_{\mathrm{SPX}},\beta_{\mathrm{VIX}},\beta_{\mathrm{VXX}})=(0.98,0.1,0.5).$$
The primary-process hyperparameters remain fixed, while the signature coefficients and the three correlation parameters $\rho$ are optimized jointly, the latter subject to $\rho^\T R^{-1}\rho<1$.

\paragraph{Parsimonious level-three enrichment.}
We use the time-extended process $\widehat X_t=(t,X_t^1,X_t^2,Y_t)$, with letters $0,1,2,3$ corresponding to time, the slow OU, the fast OU, and the CIR factor, respectively. The baseline functional retains all words of length at most two, including the empty word, and therefore has $1+4+4^2=21$ coefficients. Following again the idea proposed in Remark~\ref{dimRed}, the specification evaluated in the numerical implementation adds the three mixed time--CIR words and the repeated slow-OU word:
\begin{equation}
 \ell=\ell^{\leq2}
       +\ell_{033}e_{(0,3,3)}
       +\ell_{111}e_{(1,1,1)}
       +\ell_{303}e_{(3,0,3)}
       +\ell_{330}e_{(3,3,0)}.
\label{eq:ou-cir-enriched-ell}
\end{equation}
Consequently, the enriched VIX functional has $21+4=25$ coefficients, rather than the 85 coefficients of a complete level-three truncation. The basis keeps indices $0,\ldots,20$ and appends $36,42,72,81$ in the order $(033,111,303,330)$.  Note that the three mixed words encode time-integrated quadratic CIR information, while the repeated slow-OU word is a cubic of its terminal increment:
\begin{equation}
\begin{aligned}
    \langle e_{(0,3,3)},\wh{\X}_t\rangle&=\int_0^t\int_0^{s_3}\int_0^{s_2}\dd{s}_1\circ\dd{Y}_{s_2}\circ\dd{Y}_{s_3}\quad&
    \langle e_{(3,3,0)},\wh{\X}_t\rangle&=\int_0^t\int_0^{s_3}\int_0^{s_2}\dd{Y}_{s_1}\circ\dd{Y}_{s_2}\dd{s}_3\\
    \langle e_{(1,1,1)},\wh{\X}_t\rangle&=\frac{(X_t^1-X_0^1)^3}{6},\quad&
    \langle e_{(3,0,3)},\wh{\X}_t\rangle&=\int_0^t\int_0^{s_3}\int_0^{s_2}\dd{Y}_{s_1}\dd{s}_2\circ\dd{Y}_{s_3}
\end{aligned}
\label{eq:ou-cir-cubic-coordinates}
\end{equation}
Observing that
\begin{equation}
\begin{aligned}
 L e_{(0,3,3)}
  &=-2\kappa_3e_{(0,3,3)}-\kappa_3e_{(3,0,3)}
    +\delta_3e_{(0,3)}+\frac{\eta_3^2}{2}
       \bigl(Y_0e_{(0)}+e_{(0,3)}+e_{(3,0)}\bigr),\quad
        L e_{(3,3,0)}=e_{(3,3)},\\
 L e_{(1,1,1)}
  &=-3\kappa_1e_{(1,1,1)}+\delta_1e_{(1,1)}
    +\frac{\eta_1^2}{2}e_{(1)},\quad
 L e_{(3,0,3)}
  =-\kappa_3e_{(3,0,3)}-2\kappa_3e_{(3,3,0)}
    +\delta_3e_{(3,0)},
\end{aligned}
\label{eq:ou-cir-symbolic-cubic-closure}
\end{equation}
for $\delta_1=\kappa_1(\theta_1-X_0^1)$ and
$\delta_3=\kappa_3(\theta_3-Y_0)$, we can see that the space spanned by the complete level-two basis and the four additional words is invariant under $L$.

\begin{figure}[h]
    \centering
    \includegraphics[width=0.95\linewidth]{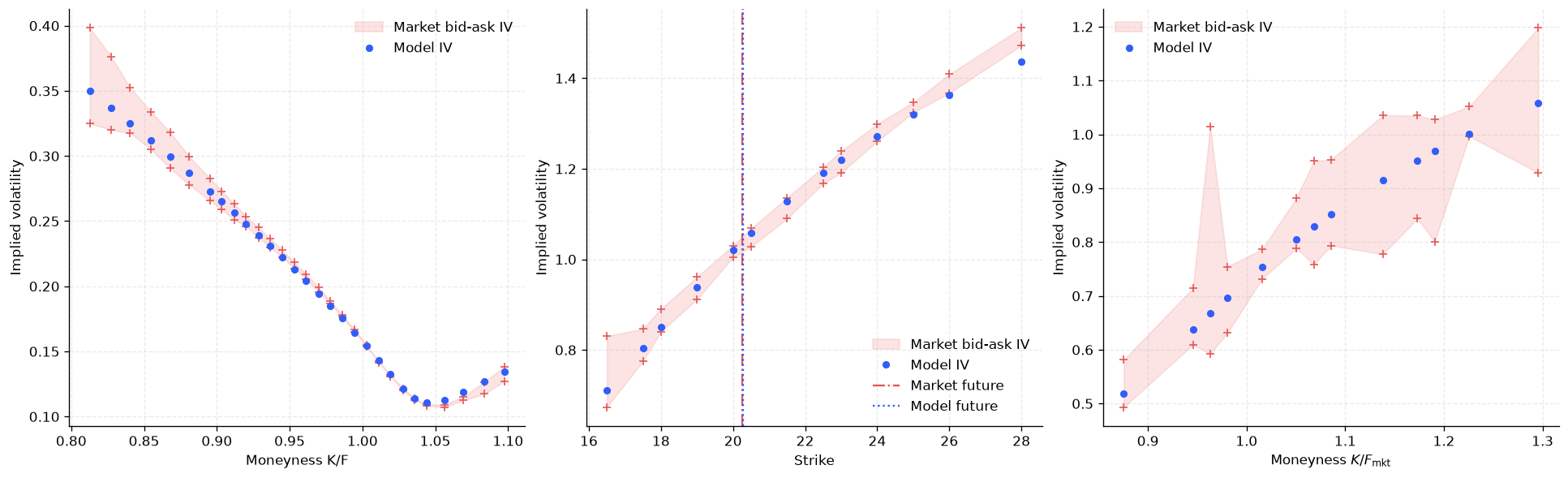}
    \caption{SPX, VIX, and VXX implied volatilities for the selected level-three OU--CIR specification, using $10^6$ independent Monte Carlo paths. The VIX functional contains the complete level-two basis together with the four additional words $(0,3,3)$, $(1,1,1)$, $(3,0,3)$, and $(3,3,0)$, for a total of 25 coefficients. Shaded regions show the Black–76 implied-volatility bands obtained from call bid and ask prices. The VIX panel also reports the model and market futures levels.}
    \label{fig:ou-cir-selected-level3-oos}
\end{figure}

 Figure~\ref{fig:ou-cir-selected-level3-oos} reports implied-volatility smiles computed with the calibrated parameters held fixed on $10^6$ independent Monte Carlo trajectories.
The optimal coefficients and the three SPX-factor correlation parameters are
\begin{equation*}
\begin{aligned}
\ell^\star
&= \left(
\begin{gathered}
    0.180019427,\ 0.461950053,\ -0.108385853,\ -0.0302125301,\\
    0.156072432,\ 6.17876383,\ 0.842840799,\ -0.000707860290,\\
    -0.863394392,\ -0.782715752,\ 0.0551015859,\ 0.0151765465,\\
    -0.0816255771,\ 0.0331908723,\ 0.00654638618,\ 0.00687084954,\\
    -0.0218291533,\ -0.209107658,\ -0.0967873223,\ -0.0217794848,\\
    0.247639880,\ 0.908004241,\ 0.345714490,\ -0.394356075,\ 1.74820083
\end{gathered}
\right)^{\mathrm{T}} \in \mathbb{R}^{25},\\[2pt]
\rho^\star
&= \bigl(0.31357466,\ 0.87752185,\ -0.46081810\bigr)^{\mathrm{T}}.
\end{aligned}
\end{equation*}
Observe that ${\rho^\star}^\T R^{-1}{\rho^\star}\approx0.9992\leq 1$,  as required by the conditions of Proposition~\ref{prop:true-martingale}.
 Finally,
 Table~\ref{tab:ou-cir-call-price-error-bounds} reports Monte Carlo estimates of the VIX and VXX call-price errors introduced by working with tractable proxies, together with the corresponding market bid--ask spreads.
\begin{table}[htbp]
\centering
\begin{tabular}{lccc}
\toprule
Underlying & Maturity (days) & Error bound & Spread (min--max)\\
\midrule
VIX & 22 & $1.7428\cdot 10^{-6}$ & 0.04 - 0.15\\
VXX & 24 & $1.01575\cdot 10^{-5}$ & 0.13 - 1.18\\
\bottomrule
\end{tabular}
\caption{Monte Carlo estimates of upper bounds on the absolute approximation errors in European call prices, expressed in price points. The VIX bound follows from equation \eqref{eq:VIX-option-price-error}, while the VXX bound follows from equation \eqref{eq:VXX-option-price-error}. Both bounds are estimated using $10^6$ trajectories with $i^*=2$. Also in this case none of the relevant VIX futures proxies reached the floor $m=0.01$ on any simulated path or grid point.}
\label{tab:ou-cir-call-price-error-bounds}
\end{table}
Figure~\ref{fig:ou-cir-selected-level3-oos} shows that the OU--CIR specification preserves the SPX downside skew obtained with the enriched three-OU model, while producing a more pronounced right-wing upturn. The VIX fit improves at several intermediate strikes, where the overestimation observed in the OU specification is reduced. The upper wing also moves closer to the market bands. The model and market VIX futures levels are nearly indistinguishable. For VXX, the implied-volatility curve shifts upwards, with all displayed model points lying within or at the boundaries of the quoted bid--ask bands.

\section{Conclusions}
\label{sec:conclusions}

The joint modeling of SPX, VIX and VXX exposes a tension between flexibility and the preservation of the financial relations linking the three markets. A specification that is convenient for VIX futures need not correspond to a nonnegative equity variance, while its implications for VXX depend on the dynamics of the futures entering the roll. Starting from the VIX makes these relations accessible through finite-dimensional operators, and regularization provides a way to retain flexible specifications when positivity cannot be established globally. The accompanying error bounds give this compromise a financial interpretation: the gain in tractability can be assessed against the pricing discrepancy it introduces. The Ornstein--Uhlenbeck and mixed Ornstein--Uhlenbeck/CIR examples also illustrate that the primary process is a substantive modeling choice. Mean-reversion scales and state-dependent covariance shape the distribution of the paths on which the signature acts, so that the signature coordinates then select which nonlinear and temporal features of those paths enter the model. This suggests developing factor dynamics and signature representations together. Alternative polynomial factors, or combinations of factors operating on different time scales, could be selected to address the maturity structure of equity and volatility smiles, while suitable invariant subspaces preserve the conditional-expectation calculations needed for variance recovery and the VXX roll. Allowing the factor parameters to participate in calibration would extend this flexibility, although it would also require adapting the current precomputation strategy. A natural objective for such developments is to obtain richer smile and term-structure behavior with a compact invariant representation and a quantitatively controlled pricing impact of regularization.
\paragraph{Data availability.}  
The historical market quotes used in this study were obtained through Cboe DataShop under license and cannot be redistributed by the authors. Calibration code, input specifications, and a synthetic example dataset are available at \href{https://github.com/andreastanghellini/Joint-calibration-SPX-VIX-VXX}{https://github.com/andreastanghellini/Joint-calibration-SPX-VIX-VXX}. Researchers with appropriate access to the specified Cboe product can use the code to reproduce the analyses.
\paragraph{AI usage disclosure} Large language models were used solely to refine language and improve the clarity and presentation of the code. All scientific content, mathematical developments, analyses, and conclusions remain entirely the work of the authors.

\begin{appendix}

\section{Square root shuffle}\label{srshuffle}
In this section, we develop a tractable method for finding an element
$\ell\in T((\mathbb{R}^d))$ whose shuffle square equals a given
$h\in T((\mathbb{R}^d))$.
Let $S^{n,m}$ be the set of permutations introduced in~\eqref{shuffleset}.

\begin{lemma}\label{shuffle of l with itself}
 For $\ell\in T(\R^d)$, $I=(i_1,\ldots ,i_n)$, and $k\in\{1,\ldots, n-1\}$ set
    \begin{equation}\label{shuffle as index}
        \ell_{I}^{k,n-k}=\sum_{\sigma\in S^{k,n-k}}\ell_{(i_{\sigma(1)},\ldots,i_{\sigma(k)})}\ell_{(i_{\sigma(k+1)},\ldots,i_{\sigma(n)})}
    \end{equation}
    and 
    $
        \ell_I^{0,n}:=\ell_\emptyset\ell_I=:\ell_I^{n,0}.
  $
     Then
    $\langle e_I,\ell\shuffle \ell\rangle=\sum_{k=0}^n \ell_I^{n-k,k}$.
\end{lemma}
\begin{proof}
By bilinearity of the shuffle product, we have that $\langle e_I,\ell\shuffle \ell\rangle =
    \sum_{m_1+m_2= n}\sum_{|J_1|=m_1, |J_2|=m_2}\ell_{J_1}\ell_{J_2} \langle e_I,e_{J_1}\shuffle e_{J_2}\rangle$. Setting $J=J_1J_2$, observe that by the definition of the shuffle product it holds
    $$\langle e_I,e_{J_1}\shuffle e_{J_2}\rangle=\sum_{\sigma\in S^{m_1,m_2}}
        \langle e_I, e_{\sigma^{-1}(J)}\rangle=\sum_{\sigma\in S^{m_1,m_2}}\langle e_{\sigma(I)}, e_{J}\rangle=\sum_{\sigma\in S^{m_1,m_2}}\mathbbm{1}_{\{\sigma(I)=J_1J_2\}}.$$
    Substituting this into the original expression, we can then conclude that
\begin{align*}
    \langle e_I,\ell\shuffle \ell\rangle 
     &=\sum_{m_1+m_2= n}\sum_{|J_1|=m_1, |J_2|=m_2}\sum_{\sigma\in S^{m_1,m_2}} \ell_{J_1 }\ell_{J_2}
   \mathbbm{1}_{\{\sigma(I)=J_1J_2\}}\\
    &=\sum_{m_1+m_2= n}\sum_{|J_1|=m_1, |J_2|=m_2}\sum_{\sigma\in S^{m_1,m_2}} \ell_{(\sigma(I)_1,\ldots,\sigma(I)_{m_1}) }\ell_{(\sigma(I)_{m_1+1},\ldots,\sigma(I)_{n} )}
   \mathbbm{1}_{\{\sigma(I)=J_1J_2\}}\\
   &=\sum_{m_1+m_2= n}\sum_{\sigma\in S^{m_1,m_2}} \ell_{(\sigma(I)_1,\ldots,\sigma(I)_{m_1}) }\ell_{(\sigma(I)_{m_1+1},\ldots,\sigma(I)_{n} )}
   \sum_{|J_1|=m_1, |J_2|=m_2}\mathbbm{1}_{\{\sigma(I)=J_1J_2\}}\\
   &=\sum_{m_1+m_2= n}\ell_I^{m_1,m_2},
\end{align*}
proving the claim.
\end{proof}
 We give some additional examples.
\begin{example}
    Let $I=(i_1,i_2,i_3)$ and $k=2$. We know that there are $3$ permutations in $S_n$, with $n=|I|=3$, that preserve the order of the first $k$ elements, i.e.~$\sigma(I)\in\{(i_1,i_2,i_3),(i_1,i_3,i_2),(i_2,i_3,i_1)\}$.
    We thus get that
    \begin{equation*}
        \ell_I^{2,1}=\sum_{\sigma\in S^{2,1}}\ell_{(i_{\sigma(1)},i_{\sigma(2)})}\ell_{(i_{\sigma(3)})}=\ell_{(i_1,i_2)}\ell_{i_3}+\ell_{(i_1,i_3)}\ell_{i_2}+\ell_{(i_2,i_3)}\ell_{i_1}.
    \end{equation*}
    For $I=(i_1,i_2,i_3,i_4)$ we would similarly get
    \begin{align*}
        \ell_{I}^{3,1}&=\sum_{\sigma\in S^{3,1}}\ell_{(i_{\sigma(1)},i_{\sigma(2)},i_{\sigma(3)})}\ell_{(i_{\sigma(4)})}=\ell_{(i_1,i_2,i_3)}\ell_{i_4}+\ell_{(i_1,i_2,i_4)}\ell_{i_3}+\ell_{(i_1,i_3,i_4)}\ell_{i_2}+\ell_{(i_2,i_3,i_4)}\ell_{i_1}\\
            \ell_{I}^{2,2}&=\sum_{\sigma\in S^{2,2}}\ell_{(i_{\sigma(1)},i_{\sigma(2)})}\ell_{(i_{\sigma(3)},i_{\sigma(4)})}\\
            &=\ell_{(i_1,i_2)}\ell_{(i_3,i_4)}+\ell_{(i_1,i_3)}\ell_{(i_2,i_4)}+\ell_{(i_1,i_4)}\ell_{(i_2,i_3)}+\ell_{(i_2,i_3)}\ell_{(i_1,i_4)}+\ell_{(i_2,i_4)}\ell_{(i_1,i_3)}+\ell_{(i_3,i_4)}\ell_{(i_1,i_2)}.   
    \end{align*}
\end{example}

\begin{proposition}\label{propsqrt}
Fix $h\in T((\R^d))$  { and assume that} $h_\emptyset>0$ and there exists  $\ell\in T((\R^d))$ such that $\ell\shuffle\ell=h$. { Then, }  $\ell$ satisfies $\ell_\emptyset^2={h_\emptyset}$
    and for each $I=(i_1,\ldots,i_n)$  it holds
    \begin{equation}\label{sqrtsh}
        \ell_I
         =
        \dfrac{h_I-\langle e_I,\ell^{\leq n-1}\shuffle \ell^{\leq n-1}\rangle}{2\ell_\emptyset}.
    \end{equation}
\end{proposition}
\begin{proof}
Recall that for each $I$ and $J$ it holds that $\langle e_H,e_I\shuffle e_J\rangle=0$ for each $H$ such that $|H|\neq |I|+|J|$.
Since $\ell\shuffle\ell=h$, we know that
$$\ell_\emptyset^2=\langle e_\emptyset,\ell\shuffle \ell\rangle=\langle e_\emptyset,h\rangle=h_\emptyset,$$
showing that $\ell_\emptyset^2=h_\emptyset$. 
Moreover, for each multi-index $I=(i_1,\ldots, i_n)$ 
$$h_I=\langle e_I,\ell\shuffle \ell\rangle
=\langle e_I,\ell^{\le n}\shuffle \ell^{\leq n}\rangle
=\langle e_I,(\ell^{\leq n-1}+\sum_{|J|=n}\ell_Je_J)\shuffle (\ell^{\leq n-1}+\sum_{|J|=n}\ell_Je_J)\rangle$$
and by bilinearity, this yields
$$2\langle e_I,\ell_\emptyset e_\emptyset\shuffle\sum_{|J|=n}\ell_Je_J\rangle
+\langle e_I,\ell^{\leq n-1}\shuffle \ell^{\leq n-1}\rangle
=
2\ell_\emptyset \ell_I\langle e_I,e_I\rangle
+\langle e_I,\ell^{\leq n-1}\shuffle \ell^{\leq n-1}\rangle=2\ell_I \ell_\emptyset+\langle e_I,\ell^{\leq n-1}\shuffle \ell^{\leq n-1}\rangle,
$$
proving the claim.
\end{proof}

\begin{remark}\label{shsqrt}
    Observe that the map $h\mapsto \ell$ induced by \eqref{sqrtsh}  is well defined on $\{h\in T((\R^d))\colon h_\emptyset> 0\}$. We can indeed recursively define $h^{\shuffle/2}\in T((\R^d))$ as $h^{\shuffle/2}_\emptyset=\sqrt{h_\emptyset}$ and
$$h^{\shuffle/2}_I:=\Big(h_{I}-\langle e_I,h^{\shuffle/2,\leq |I|-1}\shuffle h^{\shuffle/2,\leq |I|-1}\rangle\Big)
(2h^{\shuffle/2}_\emptyset)^{-1},$$
for each $|I|> 0$. By construction it holds
$$h^{\shuffle/2}\shuffle h^{\shuffle/2}=h.$$
What is not clear is whether or under which conditions
$\langle h^{\shuffle/2,\leq m},\X_T\rangle^2$ provides a good approximation of $\langle h,\X_T\rangle$.
\end{remark}

\begin{corollary}
    Fix $h\in T^n(\R^d)$  { and assume that} $h_\emptyset>0$. Define $\ell\in T^{n}(\R^d)$ through \eqref{sqrtsh} and assume that for each $I$ with $|I|\in (n/2,n]$ it holds
    $$
        h_I=\langle e_I,\ell^{\leq \lfloor n/2\rfloor}\shuffle \ell^{\leq \lfloor n/2\rfloor}\rangle.$$
        Then $\ell^{\leq \lfloor n/2\rfloor}\shuffle \ell^{\leq \lfloor n/2\rfloor}=h$.
\end{corollary}
\begin{proof}
    The provided assumption guarantees that $\ell_I=0$ for each $|I|\in ( n/2,n]$ and thus that
    $\langle e_I, \ell\shuffle \ell\rangle=0$
    for each $|I|>n$.
\end{proof}

\begin{remark}
    Observe that condition \eqref{sqrtsh} can be checked recursively on the length of $I$. Observe that if $\ell$ satisfies \eqref{sqrtsh}, so does $-\ell$; hence, assume without loss of generality that $\ell_\emptyset>0$. Moreover, $\langle e_I,\ell^{\leq n-1}\shuffle \ell^{\leq n-1}\rangle$ does not depend on $\ell_\emptyset$, which can thus be treated as $0$.
    \begin{align*}
        \ell_\emptyset&=\sqrt{h_\emptyset},\\
        \ell_i&=(h_i -0)/(2\ell_\emptyset)=h_i/(2\ell_\emptyset),\\
\ell_{(i_1,i_2)}&=(h_{(i_1,i_2)} -\langle e_{(i_1,i_2)},\sum_{j_1,j_2}\ell_{j_1}\ell_{j_2}
e_{j_1}\shuffle e_{j_2}\rangle)/(2\ell_\emptyset)
=(h_{(i_1,i_2)} -\sum_{j_1,j_2}\ell_{j_1}\ell_{j_2}\langle e_{(i_1,i_2)},
e_{(j_1,j_2)}+e_{(j_2,j_1)}\rangle)/(2\ell_\emptyset)\\
&=(h_{(i_1,i_2)}-2\ell_{i_1}\ell_{i_2})/(2\ell_\emptyset),\ldots
\end{align*}
\end{remark}
We conclude the section by constructing a more explicit representation of the map $h^{\shuffle/2}:h\mapsto\ell$ induced by \eqref{sqrtsh}. Let $\ell^{n,k}_I$ be as in Lemma~\ref{shuffle of l with itself}.

\begin{corollary}
    Let $h\in T((\R^d))$ satisfy $h_\emptyset>0$ { and assume that} there exists  $\ell\in T((\R^d))$ such that $\ell\shuffle\ell=h$. { Then,} this $\ell$ satisfies $\ell_\emptyset^2=h_\emptyset$
    and for a given multi-index $I=(i_1,\ldots,i_n)$  it holds
    \begin{equation}\label{square_root_shuffle}
        \ell_I=\dfrac{h_I-\sum_{k=1}^{n-1}\ell_I^{k,n-k}}{2{\ell_\emptyset}}.
    \end{equation}
\end{corollary}
\begin{proof}
    The proof follows directly from Proposition~\ref{propsqrt} and Lemma~\ref{shuffle of l with itself}.
\end{proof}

\section{Proofs of results in the one-dimensional case}\label{app1d}

\begin{proof}[Proof of Lemma~\ref{lem1d}]
Observe that the dual operator $L$ corresponding to $\X$ is given by
\[
L e_1^{\otimes k}
=
b e_1^{\otimes (k-1)}\mathbbm{1}_{\{k\geq 1\}}
+\frac{a}{2}e_1^{\otimes (k-2)}\mathbbm{1}_{\{k\geq 2\}}
+\frac{(k-1)\alpha}{2}e_1^{\otimes (k-1)}
\mathbbm{1}_{\{k\geq 1\}}.
\]
This implies that the corresponding matrix representative $G$ reads
\[
G =
\begin{pmatrix}
0 & b & a/2 & 0 & 0 \\
0 & 0 & b + \alpha/2 & a/2 & 0 \\
0 & 0 & 0 & b + \alpha & a/2 \\
0 & 0 & 0 & 0 & b + 3\alpha/2 \\
0 & 0 & 0 & 0 & 0
\end{pmatrix}.
\]
Since $G$ is strictly upper triangular, it is nilpotent and all its
eigenvalues are equal to $0$ and hence
$
A=\int_0^\Delta e^{tG^\T}\dd t\in\R^{5\times 5}
$
 is invertible.
Moreover, since $(G^\T)^5=0$, we have
\[
A^{-1}
=
\left(\int_0^\Delta \exp(tG^\T)\dd t\right)^{-1}
=
\frac{1}{\Delta}
\sum_{n=0}^{4}
\frac{B_n}{n!}(\Delta G^\T)^n,
\]
where $B_n$ denotes the $n$-th Bernoulli number,
\[
B_0=1,\qquad
B_1=-\frac12,\qquad
B_2=\frac16,\qquad
B_3=0,\qquad
B_4=-\frac1{30}.
\]
Explicitly,
\[
A^{-1}
=
\begin{pmatrix}
\frac{1}{\Delta}
& 0
& 0
& 0
& 0
\\[1mm]
-\frac{b}{2}
& \frac{1}{\Delta}
& 0
& 0
& 0
\\[1mm]
\frac{\Delta\alpha b}{24}
+\frac{\Delta b^2}{12}
-\frac{a}{4}
&
-\frac{\alpha}{4}-\frac{b}{2}
&
\frac{1}{\Delta}
& 0
& 0
\\[1mm]
\frac{\Delta a(\alpha+2b)}{24}
&
\frac{\Delta\alpha^2}{24}
+\frac{\Delta\alpha b}{8}
+\frac{\Delta b^2}{12}
-\frac{a}{4}
&
-\frac{\alpha}{2}-\frac{b}{2}
&
\frac{1}{\Delta}
& 0
\\[1mm]
\frac{\Delta\left(
-3\Delta^2\alpha^3b
-11\Delta^2\alpha^2b^2
-12\Delta^2\alpha b^3
-4\Delta^2b^4
+60a^2
\right)}{2880}
&
\frac{\Delta a(\alpha+b)}{12}
&
\frac{\Delta\alpha^2}{8}
+\frac{5\Delta\alpha b}{24}
+\frac{\Delta b^2}{12}
-\frac{a}{4}
&
-\frac{3\alpha}{4}-\frac{b}{2}
&
\frac{1}{\Delta}
\end{pmatrix}.
\]

Next, fix $\ell\in T^2(\R)$ and note that
\[
\ell\shuffle\ell
=
\left(
\ell_\emptyset^2,
2\ell_\emptyset\ell_1,
2\ell_1^2+2\ell_\emptyset\ell_{11},
6\ell_1\ell_{11},
6\ell_{11}^2
\right).
\]
The explicit representations of $\VIX_t(\ell)$, $F_t(T,\ell)$ and
$\sigma_t^S(\ell)^2$ are given by
\begin{align*}
\VIX_t
&=
\langle\ell,\mathbb{X}^{\leq 2}_t\rangle
=
\ell_\emptyset
+\ell_1 (X_t-X_0)
+\ell_{11}\frac{(X_t-X_0)^2}{2},
\\
F_t(T,\ell)
&=
(\ell_\emptyset,\ell_1,\ell_{11},0,0)
e^{(T-t)G^\T}
\ve(\mathbb{X}^{\leq 4}_t)
\\
&=
(\ell_\emptyset,\ell_1,\ell_{11})
\begin{pmatrix}
1 & 0 & 0
\\
b(T-t) & 1 & 0
\\
\frac{(T-t)\left(2a+b(T-t)(\alpha+2b)\right)}{4}
&
\frac{(T-t)\alpha}{2}+(T-t)b
&
1
\end{pmatrix}
\ve(\mathbb{X}_t^{\leq 2}),
\\
\sigma_t^S(\ell)^2
&=
\Delta\,
\ve(\ell\shuffle\ell)^\T
A^{-1}
\ve(\mathbb X^{\leq 4}_t)
\\
&=
\left(
\ell_\emptyset^2,
2\ell_\emptyset\ell_1,
2\ell_1^2+2\ell_\emptyset\ell_{11},
6\ell_1\ell_{11},
6\ell_{11}^2
\right)
\Delta A^{-1}\ve(\mathbb X^{\leq 4}_t).
\end{align*}
\end{proof}

\begin{proof}[Proof of Corollary~\ref{corPippo}]
    The representations of $F_t(T,\ell)$ and $\sigma_t^S(\ell)^2$ follow from Lemma~\ref{lem1d}. Concerning  $\VXX$, choosing $b=0$ and inserting the explicit representation of $F_t(T,\ell)$ into the formulas in Proposition~\ref{corVXX} we get
\begin{align*}
    P_t(\ell)&:=\sum_{i=1}^\infty\Big((t-T_{i-1})F_t(T_{i+1},\ell)+(T_i-t)F_t(T_i,\ell)\Big)\mathbbm{1}_{\{t\in(T_{i-1},T_i]\}}
    =\tau\VIX_t+\frac{\ell_{11}(a+\alpha (X_t-X_0))\tau^2}{2}
\end{align*}
and hence
\begin{align*}
    \VXX_t(\ell)
    &=\VXX_0\times\frac{P_t(\ell)}{P_0(\ell)}\times\exp\bigg(\sum_{i=1}^\infty\int_{(t\land T_{i-1})\lor 0}^{t\land T_i} \frac{F_s(T_i,\ell)-F_s(T_{i+1},\ell)}{(s-T_{i-1})F_{s}(T_{i+1},\ell)+(T_i-s)F_{s}(T_i,\ell)}\dd{s}\bigg)\\
    &=\VXX_0\times\frac{P_t(\ell)}{P_0(\ell)}\times\exp\bigg(\sum_{i=1}^\infty\int_{(t\land T_{i-1})\lor 0}^{t\land T_i}\frac{-\frac{\ell_{11}(a+\alpha (X_s-X_0))}{2}\tau}{\tau\VIX_s+\frac{\ell_{11}(a+\alpha (X_s-X_0))\tau^2}{2}}ds\bigg)\\
    &=\VXX_0\times\frac{2\VIX_t+\ell_{11}(a+\alpha (X_t-X_0))\tau}{2\ell_\emptyset+\ell_{11}a\tau}\times\exp\bigg(-\int_0^t \frac{\ell_{11}(a+\alpha (X_s-X_0))}{2\VIX_s+\ell_{11}(a+\alpha (X_s-X_0))\tau}\dd{s}\bigg),
\end{align*}
and the claim follows. For nonnegativity, we study the two cases separately, starting with the Brownian case.
Without loss of generality, assume that $\ell_{11}=1$. To guarantee that $\VIX_t$ is nonnegative it is sufficient to require that $ \ell_\emptyset\geq\ell_1^2/2$, whereas, for $\sigma^S_t(\ell)^2$, we need to study the nonnegativity of the polynomial on $\R$
$$p(x)=\ell_\emptyset^2-\frac \Delta 2(\ell_1^2+\ell_\emptyset)+\frac{\Delta^2}{8}+\bigg(2\ell_\emptyset\ell_1-\frac{\Delta}{2}3\ell_1\bigg)x+\bigg(\ell_1^2+\ell_\emptyset-\frac{3\Delta}{4}\bigg)x^2+\ell_1x^3+\frac{x^4}{4}.$$
Its derivative reads 
$$(\ell_1+x)\bigg(2\ell_\emptyset+2\ell_1x+x^2-\dfrac{3{ \Delta} }{2}\bigg)$$
and its zeros are thus given by $x_1=-\ell_1$ and $x_{2,3}=-\ell_1\pm z$ for
$z=\sqrt{\ell_1^2-\Big(2\ell_\emptyset-\frac{3\Delta}{2}\Big)}.$
By symmetry, it is sufficient to find conditions on $\ell_\emptyset$ such that $p(x_1), p(x_2)\geq 0$. As
\begin{align*}
    p(-\ell_1)
    &=\ell_\emptyset^2-\bigg(\frac{\Delta}{2}+\ell_1^2\bigg)\ell_\emptyset+\bigg(\frac{\Delta \ell_1^2}{4}+\frac{\Delta^2}{8}+\frac{\ell_1^4}{4}\bigg)
\end{align*}
we can conclude that, since
$$\bigg(\frac{\Delta}{2}+\ell_1^2\bigg)^2-\bigg(\Delta \ell_1^2+\frac{\Delta^2}{2}+\ell_1^4\bigg)=-\frac{\Delta^2}{4}\leq 0,$$
$p(x_1)$ is nonnegative for each $\ell_\emptyset$ and $\ell_1$.
For $p(x_2)$, first observe that since
$$p(x)=p(0)+\bigg(\dfrac{1}{4}(x+\ell_1)^4-\dfrac{z^2}{2}(x+\ell_1)^2-\dfrac{\ell_1^4}{4}+\dfrac{z^2\ell_1^2}{2}\bigg)$$
we get that
\begin{align*}
    p(x_2)&=p(0)+\bigg(\dfrac{1}{4} z^4-\dfrac{ z^2}{2} z^2-\dfrac{\ell_1^4}{4}+\dfrac{ z^2\ell_1^2}{2}\bigg)\\
    &=p(0)+\bigg(-\dfrac{1}{4}\bigg(\ell_1^2-\Big(2\ell_\emptyset-\frac{3\Delta}{2}\Big)\bigg)^2-\dfrac{\ell_1^4}{4}+\dfrac{\ell_1^4-\ell_1^2\big(2\ell_\emptyset-\frac{3\Delta}{2}\big)}{2}\bigg)\\
    &=\ell_\emptyset^2-\frac \Delta 2(\ell_1^2+\ell_\emptyset)+\frac{\Delta^2}{8}+\bigg(-\dfrac{1}{4}\bigg(\ell_1^2-\Big(2\ell_\emptyset-\frac{3\Delta}{2}\Big)\bigg)^2-\dfrac{\ell_1^4}{4}+\dfrac{\ell_1^4-\ell_1^2\big(2\ell_\emptyset-\frac{3\Delta}{2}\big)}{2}\bigg)\\
    &=\Delta\bigg(\ell_\emptyset-\frac{\ell_1^2}{2}-\frac{7\Delta}{16}\bigg),
\end{align*}
which is nonnegative if and only if $\ell_\emptyset\geq \frac {\ell_1^2}2 +\frac{7\Delta}{16}$.

For the second choice of the parameters, recall that $X_t\geq0$. We distinguish the
cases $\ell_{11}=0$ and $\ell_{11}>0$.

First, let $\ell_{11}=0$. If $\ell_1=0$, then
$\VIX_t=\ell_\emptyset$ and
$\sigma_t^S(\ell)^2=\ell_\emptyset^2$, so it is sufficient to require
$\ell_\emptyset\geq0$. Suppose now that $\ell_1>0$. In this case,
$
\VIX_t
=
\ell_\emptyset+\ell_1(X_t-X_0)
$
and, writing $\sigma_t^S(\ell)^2=p(X_t)$, we have
\[
\frac{p(x)}{\ell_1^2}
=
\left(
x+\frac{\ell_\emptyset}{\ell_1}-X_0
\right)^2
-\frac{\Delta x}{2}.
\]
Its minimum over $\mathbb R$ is
\[
\min_{x\in\mathbb R}\frac{p(x)}{\ell_1^2}
=
\frac{\Delta}{16}
\left(
8\left(\frac{\ell_\emptyset}{\ell_1}-X_0\right)-\Delta
\right).
\]
Consequently, $p(x)\geq0$ for every $x\geq0$ whenever
\[
\ell_\emptyset
\geq
\ell_1X_0+\frac{\Delta\ell_1}{8}.
\]
Under the same condition,
$
\VIX_t
=
\ell_\emptyset-\ell_1X_0+\ell_1X_t
\geq0,
$
because $\ell_1\geq0$ and $X_t\geq0$.

Consider now the case $\ell_{11}>0$. Since both $\VIX$ and
$\sigma^S(\ell)^2$ are homogeneous in $\ell$, with degrees one and two,
respectively, divide the coefficients by $\ell_{11}$ and set
$
r:=\frac{\ell_1}{\ell_{11}}-X_0,
$ and $
c:=
\frac{\ell_\emptyset}{\ell_{11}}
-\frac{\ell_1}{\ell_{11}}X_0
+\frac{X_0^2}{2}.
$
Then
$
\frac{\VIX_t}{\ell_{11}}
=
c+rX_t+\frac{X_t^2}{2}.
$
Moreover, writing
$p(X_t)=
\frac{\sigma_t^S(\ell)^2}{\ell_{11}^2},
$
we obtain
\[
\begin{aligned}
p(x)
&=
c^2
+\left(
\frac{\Delta^2r}{4}
-\frac{\Delta}{2}(r^2+c)
+2rc
\right)x
\\
&\quad
+\left(
\frac{3\Delta^2}{4}
-3\Delta r
+2r^2+2c
\right)\frac{x^2}{2}
+\left(
-\frac{9\Delta}{2}+6r
\right)\frac{x^3}{6}
+6\frac{x^4}{24}.
\end{aligned}
\]
Since $x\geq0$, a sufficient condition for $p(x)\geq0$ is that all
its monomial coefficients be nonnegative. The coefficient of $x^3$
is nonnegative if and only if
$
r\geq\frac{3\Delta}{4}
$ and the coefficient of $x$ is nonnegative
if and only if
$
\frac{\Delta^2r}{4}
-\frac{\Delta}{2}(r^2+c)
+2rc
\geq0.
$
Similarly, the coefficient of $x^2$ is nonnegative if and only if
\[
c
\geq
-\frac{3\Delta^2}{8}
+\frac{3\Delta r}{2}
-r^2
=
\frac{3\Delta^2}{16}
-\left(r-\frac{3\Delta}{4}\right)^2.
\]
Therefore, it is sufficient to require
$
r\geq\frac{3\Delta}{4}
$
and
\[
c
\geq
\max\left\{
\frac{\Delta r(2r-\Delta)}
     {2(4r-\Delta)},
\frac{3\Delta^2}{16}
-\left(r-\frac{3\Delta}{4}\right)^2
\right\}.
\]
Notice that the first quantity in the maximum is nonnegative whenever
$r\geq3\Delta/4$. Hence $c\geq0$, and consequently
\[
\frac{\VIX_t}{\ell_{11}}
=
c+rX_t+\frac{X_t^2}{2}
\geq0
\qquad\text{for every }X_t\geq0.
\]
Substituting the definitions of $r$ and $c$ gives precisely
$
\frac{\ell_1}{\ell_{11}}
\geq
X_0+\frac{3\Delta}{4}
$
and
\[
\begin{aligned}
\frac{\ell_\emptyset}{\ell_{11}}
&\geq
\frac{\ell_1}{\ell_{11}}X_0-\frac{X_0^2}{2}
+
\max\Bigg\{
\frac{
\Delta\left(\frac{\ell_1}{\ell_{11}}-X_0\right)
\left[
2\left(\frac{\ell_1}{\ell_{11}}-X_0\right)-\Delta
\right]
}{
2\left[
4\left(\frac{\ell_1}{\ell_{11}}-X_0\right)-\Delta
\right]
},
\frac{3\Delta^2}{16}
-
\left(
\frac{\ell_1}{\ell_{11}}-X_0-\frac{3\Delta}{4}
\right)^2
\Bigg\}.
\end{aligned}
\]
These conditions therefore guarantee the nonnegativity of both
$\VIX_t$ and $\sigma_t^S(\ell)^2$.
\end{proof}

\section{Computational aspects}\label{QuantCorner}
This section is dedicated to the computational aspects needed for the calibration to the prices of the contracts introduced in Section~\ref{Calibration to option prices}. Firstly, following \cite{CGMS:24}, we  introduce the corresponding loss function.  Motivated by Remark 5.9 in \cite{CGMS:24}, we propose, for a fixed maturity $T$ and strike price $K$, the following loss function for $\beta\in [0,1]$
\begin{align}\label{greek_loss}
    \mathcal{L}^{\beta,T,K}_{ \pi^{\mathrm{mkt},b,a}, r,F^{\mathrm{mkt}}}(\pi, F)= \left(\frac{\beta\, d(\pi;\pi^{\mathrm{mkt},b},\pi^{\mathrm{mkt},a})+(1-\beta)\big|\pi-\pi^{\mathrm{mkt,mid}}\big|+\big|\delta_{T,K}^{F,\mathrm{mkt}}(F-F^{\mathrm{mkt}})\big|}{D_{T,K}^{\mathrm{mkt}}}\right)^2,
\end{align}
where
\begin{itemize}
    \item $\pi$ denotes the candidate price of a call option with maturity $T$ and strike $K$, while $F$ denotes the candidate forward level of the underlying at maturity $T$. 
    \item $\pi^{\mathrm{mkt},b,a}=(\pi^{\mathrm{mkt},b},\pi^{\mathrm{mkt},a})$ denotes   the  reference bid and ask prices of the same call option. $F^{\mathrm{mkt}}>0$ denotes the corresponding reference forward level. We also define the mid price as
    $$\pi^{\mathrm{mkt,mid}}=\frac{\pi^{\mathrm{mkt},a}+\pi^{\mathrm{mkt},b}}{2}.$$
    All these reference values need to be extracted from the market. 
    \item $\sigma_{T,K}^{\mathrm{mkt},b}$ and $\sigma_{T,K}^{\mathrm{mkt},a}$ denote the Black--76 implied volatilities corresponding to $\pi^{\mathrm{mkt},b}$ and $\pi^{\mathrm{mkt},a}$, respectively, computed using the same reference forward $F^{\mathrm{mkt}}$, interest rate $r$, maturity $T$ and strike $K$. Moreover, we set
    $$\bar\sigma_{T,K}^{\mathrm{mkt}}=\frac{\sigma_{T,K}^{\mathrm{mkt},b}+\sigma_{T,K}^{\mathrm{mkt},a}}{2}.$$
    \item The reference Vega and forward Delta are computed under Black--76 at the market forward and at $\bar\sigma_{T,K}^{\mathrm{mkt}}$:
    $$\nu_{T,K}^{\mathrm{mkt}}=e^{-rT}F^{\mathrm{mkt}}\phi(d_1)\sqrt{T},\qquad\delta_{T,K}^{F,\mathrm{mkt}}=e^{-rT}\Phi(d_1)$$
    where
    $$d_1=\frac{\log(F^{\mathrm{mkt}}/K)+\frac12(\bar\sigma_{T,K}^{\mathrm{mkt}})^2T}{\bar\sigma_{T,K}^{\mathrm{mkt}}\sqrt{T}},$$
    and $\Phi$ and $\phi$ denote the standard normal distribution function and density. 
    Finally, we set
    $$D_{T,K}^{\mathrm{mkt}}=\max\left\{\nu_{T,K}^{\mathrm{mkt}}\bigl(\sigma_{T,K}^{\mathrm{mkt},a}-\sigma_{T,K}^{\mathrm{mkt},b}\bigr),10^{-12}\right\}.$$
    \item The distance $d$ from the bid--ask band is defined by
    $d(\pi;b,a):=(b-\pi)^++(\pi-a)^+.$
    \item$\beta\in[0,1]$ balances the distance from the bid--ask interval versus the distance from the mid price.
\end{itemize}

Observe that when $F=F^{\mathrm{mkt}}$, the forward-error term vanishes; in this case, we make use of the abbreviated notation
$$\begin{aligned}
    \mathcal{L}_{\pi^{\mathrm{mkt},b,a},r,F^{\mathrm{mkt}}}^{\beta,T,K}(\pi)&:=\mathcal L_{\pi^{\mathrm{mkt},b,a},r,F^{\mathrm{mkt}}}^{\beta,T,K}(\pi,F^{\mathrm{mkt}})
    =\left(\frac{(1-\beta)|\pi-\pi^{\mathrm{mkt,mid}}|+\beta\,d(\pi;\pi^{\mathrm{mkt},b},\pi^{\mathrm{mkt},a})}{D_{T,K}^{\mathrm{mkt}}}\right)^2.
\end{aligned}$$
\noindent Let $\vartheta=(\ell,\rho)$ collect the signature coefficients and the three parameters correlating the pricing Brownian motion to the primary-process Brownian drivers. The joint calibration loss function is
\begin{equation}
    \mathcal{L}_{\mathrm{joint}}(\vartheta)=\sum_{A\in\{\SPX,\VIX,\VXX\}}\lambda_A\frac{\mathcal B_A(\vartheta)}{s_A}.
    \label{eq:joint-calibration-loss}
\end{equation}
Here $\lambda_A>0$ and $\sum_A\lambda_A=1$. Writing $n_A=|\mathcal K^A|$ to denote the number of calibrated maturity–strike pairs for product $A$, the fixed block scales are
$$s_A=n_A\left[\max\left\{\frac{\mathcal B_A(\vartheta^{(0)})}{n_A},10^{-12}\right\}\right]^{1/2},$$
where $\vartheta^{(0)}$ is the feasible initial parameter vector. The weights and scales are computed before optimization and remain fixed throughout it. The three product blocks appearing in the joint objective are defined by
\begin{align*}
    \mathcal{B}_{\SPX}(\vartheta)&=\sum_{(T,K)\in\Kcal^{\SPX}}\Lcal^{\beta_\SPX,T,K}_{\pi_{\SPX}^{\mathrm{mkt},b,a}(T,K),r,F^{\mkt}_\SPX(T)}(\pi_{\SPX}^{\text{model}}(\ell,\rho;T,K)),\\
    \mathcal{B}_{\text{VIX}}(\vartheta)&=\sum_{(T,K)\in\Kcal^{\VIX}}\Lcal^{\beta_\VIX,T,K}_{\pi_{\VIX}^{\mathrm{mkt},b,a}(T,K),r,F^{\mkt}_\VIX(T)}\left(\pi_{\VIX}^{\text{model}}(\ell,T,K),F_{\VIX}^{\text{model}}(\ell,T)\right),\\
    \mathcal{B}_{\VXX}(\vartheta)&=\sum_{(T,K)\in\Kcal^{\VXX}}\Lcal^{\beta_\VXX,T,K}_{\pi_{\VXX}^{\mathrm{mkt},b,a}(T,K),r,F^{\mkt}_\VXX(T)}(\pi_{\VXX}^{\text{model}}(\ell,T,K)).
\end{align*}
Here $\pi_A^{\mathrm{model}}$ denotes the model call price, evaluated using the pricing proxies and numerical scheme described below, and $F_{\mathrm{VIX}}^{\mathrm{model}}$ denotes the VIX futures proxy. With the factor dynamics held fixed, SPX call prices depend on both $\ell$ and $\rho$, whereas VIX and VXX call prices and the VIX futures proxy depend only on $\ell$. The sets $\mathcal{K}^A$ contain the maturity--strike pairs used for calibration.

After calibration, all fitted parameters are held fixed and option
prices are recomputed on $N_{\mathrm{OOS}}=10^6$ independent paths.
This validation uses independent Monte Carlo paths with the calibration date and market quotes held fixed.
To limit memory use, the sample is processed in batches. If batch $b$
contains $m_b$ observations of a discounted payoff $H$, its estimates
are combined as
\begin{equation}
\widehat{\mathbb E}[H]=\frac{1}{N_{\mathrm{OOS}}}\sum_{b=1}^{B_{\mathrm{MC}}}\sum_{i=1}^{m_b}H(\omega_{b,i})=\sum_{b=1}^{B_{\mathrm{MC}}}\frac{m_b}{N_{\mathrm{OOS}}}\widehat{\mathbb E}_b[H],\qquad\sum_bm_b=N_{\mathrm{OOS}}.
 \label{eq:three-ou-batched-monte-carlo}
\end{equation}

Consider now a regularized VIX-based signature model as in Definition~\ref{regVIXbased}. Fix $R$ and $\rho$ as in Proposition~\ref{prop:true-martingale}, and suppose that the Brownian motion $B$ driving the price is given by \eqref{eq:martingale-preserving-correlation} for some $L>0$. In the numerical implementation, we use the absolute-value
VIX proxy
$$\overline{\mathrm{VIX}}_t(\ell):=\left|\langle\ell,\widehat{\mathbb X}_t^{\le n}\rangle\right|$$
and the signed $\VIX$ futures proxy
$$\widetilde F_t(T):=\mathbb E_{\mathbb Q}\left[\langle\ell,\widehat{\mathbb X}_T^{\le n}\rangle\mid\mathcal F_t\right].$$
For every relevant futures maturity $T$, we require $0<m<\widetilde F_0(T)$. Let $\tau_m(T)$ be the stopping time defined in Definition~\ref{deflowm} and set
$$\widetilde F_t^{m}(T):=\widetilde F_{t\wedge\tau_m(T)}(T).$$
The VXX proxy is therefore constructed from these stopped futures and we retain the superscript $m$ throughout this subsection to identify stopped quantities. On the simulation grid, continuous-time stopping is approximated by freezing each contract at $m$ after its first observed value at or below $m$. So far we have introduced tractable proxies for the corresponding $\VIX$, $\VIX$ futures and $\VXX$ quantities generated by the regularized model. The approximation gap between the continuous-time proxies and the exact model values is controlled by the bounds derived in Section~\ref{relaxing}.

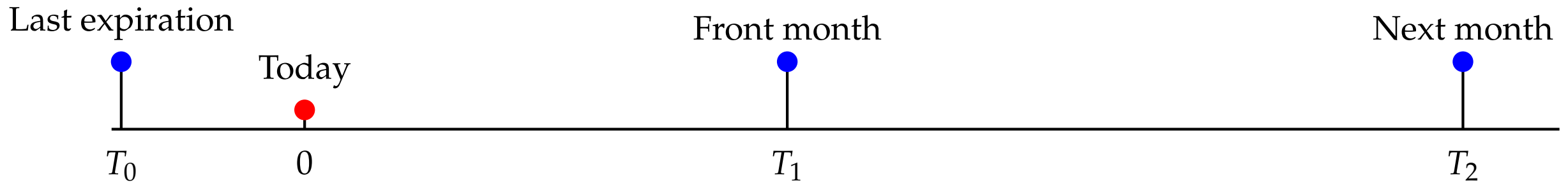
\begin{figure}
    \begin{tikzpicture}
        \draw[thick] (0,0) -- (15,0);
        \draw[thick] (0.1,0) -- (0.1,0.7);       
        \filldraw[blue] (0.1,0.7) circle (0.1); 
        \node[below=0.1cm] at (0.1,0) {$T_{0}$};  
        \node[above=0.1cm] at (0.1,0.7) {Last expiration}; 
        \draw[thick] (2,0) -- (2,0.2);       
        \filldraw[red] (2,0.2) circle (0.1); 
        \node[below=0.1cm] at (2,0) {$0$};  
        \node[above=0.1cm] at (2,0.2) {Today}; 
        \draw[thick] (7,0) -- (7,0.7);       
        \filldraw[blue] (7,0.7) circle (0.1); 
        \node[below=0.1cm] at (7,0) {$T_{1}$};  
        \node[above=0.1cm] at (7,0.7) {Front month}; 
        \draw[thick] (14,0) -- (14,0.7);       
        \filldraw[blue] (14,0.7) circle (0.1); 
        \node[below=0.1cm] at (14,0) {$T_{2}$};  
        \node[above=0.1cm] at (14,0.7) {Next month}; 
    \end{tikzpicture}
            \caption{Enumeration of VIX futures maturities}\label{fig2}
    \end{figure}

Fix an injective labeling function $\mathcal{L}:\{I:|I|\leq 2n\}\to\{1,\ldots,(d+1)_{2n}\}$, and let $G$ be the $(d+1)_{2n}$-dimensional matrix representative of the dual operator corresponding to $\wh{\X}^{\leq 2n}$. Set 
$$A:=\int_0^\Delta e^{sG^\T}\dd{s}$$
and verify that it is invertible, for instance by applying Lemma~\ref{leminv}; here $G$ acts on the level-$2n$ signature. With a small abuse of notation,  denote by $G$ also the corresponding dual-generator matrix on the level-$n$ signature. Let $T_0\leq 0<T_1<T_2<\cdots$ denote the maturities of $\VIX$ futures, where $T_0$ is the most recently expired maturity, $0$ represents ``today'', that is, the calibration date, and $T_1$ and $T_2$ are, respectively, the first and second upcoming expiration dates; see Figure~\ref{fig2}.
The tractable proxy for the VIX process is given by
$$\overline\VIX_T(\ell)=|\ve(\ell)^\T \ve(\widehat{\X}^{\leq n}_T)|$$
while the instantaneous variance is given by
$$\sigma_t^S(\ell)^2=\left(\ve(h(\ell))^\T \ve(\widehat{\X}^{\leq 2n}_t)\right)^+,\qquad \text{with} \qquad\ve(h(\ell))^\T = \Delta \cdot \ve(\ell \shuffle \ell)^\T A^{-1}.$$
As a direct consequence, the stock price process admits the exponential representation
$$S_T(\ell)=S_0\exp\Bigg(\int_0^T \sqrt{(\ve(h(\ell))^\T \ve(\widehat{\X}^{\leq 2n}_t))^+}\dd{B}_t-\frac{1}{2}\int_0^T (\ve(h(\ell))^\T \ve(\widehat{\X}^{\leq 2n}_t))^+\dd t\Bigg).$$
The signed futures proxy admits the representation
$$\widetilde{F}_t(T)=\ve(\ell)^\T e^{(T-t)G^\T}\ve(\widehat{\mathbb X}_t^{\le n}).$$
The futures used in the $\VXX$ roll are obtained by applying the stopping rule of Definition~\ref{deflowm} to these futures paths.
Let $\tau_i:=T_i-T_{i-1},$ for each  $i\geq 1$. The proxy for the VXX ETN evaluated at time $T\leq T_2$ is then given by
$$\widetilde \VXX^m_T=\VXX_0\,\frac{(T-T_0)\widetilde F^m_T(T_2)+(T_1-T)\widetilde F^m_T(T_1)}{-T_0\widetilde F^m_0(T_2)+T_1\widetilde F^m_0(T_1)}\exp\!\left(\int_0^T\frac{\widetilde F^m_s(T_1)-\widetilde F^m_s(T_2)}{(s-T_0)\widetilde F^m_s(T_2)+(T_1-s)\widetilde F^m_s(T_1)}\dd{s}\right)$$
for $0\leq T\leq T_1$, whereas, for $T\in(T_1,T_2]$,
\begin{align*}
    \widetilde \VXX^m_T&=\VXX_0\,\frac{\tau_1}{\tau_2}\,\frac{(T-T_1)\widetilde F^m_T(T_3)+(T_2-T)\widetilde F^m_T(T_2)}{-T_0\widetilde F^m_0(T_2)+T_1\widetilde F^m_0(T_1)}\\
    &\qquad\times\exp\!\left(\int_0^{T_1}\frac{\widetilde F^m_s(T_1)-\widetilde F^m_s(T_2)}{(s-T_0)\widetilde F^m_s(T_2)+(T_1-s)\widetilde F^m_s(T_1)}\dd{s}+\int_{T_1}^{T}\frac{\widetilde F^m_s(T_2) -\widetilde F^m_s(T_3)}{(s-T_1)\widetilde F^m_s(T_3)+(T_2-s)\widetilde F^m_s(T_2)}\dd{s}\right).
\end{align*}
Approximation errors for futures and VXX values are bounded in Lemmas~\ref{errFut} and \ref{errVXX}, respectively; the corresponding call-price bounds are given in Corollary~\ref{cor:option-price-errors}.
\begin{remark}
To price a VXX option with maturity $T$, each required futures proxy is tracked from time zero up to the earlier of its expiry and $T$ and its stopping history is retained across roll dates. The required futures maturities are identified as follows: we define $\mathcal M^{\VXX}:=\left\{T:\ (T,K)\in\mathcal K^{\VXX}\text{ for some }K\right\}$ and then set
\begin{equation}\label{Tcal}
    \Tcal:=\bigcup_{T\in\mathcal M^{\VXX}}\{T_1,\ldots,T_{m(T)+1}\},\qquad m(T):=\min\{i\geq1:T\leq T_i\}.
\end{equation}
\end{remark}
\begin{remark}
We remark that the signature coefficients are expressed in decimal volatility units, while VIX option strikes and prices are quoted in index points. In a (regularized) VIX-based signature model, pricing is performed through Monte Carlo. Let $N_{MC}$ denote the number of simulated paths used for a pricing evaluation; we use $N_{MC}=10\,000$ during calibration and $N_{MC}=10^6$ for independent-path validation. The pricing formulas are
\begin{align*}
    \pi_{\SPX}^{\text{model}}(\ell,\rho,T,K)&=\E_\Q[e^{-rT}(e^{(r-q)T} S_T(\ell,\rho)-K)^+]\approx\frac{e^{-rT}}{N_{MC}}\sum_{i=1}^{N_{MC}}(e^{(r-q)T}S_T(\ell,\rho,\omega_i)-K)^+\\
    \pi^\mo_\VIX(\ell,T,K)&=\E_\Q[e^{-rT}(100\cdot{\overline \VIX}_T-K)^+]\approx\frac{e^{-rT}}{N_{MC}}\sum_{i=1}^{N_{MC}}(100\cdot{\overline \VIX}_T(\ell,\omega_i)-K)^+\\
    \pi^\mo_\VXX(\ell,T,K)&=\E_\Q[e^{-rT}(e^{(r-c)T} {\widetilde \VXX}^m_T(\ell)-K)^+]\approx\frac{e^{-rT}}{N_{MC}}\sum_{i=1}^{N_{MC}}(e^{(r-c)T}{\widetilde \VXX}^m_T(\ell,\omega_i)-K)^+\\
    F^\mo_\VIX(\ell,T)&=100\cdot\ve(\ell)^\T e^{T G^\T}\ve(e_\emptyset)
\end{align*}
The raw signature functionals are evaluated through scalar products; the positive part, absolute value and stopping operations are subsequently applied pathwise. We use a uniform grid with $n_{\mathrm{day}}$ time steps per day:
$$ t_j=j\Delta t,\qquad \Delta t=\frac{1}{365n_{\mathrm{day}}},\qquad N_T=\frac{T}{\Delta t}\in\mathbb N\quad\Rightarrow\quad t_{N_T}=T.$$
In the numerical examples, we set $n_{\mathrm{day}}=4$.
\end{remark}
\noindent Summarizing, for the joint calibration of a VIX-based signature model of the three implied volatility surfaces we proceed as follows.
\begin{itemize}
    \item For each underlying $A$, collect the bid and ask prices of calls at the selected maturity--strike pairs in $\mathcal{K}^A$. Compute the bid and ask implied volatilities by inverting these prices under Black--76 using the reference forward $F_A^{\mathrm{mkt}}(T)$ and the discount factor $e^{-rT}$. For SPX, infer $F_{\mathrm{SPX}}^{\mathrm{mkt}}(T)$ from European put--call parity. Let $C_{T,K}^{\mathrm{mid}}$ and $P_{T,K}^{\mathrm{mid}}$ denote the call and put midquotes at maturity $T$ and strike $K$. For each maturity, select the three strikes with the smallest values of $\lvert C_{T,K}^{\mathrm{mid}}-P_{T,K}^{\mathrm{mid}}\rvert$ and take the median of the corresponding parity-implied forwards $K+e^{rT}(C_{T,K}^{\mathrm{mid}}-P_{T,K}^{\mathrm{mid}})$. The implied dividend yield used in the SPX pricing formula is
    $$q(T)=r-\frac{1}{T}\log\!\left(\frac{F_{\mathrm{SPX}}^{\mathrm{mkt}}(T)}{S_0}\right),$$
    so that 
    $$S_0e^{(r-q(T))T}=F_{\mathrm{SPX}}^{\mathrm{mkt}}(T).$$
    For VIX, set $F_{\mathrm{VIX}}^{\mathrm{mkt}}(T)$ equal to the quoted price of the VIX futures contract expiring at $T$. The same quote serves as the benchmark for the futures-price error in the VIX loss. For VXX, use $F_{\mathrm{VXX}}^{\mathrm{mkt}}(T)=\mathrm{VXX}_0e^{rT}$. With this reference forward, Black--76 inversion is equivalent to Black--Scholes inversion at spot $\mathrm{VXX}_0$ with $q=0$. The fee $c$ enters the model option prices through the simulated terminal value of VXX. Finally, evaluate the Black--76 vega and, where required, the call forward delta at the corresponding reference forward and at the arithmetic mean of the bid and ask implied volatilities.
    \item Choose the parameters of the primary process $X$ and sample the trajectories of its time-extended truncated signature $\widehat \X^{\leq 2n}$.  Note that to store them one needs a 3-dimensional tensor $T_X$ such that
    $$T_X[j,i,k] = \ve(\widehat \X^{\leq 2n}_{t_j}(\omega_i))_k.$$
    To construct $B$ as in \eqref{eq:martingale-preserving-correlation} we simulate the factor Brownian increments together with the increments of an auxiliary Brownian motion $W^\perp$, independent of the factor drivers. Store
    $$T_W[j,i,:]=\bigl(\Delta W_j^1(\omega_i),\ldots,\Delta W_j^d(\omega_i),\Delta W_j^\perp(\omega_i)\bigr).$$
    This tensor has time, path and component axes. The correlation vector $\rho$ is calibrated jointly with $\ell$. At each objective evaluation, the increments of $B$ are reconstructed from \eqref{eq:martingale-preserving-correlation} using the current values of $\ell$ and $\rho$. The underlying simulated drivers remain fixed throughout the optimization. Simulation and storage of the driving increments are performed offline; their parameter-dependent combination into the price driver is evaluated during optimization.
    \item Compute matrices $A^{-1}$ and $G$ and store them.
    \item Implement functions $\pi^\mo_\SPX(\ell,\rho,T,K),\pi^\mo_\VIX(\ell,T,K),\pi^\mo_\VXX(\ell,T,K),F^\mo_\VIX(\ell,T)$ for the selected maturities and strikes, taking as input $T_X,T_W,A^{-1},G,\Delta,\Tcal, r,q $ and $\ve(\ell)$.
    \item Minimize the joint loss over the signature coefficients and the correlation parameters. On the calibration sample, we impose
    $$\langle h(\ell),\widehat{\mathbb X}_{t_j}^{\le2n}(\omega_i)\rangle\ge \varepsilon_{\mathrm{var}}$$
    at every simulated node up to the chosen positivity horizon; we also verify that
    $$m<\widetilde F_0(T_i)\quad\text{for every relevant maturity},\qquad\rho^\T R^{-1}\rho<1.$$
    Futures are regularized through stopping, whereas SPX pricing uses the positive part of the variance. These sampled constraints do not imply global nonnegativity, and the model used for pricing therefore remains the regularized model. Independent-sample diagnostics and Monte Carlo estimates of the approximation bounds are computed separately for the concrete specifications of Section~\ref{Calibration to option prices}.
\end{itemize}
Set
$u_t:=\Delta A^{-1}\ve(\widehat{\mathbb X}_t^{\le2n})$ and $b_\ell:=\operatorname{vec}(\ell\shuffle\ell).$
The vectors $u_t$ can be precomputed offline. The raw variance and the pricing variance are respectively
$$v_t^{\mathrm{raw}}=b_\ell^\T u_t, \qquad (\sigma_t^S)^2=(b_\ell^\T u_t)^+.$$
On a uniform grid, the stock price is approximated by
$$\log\frac{S_T}{S_0}\approx\sum_{j=0}^{N_T-1}\sqrt{(b_\ell^\T u_{t_j})^+}\,\Delta B_j-\frac12\sum_{j=0}^{N_T-1}(b_\ell^\T u_{t_j})^+\Delta t.$$
The increments $\Delta B_j$ are reconstructed from \eqref{eq:martingale-preserving-correlation} for each candidate parameter pair $(\ell,\rho)$. When the raw variance is nonnegative at all relevant nodes, the drift sum reduces to a single scalar product; otherwise, the positive part must be applied before summing up. The propagators and signature paths used to evaluate raw
futures proxies can also be precomputed. Their stopping histories are subsequently computed pathwise for each candidate coefficient vector.

\end{appendix}


\begin{thebibliography}{40}
\providecommand{\natexlab}[1]{#1}
\providecommand{\url}[1]{\texttt{#1}}
\expandafter\ifx\csname urlstyle\endcsname\relax
  \providecommand{\doi}[1]{doi: #1}\else
  \providecommand{\doi}{doi: \begingroup \urlstyle{rm}\Url}\fi

\bibitem[Abi~Jaber et~al.(2023)Abi~Jaber, Illand, and Li]{AJIL:22}
Eduardo Abi~Jaber, Camille Illand, and Shaun~Xiaoyuan Li.
\newblock The quintic {Ornstein--Uhlenbeck} model for joint {SPX} and {VIX} calibration.
\newblock \emph{Risk Magazine}, June 2023.
\newblock URL \url{https://www.risk.net/cutting-edge/banking/7956893/the-quintic-ornstein-uhlenbeck-model-for-joint-spx-and-vix-calibration}.
\newblock Cutting Edge section, 8 June 2023.

\bibitem[Abi~Jaber et~al.(2025)Abi~Jaber, Gassiat, and Sotnikov]{AGS:25}
Eduardo Abi~Jaber, Paul Gassiat, and Dimitri Sotnikov.
\newblock Martingale property and moment explosions in signature volatility models.
\newblock \emph{arXiv preprint arXiv:2503.17103}, 2025.
\newblock URL \url{https://arxiv.org/abs/2503.17103}.

\bibitem[Ackerer et~al.(2018)Ackerer, Filipovi{\'c}, and Pulido]{AFP:18}
Damien Ackerer, Damir Filipovi{\'c}, and Sergio Pulido.
\newblock The {Jacobi} stochastic volatility model.
\newblock \emph{Finance and Stochastics}, 22\penalty0 (3):\penalty0 667--700, 2018.
\newblock \doi{10.1007/s00780-018-0364-8}.

\bibitem[Alexander and Korovilas(2013)]{AlexanderKorovilas:13}
Carol Alexander and Dimitris Korovilas.
\newblock Volatility exchange-traded notes: curse or cure?
\newblock \emph{Journal of Alternative Investments}, 16\penalty0 (2):\penalty0 52--70, 2013.
\newblock \doi{10.3905/jai.2013.16.2.052}.

\bibitem[Avellaneda and Papanicolaou(2019)]{AvellanedaPapanicolaou:17}
Marco Avellaneda and Andrew Papanicolaou.
\newblock Statistics of {VIX} futures and applications to trading volatility exchange-traded products.
\newblock \emph{International Journal of Theoretical and Applied Finance}, 22\penalty0 (1):\penalty0 1850061, 2019.
\newblock \doi{10.1142/S0219024918500619}.

\bibitem[Badran and Goldys(2015)]{BG:15}
Alexander Badran and Beniamin Goldys.
\newblock A market model for {VIX} futures.
\newblock arXiv preprint arXiv:1504.00428, 2015.
\newblock URL \url{https://arxiv.org/abs/1504.00428}.

\bibitem[Baldeaux and Badran(2014)]{BB:14}
Jan Baldeaux and Alexander Badran.
\newblock Consistent modelling of {VIX} and equity derivatives using a 3/2 plus jumps model.
\newblock \emph{Applied Mathematical Finance}, 21\penalty0 (4):\penalty0 299--312, 2014.
\newblock \doi{10.1080/1350486X.2013.868631}.

\bibitem[{Barclays Bank PLC}(2018)]{BarclaysVXX:2018}
{Barclays Bank PLC}.
\newblock Pricing supplement: ipath series b s\&p 500 vix short-term futures etn, 2018.
\newblock URL \url{https://www.ipathetn.com/us/16/en/details.app?instrumentId=341408}.
\newblock Product prospectus documentation.

\bibitem[Bayer et~al.(2023)Bayer, Hager, Riedel, and Schoenmakers]{BHR:23}
Christian Bayer, Paul~P. Hager, Sebastian Riedel, and John Schoenmakers.
\newblock Optimal stopping with signatures.
\newblock \emph{The Annals of Applied Probability}, 33\penalty0 (1):\penalty0 238--273, 2023.
\newblock \doi{10.1214/22-AAP1814}.

\bibitem[Bayer et~al.(2026)Bayer, dos Reis, Horvath, and Oberhauser]{sigbook}
Christian Bayer, Gon{\c{c}}alo dos Reis, Blanka Horvath, and Harald Oberhauser, editors.
\newblock \emph{Signature Methods in Finance: An Introduction with Computational Applications}.
\newblock Springer, 2026.
\newblock \doi{10.1007/978-3-031-97239-3}.

\bibitem[Carr and Wu(2006)]{CW:06}
Peter Carr and Liuren Wu.
\newblock A tale of two indices.
\newblock \emph{The Journal of Derivatives}, 13\penalty0 (3):\penalty0 13--29, 2006.
\newblock \doi{10.3905/jod.2006.616868}.

\bibitem[{Cboe Global Indices, LLC}(2026)]{CboeVIX2026}
{Cboe Global Indices, LLC}.
\newblock {Cboe Volatility Index Methodology}.
\newblock Technical report, Cboe Global Indices, LLC, 2026.
\newblock URL \url{https://cdn.cboe.com/resources/vix/VIX_Methodology.pdf}.
\newblock Version 6.0, February 26, Section 3.1.

\bibitem[Chen(1954)]{Chen:54}
Kuo-Tsai Chen.
\newblock Iterated integrals and exponential homomorphisms.
\newblock \emph{Proceedings of the London Mathematical Society}, s3-4\penalty0 (1):\penalty0 502--512, 1954.
\newblock \doi{10.1112/plms/s3-4.1.502}.

\bibitem[Cont and Kokholm(2013)]{CK:13}
Rama Cont and Thomas Kokholm.
\newblock A consistent pricing model for index options and volatility derivatives.
\newblock \emph{Mathematical Finance}, 23\penalty0 (2):\penalty0 248--274, 2013.
\newblock \doi{10.1111/j.1467-9965.2011.00492.x}.

\bibitem[Cuchiero et~al.(2012)Cuchiero, Keller-Ressel, and Teichmann]{CKT:12}
Christa Cuchiero, Martin Keller-Ressel, and Josef Teichmann.
\newblock Polynomial processes and their applications to mathematical finance.
\newblock \emph{Finance and Stochastics}, 16\penalty0 (4):\penalty0 711--740, 2012.
\newblock \doi{10.1007/s00780-012-0188-x}.

\bibitem[Cuchiero et~al.(2023)Cuchiero, Gazzani, and Svaluto-Ferro]{CGS:23}
Christa Cuchiero, Guido Gazzani, and Sara Svaluto-Ferro.
\newblock Signature-based models: Theory and calibration.
\newblock \emph{SIAM Journal on Financial Mathematics}, 14\penalty0 (3):\penalty0 910--957, 2023.
\newblock \doi{10.1137/22M1512338}.

\bibitem[Cuchiero et~al.(2025)Cuchiero, Gazzani, M{\"o}ller, and Svaluto-Ferro]{CGMS:24}
Christa Cuchiero, Guido Gazzani, Janka M{\"o}ller, and Sara Svaluto-Ferro.
\newblock Joint calibration to {SPX} and {VIX} options with signature-based models.
\newblock \emph{Mathematical Finance}, 35\penalty0 (1):\penalty0 161--213, 2025.
\newblock \doi{10.1111/mafi.12442}.

\bibitem[Cuchiero et~al.(2026)Cuchiero, Svaluto-Ferro, and Teichmann]{CST:23}
Christa Cuchiero, Sara Svaluto-Ferro, and Josef Teichmann.
\newblock Signature {SDEs} from an affine and polynomial perspective.
\newblock \emph{Electronic Journal of Probability}, 31:\penalty0 1--57, 2026.
\newblock \doi{10.1214/26-EJP1553}.

\bibitem[Duffie et~al.(2003)Duffie, Filipovi{\'c}, and Schachermayer]{DFS:03}
Darrell Duffie, Damir Filipovi{\'c}, and Walter Schachermayer.
\newblock Affine processes and applications in finance.
\newblock \emph{The Annals of Applied Probability}, 13\penalty0 (3):\penalty0 984--1053, 2003.
\newblock \doi{10.1214/aoap/1060202833}.

\bibitem[Filipovi{\'c} and Larsson(2016)]{FL:16}
Damir Filipovi{\'c} and Martin Larsson.
\newblock Polynomial diffusions and applications in finance.
\newblock \emph{Finance and Stochastics}, 20\penalty0 (4):\penalty0 931--972, 2016.
\newblock \doi{10.1007/s00780-016-0304-4}.

\bibitem[Forde and Smith(2023)]{FS:23}
Martin Forde and Benjamin Smith.
\newblock Markovian stochastic volatility with stochastic correlation --- joint calibration and consistency of {SPX/VIX} short-maturity smiles.
\newblock \emph{International Journal of Theoretical and Applied Finance}, 26\penalty0 (2--3):\penalty0 2350007, 2023.
\newblock \doi{10.1142/S0219024923500073}.

\bibitem[Fouque and Saporito(2018)]{FS:18}
Jean-Pierre Fouque and Yuri~F. Saporito.
\newblock Heston stochastic vol-of-vol model for joint calibration of {VIX} and {S\&P 500} options.
\newblock \emph{Quantitative Finance}, 18\penalty0 (6):\penalty0 1003--1016, 2018.
\newblock \doi{10.1080/14697688.2017.1412493}.

\bibitem[Friz and Victoir(2010)]{FV:10}
Peter~K. Friz and Nicolas~B. Victoir.
\newblock \emph{Multidimensional Stochastic Processes as Rough Paths: Theory and Applications}, volume 120 of \emph{Cambridge Studies in Advanced Mathematics}.
\newblock Cambridge University Press, 2010.
\newblock \doi{10.1017/CBO9780511845079}.

\bibitem[Gatheral et~al.(2020)Gatheral, Jusselin, and Rosenbaum]{GJR:20}
Jim Gatheral, Paul Jusselin, and Mathieu Rosenbaum.
\newblock The quadratic rough {Heston} model and the joint {S\&P} 500/{VIX} smile calibration problem.
\newblock \emph{Risk}, May 2020.
\newblock URL \url{https://www.risk.net/cutting-edge/banking/7530461/the-quadratic-rough-heston-model-and-the-joint-sp-500vix-smile-calibration-problem}.

\bibitem[Grasselli and Wagalath(2020)]{GW:20}
Martino Grasselli and Lakshithe Wagalath.
\newblock {VIX} versus {VXX}: A joint analytical framework.
\newblock \emph{International Journal of Theoretical and Applied Finance}, 23\penalty0 (05):\penalty0 2050033, 2020.
\newblock \doi{10.1142/S0219024920500338}.

\bibitem[Grasselli et~al.(2024)Grasselli, Mazzoran, and Pallavicini]{GMP:24}
Martino Grasselli, Andrea Mazzoran, and Andrea Pallavicini.
\newblock A general framework for a joint calibration of {VIX} and {VXX} options.
\newblock \emph{Annals of Operations Research}, 336\penalty0 (1):\penalty0 3--26, 2024.
\newblock \doi{10.1007/s10479-023-05205-9}.

\bibitem[Guo et~al.(2022)Guo, Loeper, Obl{\'o}j, and Wang]{GLOW:22}
Ivan Guo, Gr{\'e}goire Loeper, Jan Obl{\'o}j, and Shiyi Wang.
\newblock Joint modeling and calibration of {SPX} and {VIX} by optimal transport.
\newblock \emph{SIAM Journal on Financial Mathematics}, 13\penalty0 (1):\penalty0 1--31, 2022.
\newblock \doi{10.1137/20M1375905}.

\bibitem[Guyon(2024)]{Guyon:24}
Julien Guyon.
\newblock Dispersion-constrained martingale {Schr{\"o}dinger} problems and the exact joint {S\&P} 500/{VIX} smile calibration puzzle.
\newblock \emph{Finance and Stochastics}, 28:\penalty0 27--79, 2024.
\newblock \doi{10.1007/s00780-023-00524-y}.

\bibitem[Guyon and Lekeufack(2023)]{GL:23}
Julien Guyon and Jordan Lekeufack.
\newblock Volatility is (mostly) path-dependent.
\newblock \emph{Quantitative Finance}, 23\penalty0 (9):\penalty0 1221--1258, 2023.
\newblock \doi{10.1080/14697688.2023.2221281}.

\bibitem[Guyon and Mustapha(2023)]{GM:22}
Julien Guyon and Scander Mustapha.
\newblock Neural joint {S\&P 500}/{VIX} smile calibration.
\newblock \emph{Risk}, 2023.
\newblock URL \url{https://www.risk.net/cutting-edge/7958175/neural-joint-sp-500vix-smile-calibration}.

\bibitem[Lin and Chang(2010)]{LC:10}
Yueh-Neng Lin and Chih-Fang Chang.
\newblock A new closed-form model for {VIX} futures and options.
\newblock \emph{Journal of Banking \& Finance}, 34\penalty0 (6):\penalty0 1350--1362, 2010.
\newblock \doi{10.1016/j.jbankfin.2009.11.001}.

\bibitem[Lions and Musiela(2007)]{LM:07}
Pierre-Louis Lions and Marek Musiela.
\newblock Correlations and bounds for stochastic volatility models.
\newblock \emph{Annales de l'Institut Henri Poincar{\'e} C, Analyse Non Lin{\'e}aire}, 24\penalty0 (1):\penalty0 1--16, 2007.
\newblock \doi{10.1016/j.anihpc.2005.05.007}.

\bibitem[Lu(2023)]{Lu:23}
Shan Lu.
\newblock Joint calibration of {VIX} and {VXX} options: does volatility clustering matter?
\newblock \emph{The European Journal of Finance}, pages 1--32, 2023.
\newblock \doi{10.1080/1351847X.2023.2297042}.
\newblock Published online 30 December 2023.

\bibitem[Lu(2025)]{Lu:25}
Shan Lu.
\newblock Pricing {VXX} options with observable volatility dynamics from high-frequency {VIX} index.
\newblock \emph{Journal of Futures Markets}, 45\penalty0 (7):\penalty0 771--801, 2025.
\newblock \doi{10.1002/fut.22591}.

\bibitem[Lyons et~al.(2020)Lyons, Nejad, and Perez~Arribas]{LNA:20}
Terry Lyons, Sina Nejad, and Imanol Perez~Arribas.
\newblock Non-parametric pricing and hedging of exotic derivatives.
\newblock \emph{Applied Mathematical Finance}, 27\penalty0 (6):\penalty0 457--494, 2020.
\newblock \doi{10.1080/1350486X.2021.1891555}.

\bibitem[Lyons(1998)]{Lyons:98}
Terry~J. Lyons.
\newblock Differential equations driven by rough signals.
\newblock \emph{Revista Matem{\'a}tica Iberoamericana}, 14\penalty0 (2):\penalty0 215--310, 1998.
\newblock \doi{10.4171/RMI/240}.

\bibitem[Lyons et~al.(2007)Lyons, Caruana, and L{\'e}vy]{LCL:07}
Terry~J. Lyons, Michael Caruana, and Thierry L{\'e}vy.
\newblock \emph{Differential Equations Driven by Rough Paths}, volume 1908 of \emph{Lecture Notes in Mathematics}.
\newblock Springer, Berlin, 2007.
\newblock \doi{10.1007/978-3-540-71285-5}.

\bibitem[Menc{\'i}a and Sentana(2013)]{MS:13}
Javier Menc{\'i}a and Enrique Sentana.
\newblock Valuation of {VIX} derivatives.
\newblock \emph{Journal of Financial Economics}, 108\penalty0 (2):\penalty0 367--391, 2013.
\newblock \doi{10.1016/j.jfineco.2012.11.001}.

\bibitem[Papanicolaou(2022)]{P:22}
Andrew Papanicolaou.
\newblock {Consistent time-homogeneous modeling of SPX and VIX derivatives}.
\newblock \emph{Mathematical Finance}, 32\penalty0 (3):\penalty0 907--940, 2022.
\newblock \doi{https://doi.org/10.1111/mafi.12348}.
\newblock URL \url{https://onlinelibrary.wiley.com/doi/abs/10.1111/mafi.12348}.

\bibitem[Papanicolaou and Sircar(2014)]{PS:14}
Andrew Papanicolaou and Ronnie Sircar.
\newblock A regime-switching {Heston} model for {VIX} and {S\&P 500} implied volatilities.
\newblock \emph{Quantitative Finance}, 14\penalty0 (10):\penalty0 1811--1827, 2014.
\newblock \doi{10.1080/14697688.2013.814923}.

\end{thebibliography}
\end{document}